\documentclass[equations,12pt]{article}
\usepackage{geometry}
\usepackage{subfigure}
\makeatletter
\newcommand\ackname{Acknowledgements}
\if@titlepage
  \newenvironment{acknowledgements}{
      \titlepage
     \null\vfil
     \@beginparpenalty\@lowpenalty
      \begin{center}
        \bfseries \ackname
        \@endparpenalty\@M1
      \end{center}}
     {\par\vfil\null\endtitlepage}
\else
  \newenvironment{acknowledgements}{
      \if@twocolumn
        \section*{\abstractname}
     \else
        \small
       \begin{center}
         {\bfseries \ackname\vspace{-.5em}\vspace{\z@}}
       \end{center}
        \quotation
      \fi}
      {\if@twocolumn\else\endquotation\fi}
\usepackage{mathrsfs}
\usepackage{eucal}
\usepackage{amssymb,amsmath}

\def\undertilde#1{\mathop{\vtop{\ialign{##\crcr
    \upbrall\crcr\noalign{\kern1pt
   }}}}\limits}
\def\underhat#1{\mathop{\vtop{\ialign{##\crcr
     \upbrall\crcr\noalign{\kern1pt
   }}}}\limits}
\usepackage{pgf}
\usepackage{tikz}
\usepackage{cite}
\usepackage{graphicx}
\usepackage{color}
\usepackage{amssymb}
\usepgflibrary{arrows}
\usetikzlibrary{calc,angles,positioning,intersections,quotes,backgrounds,decorations.pathreplacing}
\usetikzlibrary{decorations.markings}
\usepackage{amsmath}
\usepackage{amsthm}
\newtheorem{theorem}{Theorem}
\newtheorem*{remark}{Remark}
\theoremstyle{remark}
\newcommand{\nn}{\nonumber}
\begin{document}
\title{\textbf{ \textsf{Integrable Hamiltonian Hierarchies and Lagrangian 1-Forms}}}
\author{Chisanupong Puttarprom $^{\dagger, +}$, Worapat Piensuk$^{-}$,  Sikarin Yoo-Kong$^{* } $ \\
\small $^\dagger $\emph{Theoretical and Computational Physics (TCP) Group, Department of Physics,}\\ 
\small \emph{Faculty of Science, King Mongkut's University of Technology Thonburi, Thailand, 10140.}\\
\small $-$ \emph{Department of Mechanical Engineering, Faculty of Engineering, }\\
 \small \emph{King Mongkut's University of Technology Thonburi, Thailand, 10140}\\
\small $^+$\emph{Theoretical and Computational Science Center (TaCS),}\\ \small\emph{Faculty of Science, King Mongkut's University of Technology Thonburi, Thailand, 10140.}\\
\small $^* $ \emph{The Institute for Fundamental Study (IF),} \\ \small\emph{Naresuan University, Phitsanulok, Thailand, 65000.}\\
}
\maketitle
\abstract
We present further developments on the Lagrangian 1-form description for one-dimensional integrable systems in both discrete and continuous levels. A key feature of integrability in this context called a closure relation will be derived from the local variation of the action on the space of independent variables. The generalised Euler-Lagrange equations and constraint equations are derived directly from the variation of the action on the space of dependent variables. This set of Lagrangian equations gives rise to a crucial property of integrable systems known as the multidimensional consistency. Alternatively, the closure relation can be obtained from generalised Stokes' theorem exhibiting a path independent property of the systems on the space of independent variables. The homotopy structure of paths suggests that the space of independent variables is simply connected. Furthermore, the N\"{o}ether charges, invariants in the context of Liouville integrability, can be obtained directly from the non-local variation of the action on the space of dependent variables.
\section{Introduction}
It is well-known that integrable systems are exceptionally rare since most (nonlinear) differential equations exhibit chaotic behaviour and no explicit solutions can be obtained. Informally, integrability is the property of a system that allows the solution to be solved in finite steps of operations (or integrations). In more simply speaking, integrability enables us to solve the set of equations in a closed form or in terms of quadratures (ordinary integrals). In classical mechanics, especially Hamiltonian systems with $N$ degrees of freedom, integrability has a direct connection to action-angle variables. It is well-understood that the choice of coordinates on phase space is not unique since one can transform an old set of coordinates to a new set of coordinates through the canonical transformation preserving the Hamilton's equations. Of course, action-angle coordinates are special since the Hamiltonian depends solely on action variables which are constants of motion and, consequently, the angle variables are cyclic, evolving linearly in time. With this feature, the complexity of the problem will be reduced and explicit solutions can be determined by quadratures. In this sense, the existence of the action-angle variables guarantees the integrability of the system. However, finding action-angle coordinates is not trivial in practice. Then, the notion of integrability is rather defined in terms of existence of the invariants known as Liouville integrability \cite{ARNO}. In this notion, the Hamiltonian system with $N$ degrees of freedom, whose evolution is on 2$N$-dimensional manifold embedded in phase space, is integrable if there exists $N$ invariants, which are normally treated as Hamiltonians, that are independent and in involution as the Poisson brackets for every pair of invariants vanish. With involution of invariants, all evolutions belong to the same level set and are mutually commute, known as commuting Hamiltonian flows which are considered to be the main feature for integrable Hamiltonian systems. Furthermore, there exists a canonical transformation such that a set of invariants is nothing but a set of action variables. This means that 2$N$-dimensional manifold can be foliated into a $N$-dimensional invariant torus in which the angle variables are naturally the periodic coordinates on this torus.
\\
\\
In discrete world, the differential equations are generalised to their discrete counterparts known as the difference equations. Under the continuum limit, it happens that many distinct difference equations could be reduced to the same differential equation. In other words, the discrete analogue for a single differential equation is not unique. This would make a difficulty to sort out what is a preferable difference equation for a particular physical system. At this point, the notion of integrability may help to choose the right equation, which is a solvable equation, from many non-solvable ones. For the Hamiltonian systems, the discrete analogue of the Liouville integrability can be naturally constructed \cite{IN1}. However, there are several notions on what integrability would mean in the discrete level, e.g. existence of a Lax pair \cite{IN2}, singularity confinement\cite{IN3} and algebraic entropy \cite{IN4}. One of the remarkable aspects of integrable many-dimensional discrete systems is a multidimensional consistency \cite{FrankB}. This feature allows us to embed the difference equation in a multidimensional lattice consistently, i.e., there exists a set of (infinite) compatible equations defined in each subspace corresponding to the number of independent variables \cite{F20011,F20012,F20021,RB,RB2}. Two-dimensional lattice system embedded in a three-dimensional space is said to be consistent in such way if the quadrilateral equations describing three side-to-side connected surfaces of a cube can be solved and yield the coincide result with given initial values, as discussed in \cite{SF1}. This consistency is called the consistency-around-the-cube (CAC) or 3D-consistency, which has led Adler, Bobenko and Suris to classify the quadrilateral equations of the two-dimensional lattice systems into the remarkable ABS list, see \cite{FSU1}.
\\
\\
In Liouville integrability, the Hamiltonian is a main object. However, alternative object called the Lagrangian can be chosen to work with resulting the same physics. According to the least action principle, the system evolves along the trajectory on the configuration space that the action functional is extremum.
It is known that most integrable discrete systems admit Lagrangian description. Therefore, it is quite natural to consider the evolution of the system through the discrete path on the space of dependent variables as well as that on the space of independent variables. Since the integrable discrete equations, exhibiting such consistency, are obtained from Lagrangian, it is essential for multidimensional consistency to be encoded on the level of Lagrangians resulting in a new notion of integrability known as the Lagrangian multiform. The pioneer works on Lagrangian 2-form and 3-form were initiated by Lobb and Nijhoff \cite{SF1}. In this context, the action functional is invariant under the local deformation of the discrete path on the space of independent variables leading to a intriguing feature called the closure relation. This means that there are many discrete paths on the space of independent variables corresponding to a single discrete path, whose the action is critical, on the space of dependent variables. The case of Lagrangian 1-form was later developed by Yoo-Kong, Lobb and Nijhoff \cite{Sikarin1} and Yoo-Kong and Nijhoff \cite{Sikarin2}. Afterwards, the variational formulation of commuting Hamiltonian flows was studied by Suris \cite{LAMS3}. The Lagrangian 1-form structure of the Toda-type systems, along with their relativistic versions, were developed by Boll, Petrera and Suris \cite{LAMS4,LAMS5}. Further investigation has been made by Jairuk, Yoo-Kong and Tanasittikosol on the Lagrangian 1-form structure of the Calogero's Goldfish and hyperbolic Calogero-Moser models \cite{Umpon, Umpon1,Umpon2}. Recently, the variational symmetries of the pluri-Lagrangian systems has been developed by Suris and Petrera \cite{LAMS6,LAMS7}.
\\
\\
In this paper, we provide further developments and a complete picture on Lagrangian 1-form structure in both discrete and continuous levels. In section 2, we give a short review on action-angle variables as well as the Liouville integrability. A key feature called the commuting Hamiltonian flows is also  discussed. An explicit example, namely rational Calogero-Moser system, of Liouville integrability is provided. In section 3, we first introduce the notion of discrete Lagrangian 1-form and the key feature called a discrete closure relation is derived from the variational principle. The multidimensional consistency on the level of discrete Lagrangians is also discussed. A sequence of continuum limits is performed resulting in the semi-discrete and full continuous Lagrangian 1-form. In the continuous level, the variational principle is considered for the path on the space of both dependent and independent variables leading to a generalised Euler-Lagrange equation, a constraint equation and a continuous closure relation. All these three equations can be considered as a set of compatible of Lagrangian equations possessing the multidimensional consistency. Furthermore, we also give a remark on the derivation of the closure relation from a point of view of the generalised Stokes' theorem as well as the homotopy of the paths on the space of independent variables. In section 4, the Legendre transformation is introduced to obtain the Hamiltonian hierarchy. The variational principle on the phase space will be considered resulting in generalised Hamilton's equations and commuting Poisson bracket which consequently gives us the commuting Hamiltonian flows. In section 5, N\"{o}ether charges are directly derived from the variation of action functional of the Lagrangian 1-form. We find that all N\"{o}ether charges are nothing but all Hamiltonians (invariants) in the system. In the last section, the conclusion together with some potential further studies are delivered. 
\section{Liouville integrability and commuting flows} \label{section2}
\subsection{Action-angle variables}
In classical mechanics, we have freedom to solve the problem with any set of coordinates and if we choose a right (good) set of coordinates, the problem can be easily solved. On the other hand, if we choose a wrong (poor) set of coordinates we possibly have to go through a tremendous work for obtaining the answer. In Hamiltonian mechanics, for a system with $N$ degrees of freedom, one can transform an old set of coordinates $(p,q)$ on the phase space to a new one $(P(p,q),Q(p,q))$, while the Hamilton's equations are still preserved, through the canonical transformation. 
However, such transformation may not guarantee analytical exact solutions. In many cases, there is a natural choice of coordinates such that
\begin{equation}
\{{p},{q} \}\Rightarrow\{ {I},\theta \}\;,\;\;\;\;\;\;\mathscr{H}({p},{q})\Rightarrow\mathcal{K}({I})\;,\nonumber
\end{equation}
where ${I}$ is a set of \emph{action variables}, $\theta$ is a set of \emph{angle variables} and $\mathcal{K}$ is a new Hamiltonian which is a function of the action variables only since the angle variables are cyclic. With a new set of coordinates, Hamilton's equations become
\begin{equation}
\dot{{I}}=\frac{\partial\mathcal{K}}{\partial \theta}=0\;,\;\;\;\;\;\;\dot{\theta}=\frac{\partial \mathcal{K}}{\partial {I}}=\omega({I})\;,\nonumber
\end{equation}
where $\omega$ is a function of action variables. It can be seen that the action variables ${I}$ are automatically constants of motion. Then we have to solve only $N$ first-order differential equations for $\theta(t)$ and, hence, the system is \emph{integrable}.
Action-angle variables define an \emph{invariant tori} as the action variables define the surface, while the angle variables provide coordinates on the tori. 
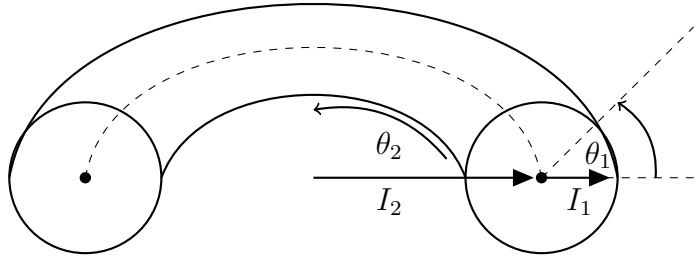
\begin{figure}[h]
\begin{center}
{
\tikzset{middlearrow/.style={
        decoration={markings,
            mark= at position 0.6 with {\arrow{#1}} ,
        },
        postaction={decorate}
    }
}
\begin{tikzpicture}[scale=0.5]
\node (G) [circle,thick,draw,inner sep=0pt,minimum width=2cm]  at (-6,0) {};
\node (A) [circle,thick,draw,inner sep=0pt,minimum width=2cm]  at (6,0) {};
\draw[-triangle 45,thick](0,0)--(5.8,0);
\draw[-triangle 45,thick](6,0)--(7.8,0);
\draw[->,thick] (3.5,0.5) to [bend right=30] (0,1.8);
\draw[dashed](6,0)--(10,0);
\draw[dashed](6,0)--(10,4);
\draw[->,thick](9,0)to [bend right=30](8,2);
  \draw[thick] (-4,0) to [bend left=70] (4,0);
  \draw[thick] (-8,0) to [bend left=80] (8,0);
  \draw[dashed] (-6,0) to [bend left=80] (6,0);
  \fill (-6,0) circle (0.15);
  \fill (6,0) circle (0.15);
\draw (7,0) node[anchor=north] {${I_1}$};
\draw (7.5,1.25) node[anchor=north] {${\theta_1}$};
\draw (2,0) node[anchor=north] {${I_2}$};
\draw (2,1.5) node[anchor=north] {${\theta_2}$};
\end{tikzpicture}
}
\caption{The cross section of invariant torus: $T^2$ in which the smaller and bigger radii are the action variables $I_1$ and $I_2$, defining the surface. Each point on the torus is defined by the angle variables which are the linear function of time.  
}\label{torus}
\end{center}
\end{figure}
\\
\\
In order to make things more transparent, we provide a simple example as follows. In the case of 2 degrees of freedom, a set of action-angle variables is $(I_1,I_2,\theta_1,\theta_2)$ and the invariant torus $T^2=S^1\times S^1$ can be visualised as a doughnut shape, see figure \ref{torus}. The coordinates on the surface of the torus are defined by the angles $\theta_1$ and $\theta_2$ which linearly increase with time $\theta_i=\omega_it+\theta_{i(0)}$. The coordinates $\theta_1$ and $\theta_2$ will be called bases of principal vectors corresponding to the translations $\theta_1\rightarrow\theta_1+2\pi$ and $\theta_2\rightarrow\theta_2+2\pi$. Then, in the case of $N$ degrees of freedom, a set of action-angle variables $(I_1,I_2,...,I_{N},\theta_1,\theta_2,...,\theta_{N})$ will form ${N}$-dimensional invariant tori $T^{N}=\underbrace{S^1\times S^1\times....\times S^1}_{N}$. There are ${N}$ similar closed loops along the basis of principal directions where the coordinates $\theta_i$ can be chosen to vary from $0$ to $2\pi$.
\subsection{Liouville integrability: The continuous time case}
From the previous section, an existence of action-angle variables is a main characteristic for integrability. However, searching for these special sets of coordinates may not be an easy task. 

\begin{theorem}[\textbf{Liouville-Arnold}\cite{ARNO}]
Suppose that there is a set of $N$ functions $(f_1,f_2,...,f_N )\equiv f$ in involution, i.e., $\{ f_i,f_j\}=0$, where $i,j=1,2,..,N$, on a symplectic $2N$-dimensional manifold. Let 
\begin{equation}
M_c := \{(p,q) \in M; \;f_k(p,q) = c_k\}, \quad c_k = \text{constant}, \quad k = 1, 2, ..., N
\end{equation}
be a level set of the functions $f_i$ which are independent. Then
\begin{itemize}
\item $M_c$ is a smooth manifold, invariant under the phase flow with Hamiltonian $H=f_1$.
\item If $M_c$ is compact and connected, there exists a diffeomorphism from $M_c$ to a torus\; $T^{N} \equiv \underbrace{S^{1} \times S^{1} \times ... \times S^{1}}_N$, and, in a vicinity of this torus, the action-angle coordinates 
\begin{eqnarray}
({I},\theta)=(I_1,I_2,...,I_{N},\theta_1,\theta_2,...,\theta_{N}),\quad 0\leq\theta_{k}\leq 2\pi,  
\end{eqnarray}
can be constructed such that angles $\theta_{k}$ are coordinates on $M_c$ and actions $I_{k}$ = $I_{k}(f_1, ..., f_N)$ are first integrals.
\item The canonical Hamilton's equations are linearised 
\begin{eqnarray}
\dot{I}_k=0\;,\;\;\;\;\;\dot{\theta}_k=\omega_k(I_1,I_2,...,I_{N})\;,\;\;\;\;k=1,2,...,{N}\;.
\end{eqnarray}
Therefore, the systems are solvable by quadratures.
\end{itemize}
\end{theorem}
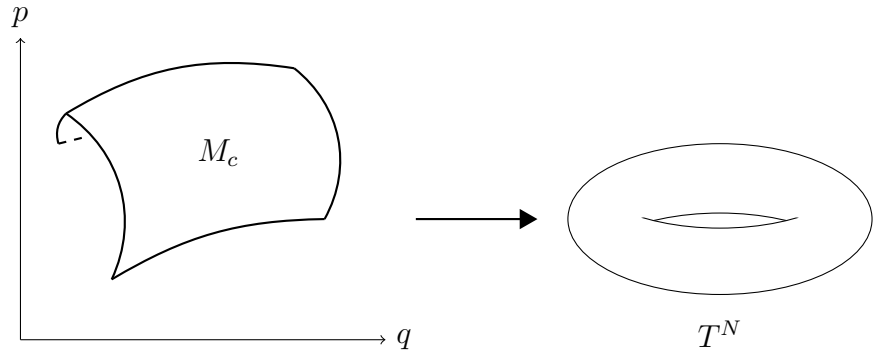
\begin{figure}[h]
\begin{center}
{
\tikzset{middlearrow/.style={
        decoration={markings,
            mark= at position 0.6 with {\arrow{#1}} ,
        },
        postaction={decorate}
    }
}
\begin{tikzpicture}[scale=0.4]
\draw[->] (-13,0) -- (-1,0) node[anchor=west] {$q$};
 \draw[->] (-13,0) -- (-13,10) node[anchor=south] {$p$};
\draw[thick](-10,2)to [bend left=15](-3,4);
\draw[thick](-10,2)to [bend right=40](-11.5,7.5);
\draw[thick](-11.5,7.5)to [bend left=20](-4,9);
\draw[thick](-3,4)to [bend right=40](-4,9);
\draw[thick](-11.5,7.5)to [bend right=30](-11.75,6.5);
\draw[dashed,thick](-11.75,6.5)--(-10.75,6.75);
  \draw (-6.5,7) node[anchor=north] {${{M}_c}$};
 \draw[-triangle 60,thick] (0,4) -- (4,4);
  \useasboundingbox (5,-1.5) rectangle (10,1.5);
  \draw (10,4) ellipse (5 and 2.5);
  \begin{scope}
    \clip (10,1.9) ellipse (5 and 2.5);
    \draw (10,6.2) ellipse (5 and 2.5);
  \end{scope}
  \begin{scope}
    \clip (10,6.2) ellipse (5 and 2.5);
    \draw (10,1.7) ellipse (5 and 2.5);
  \end{scope}
  \draw (10,1) node[anchor=north] {${T^{N}}$};
\end{tikzpicture}
}
\caption{A $2N$-dimensional phase space $M_c$ can actually be foliated to $N$-dimensional tori $T^{N}$.}\label{MM}
\end{center}
\end{figure}
The set of invariants is normally treated as a set of Hamiltonians called the Hamiltonian hierarchy: $(I_1,I_2,..,I_{N})\equiv(\mathscr{H}_1,\mathscr{H}_2,...,\mathscr{H}_{N})$. Therefore, the evolution on phase space is associated with $N$ time variables $(t_1,t_2,...,t_{N})$. For any function $F(p,q)$ defined on the phase space, we find that
\begin{equation}\label{Firt111}
\frac{\mathsf {d}F}{\mathsf {d}t_j}=\{ \mathscr{H}_j,F\}\;,
\end{equation}
where $t_j$ is the time variable associated with the Hamiltonian $\mathscr{H}_j$. Equation \eqref{Firt111} describes the time evolution (flow) of function $F$ along the surface that $\mathscr{H}_j$ is constant. We also have another Hamiltonian $\mathscr{H}_k$ such that
\begin{equation}
\frac{\mathsf {d}F}{\mathsf {d}t_k}=\{ \mathscr{H}_k,F\}\;,
\end{equation}
where $t_k$ is the time variable associated with the Hamiltonian $\mathscr{H}_k$.
We find that
\begin{eqnarray}\label{commuting}
\frac{\partial}{\partial t_j}\frac{\partial F}{\partial t_k}&=&\frac{\partial}{\partial t_k}\frac{\partial F}{\partial t_j}\nonumber\\
\frac{\partial}{\partial t_j}\{\mathscr{H}_k,F \}&=&\frac{\partial}{\partial t_k}\{ \mathscr{H}_j,F \}\nonumber\\
\{ \mathscr{H}_j,\{ \mathscr{H}_k,F\}\}-\{ \mathscr{H}_k,\{ \mathscr{H}_j,F\}\}&=&0\nonumber\\
\{\{ \mathscr{H}_j,\mathscr{H}_k\},F\}&=&0\;,
\end{eqnarray}
since $\{ \mathscr{H}_j,\mathscr{H}_k\}=0$. The relation \eqref{commuting} represents an interesting feature, known as \emph{commuting Hamiltonian flows}, which is a main feature for integrable Hamiltonian system telling us that the flows extend to all possible of time variables $t_i$ and fill the whole manifold $M_c$.
\subsection{Liouville integrability: The discrete-time case}
For the discrete case, the notion of Liouville integrability can be naturally constructed. 
\begin{theorem}[\textbf{Liouville-Arnold-Veselov}\cite{IN1,VES1}]
If a canonical map is integrable, then any compact non-singular level set $M_c = \{(p,q)\in M:\mathscr{I}_k (p,q) = c_k,\;k=1,2,..,N\}$ is a disconnected union of tori on which the dynamic takes place according to regular shifts.
\end{theorem}
A discrete system is completely integrable if there exists a set of functions $\{\mathscr{I}_1, \mathscr{I}_2,...,\mathscr{I}_{N}\}$ satisfying the following requirements:
\begin{itemize}
\item The functions are invariant, $\mathscr{I}(p,q)=\mathscr{I}(\mathsf T p, \mathsf T q)$, under discrete map: $(p,q)\rightarrow(\mathsf T p, \mathsf T q)$, where $\mathsf T$ is the discrete-time shift. The Hamiltonian is one of them. 
\item The functions are in involution with respect to the Poisson bracket
\[
\{ \mathscr{I}_i, \mathscr{I}_j \}=\sum_{k=1}^{N}\left \{ \frac{\partial \mathscr{I}_i}{\partial p_k}\frac{\partial \mathscr{I}_j}{\partial q_k} -\frac{\partial \mathscr{I}_j}{\partial p_k}\frac{\partial \mathscr{I}_i}{\partial q_k}   \right\}=0\;.
\]
\item The functions are functionally independent throughout the phase space.
\end{itemize}
In order to show that $\mathscr{I}_i (p,q)$ is invariant under discrete-time shift, suppose there exists a discrete symplectic map: $(p,q)\mapsto (\mathsf T p,\mathsf T q)$ such that
\begin{eqnarray}\label{SYM1}
\mathsf{T}p_j=g_j (q,p)\;,\;\;\; \mathsf{T}q_j=f_j (q,p)\;,\;\;\;\;j=1,2,...,{N}\;, 
\end{eqnarray} 
where $g_j$ and $f_j$ are some functions of coordinates $(q,p)$. Equation \eqref{SYM1}, equipped with the Poisson bracket structure
\begin{eqnarray}\label{SYM2}
\{ \mathsf{T}q_j,\mathsf{T}q_k\}=0\;,\;\;\;\{ \mathsf{T}p_j,\mathsf{T}p_k\}=0\;,\;\;\;\{ \mathsf{T}q_j,\mathsf{T}p_k \}=\delta_{jk}\;,\;\;\;j,k=1,2,..,N\;,
\end{eqnarray}
can be considered as a canonical transformation from an old set of coordinates $(p, q)$ to a new set of coordinates $(\mathsf{T}p, \mathsf{T}q)$. If the Jacobian $|\partial f_j/\partial p_i|$ is nonzero, we introduce a generating function $H(q, \mathsf{T}p)$ \cite{BRST} in which \eqref{SYM1} becomes
\begin{subequations}\label{SYM3}
\begin{eqnarray}
\mathsf{T}q_j - q_j &=& \frac{\partial H(q, \mathsf{T}p)}{\partial \mathsf{T}p_j}\;,\label{SYM3-1}\\
\mathsf{T}p_j - p_j &=&- \frac{\partial H(q, \mathsf{T}p)}{\partial q_j}\;,\label{SYM3-2}
\end{eqnarray}
where \eqref{SYM3} is sometimes referred to \emph{a set of discrete-time Hamilton's equations}. On the other hand, if the Jacobian $|\partial g_j/\partial q_i|$ is nonzero, we may introduce another generating function $H'( \mathsf{T}q,p)$ in which \eqref{SYM1} becomes
\begin{eqnarray}
\mathsf{T}q_j - q_j &=& \frac{\partial H'(\mathsf{T}q, p)}{\partial p_j}\;,\label{SYM3-1}\\
\mathsf{T}p_j - p_j &=&- \frac{\partial H'(\mathsf{T}q, p)}{\partial \mathsf{T}q_j}\;.\label{SYM3-22}
\end{eqnarray}
\end{subequations}
Next, we would like to consider a canonical transformation between old variables $(q,p)$ and new variables $(Q,P)$ such that
\begin{eqnarray}\label{SYM122}
P_j=P_j (q,p)\;,\;\;\; Q_j=Q_j (q,p)\;,\;\;\;\;j=1,2,...,{N}\;, 
\end{eqnarray} 
with the Poisson bracket  structure
\begin{eqnarray}\label{SYM222}
\{ Q_j, Q_k\}=0\;,\;\;\;\{ P_j, P_k\}=0\;,\;\;\;\{ Q_j, P_k \}=\delta_{jk}\;,\;\;\;j,k=1,2,..,N\;.
\end{eqnarray}
Therefore, the discrete symplectic map for new variables reads
\begin{eqnarray}\label{SYM122}
\mathsf{T}P_j=G_j (Q, P)\;,\;\;\; \mathsf{T}Q_j=F_j (Q, P)\;,\;\;\;\;j=1,2,...,{N}\;, 
\end{eqnarray} 
where $G_j$ and $F_j$ are some functions of the coordinates $(Q, P)$. Again, \eqref{SYM122}, equipped with the Poisson bracket  structure
\begin{eqnarray}\label{SYM2}
\{ \mathsf{T}Q_j,\mathsf{T}Q_k\}=0\;,\;\;\;\{ \mathsf{T}P_j,\mathsf{T}P_k\}=0\;,\;\;\;\{ \mathsf{T}Q_j,\mathsf{T}Q_k \}=\delta_{jk}\;,\;\;\;j,k=1,2,..,N\;,
\end{eqnarray}
can be considered as a canonical transformation from an old set of variables $(P, Q)$ to a new set of variables $(\mathsf{T}P, \mathsf{T}Q)$. If it happens to be that $\mathsf{T}P_j=P_j$, one can obtain $F_j(Q, P)=Q_j+\nu_j(P)$ which immediately leads to $Q_j(n)=n\nu_j(P)+Q_j(0)$. The original discrete evolution can be recovered by means of \eqref{SYM122}. Here, $\nu_j$ are the frequencies and can be obtained as follows. Suppose there is a generating function $W(q,P)$ such that
\begin{subequations}\label{SYM4}
\begin{eqnarray}\label{SYM5}
p_j = \frac{\partial W(q,P)}{\partial q_j}\;, \;\;\;\;Q_j = \frac{\partial W(q,P)}{\partial P_j}\;.
\end{eqnarray}  
Integrating the first equation of \eqref{SYM5}, we have
\begin{eqnarray}\label{SYM6}
W(q,P)=\sum_{k=1}^{N} \int_{q_{k}(0)}^{q_{k}(n)} p_k (P,q')\mathsf{d} q'_{k}\;,
\end{eqnarray} 
and then the second equation of \eqref{SYM5} gives
\begin{eqnarray}\label{SYM7}
Q_j (n)=\sum_{k=1}^{N} \int_{q_{k}(0)}^{q_{k}(n)} \frac{ \partial p_k (P,q')}{\partial P_j }\mathsf{d} q'_{k}\;.
\end{eqnarray}
\end{subequations}
Since $Q_j(n)$ is a function of the invariants $P$, the corresponding frequencies $\nu_j$ are given by
\begin{eqnarray}
\nu_j(P) = \sum_{k=1}^{N} \int_{q_{k}}^{\mathsf T q_{k}} \frac{ \partial p_k (P,q')}{\partial P_j }\mathsf{d} q'_{k} \;.
\end{eqnarray} 
An above structure gives us a symplectic map, i.e., $(Q,P)\mapsto (\mathsf T Q,\mathsf T P=P)$ such that
\begin{subequations}
\begin{eqnarray}\label{dddd} 
\mathsf T P_j-P_j &=&0 \;,\\
\mathsf T Q_j-Q_j &=&\frac{\partial S}{\partial P_j}\;,
\end{eqnarray} 
\end{subequations}
where $S$ is nothing but the action of the system given by
\begin{eqnarray}
S = \sum_{k=1}^{N} \int_{q_{k}}^{\mathsf T q_{k}}  p_k (P,q')\mathsf{d} q'_{k} \;.
\end{eqnarray} 
Then, a set of coordinates $(Q,P)$ is of course the action-angle variables in the discrete case and the existence of these variables, together with interpolation discrete map \eqref{dddd}, implies the integrability of the system. 
\subsection{Lax pair: The continuous case}
The main object in Liouville integrability is a set of invariants. As we have mentioned earlier, searching for these invariants may turn out to be a difficult task. However, with the help of Lax matrices, all invariants are possible to be computed. Let $(\boldsymbol L,\boldsymbol M)$ be a Lax pair where $\boldsymbol L$ is the spatial part and $\boldsymbol M$ is the temporal part. These two matrices are functions on the phase space of the system satisfying 
\begin{eqnarray}
\boldsymbol{L}\Psi &=& \lambda \Psi\;,\label{LC} \\
\boldsymbol{M} \Psi &=& \frac{\mathsf {d} \Psi}{\mathsf {d} t}\label{MC}\;,
\end{eqnarray}
where $\lambda$ is a fixed parameter, $t$ is a time variable and $\Psi$ is an auxiliary function. 
Compatibility between \eqref{LC} and \eqref{MC} gives
\begin{eqnarray}
\frac{\mathsf {d}}{\mathsf {d}t}( \boldsymbol L\Psi)&=&\frac{\mathsf {d}}{\mathsf {d}t}(\lambda\Psi)\nn\\
\frac{\mathsf {d}\boldsymbol L}{\mathsf {d}t}\Psi+\boldsymbol L\frac{\mathsf {d}\Psi}{\mathsf {d}t}&=&\lambda\frac{\mathsf {d}\Psi}{\mathsf {d}t}\nn\\
\frac{\mathsf {d}\boldsymbol L}{\mathsf {d}t}\Psi+\boldsymbol L\boldsymbol M\Psi&=&\boldsymbol M\boldsymbol L\Psi\nn\\
\frac{\mathsf {d}\boldsymbol L}{\mathsf {d}t}\Psi&=&(\boldsymbol L\boldsymbol M-\boldsymbol M\boldsymbol L)\Psi\nn\\
\Rightarrow \frac{\mathsf {d}\boldsymbol L}{\mathsf {d}t}&=&-[\boldsymbol L,\boldsymbol M]\;,\label{Lequation}
\end{eqnarray}
which is called the ``Lax equation" producing the equations of motion.
\\
\\
Furthermore, the solution of \eqref{Lequation} is in the form
\begin{eqnarray}
\boldsymbol{L}(t)&=&\boldsymbol{U}(t)\boldsymbol{L}(0)\boldsymbol{U}^{-1}(t)\;, \label{NewL}
\end{eqnarray} 
where $\boldsymbol{U}$ is an invertible matrix. The time derivative of \eqref{NewL} yields 
\begin{eqnarray}
\frac{\mathsf {d}\boldsymbol{L}}{\mathsf {d}t}&=&\frac{\mathsf {d}\boldsymbol{U}(t)}{\mathsf {d}t}\boldsymbol{L}(0)\boldsymbol{U}^{-1}(t) -\boldsymbol{U}(t)\boldsymbol{L}(0)\frac{\mathsf {d}\boldsymbol{U}^{-1}(t)}{\mathsf {d}t} \nn\\
&=&\frac{\mathsf {d}\boldsymbol{U}(t)}{\mathsf {d}t}\boldsymbol{U}^{-1}(t)\boldsymbol{U}(t)\boldsymbol{L}(0)\boldsymbol{U}^{-1}(t) -\boldsymbol{U}(t)\boldsymbol{L}(0)\boldsymbol{U}^{-1}(t)\frac{\mathsf {d}\boldsymbol{U}(t)}{\mathsf {d}t}\boldsymbol{U}^{-1}(t) 
\label{DtNewL}\;.
\end{eqnarray}
Comparing \eqref{DtNewL} with \eqref{Lequation}, we find that the matrix $\boldsymbol{M}(t)$ takes the form of
\begin{eqnarray}
\boldsymbol{M}(t)&=&\frac{\mathsf {d}\boldsymbol{U}(t)}{\mathsf {d}t}\boldsymbol{U}^{-1}(t) \;.
\end{eqnarray}
From structure of \eqref{NewL}, we find that $Tr(\boldsymbol L(t))=Tr(\boldsymbol L(0))$ and, of course, $Tr(\boldsymbol L^{l}(t))=Tr(\boldsymbol L^{l}(0))$. This suggests that $I_{l} \equiv Tr\boldsymbol{L}^{l}$, where $l=1,2,...,N$, are nothing but the invariants of the system. Then, the existence of Lax pair allows us to construct all invariants.
\subsection{Lax pair: The discrete case}
The same procedure on the construction of conserved quantities in continuous case can be applied in the discrete case. Suppose $\boldsymbol{L}$ and $\boldsymbol{M}$ are Lax matrices for the discrete system satisfying
\begin{eqnarray}
\boldsymbol{L}\Psi &=& \Lambda \Psi\;,\label{DisL}\\
\boldsymbol{M} \Psi &=& \mathsf{T} \Psi\;,\label{DisM}
\end{eqnarray}
where $\Lambda$ is a fixed parameter and $\Psi$ is an auxiliary function. The operator $\mathsf{T}$ is the discrete-time shift such that $\mathsf{T}\Psi(n)=\Psi(n+1)$, where $n$ is a discrete-time variable. Performing the shift operator on \eqref{DisL}, we obtain
\begin{eqnarray}
\mathsf{T}(\boldsymbol{L}\Psi)&=&\mathsf{T}(\Lambda \Psi) \nn \\
\mathsf{T}\boldsymbol{L}\mathsf{T}\Psi &=& \Lambda \mathsf{T}\Psi\;. \label{SDisL}
\end{eqnarray}
Substituting \eqref{DisM} into \eqref{SDisL}, the equation turns into
\begin{eqnarray}
(\mathsf{T}\boldsymbol{L})\boldsymbol{M}&=& \boldsymbol{ML} \;,\label{DisLax}
\end{eqnarray}
which is a discrete-time analogue of \eqref{Lequation} and it gives us the discrete-time equation of motion of the system. Suppose that Lax matrix $\boldsymbol{L}$ can be factorised as
\begin{eqnarray}
\boldsymbol{L}(n)&=&\boldsymbol{U}(n)\boldsymbol{L}(0)\boldsymbol{U}^{-1}(n)\;, \label{DisNewL}
\end{eqnarray}
where $\boldsymbol{U}$ is an invertible matrix. Applying shift operator $\mathsf{T}$ on \eqref{DisNewL} and inserting into \eqref{DisLax}, we find that the matrix $\boldsymbol{M}(n)$ can be expressed in the form of
\begin{eqnarray}
\boldsymbol{M}(n)&=&\mathsf{T}{(\boldsymbol{U}(n))}\boldsymbol{U}^{-1}(n)\;. \label{DisNewM}
\end{eqnarray}
From the structure in \eqref{DisNewL}, we can see immediately that $Tr(\boldsymbol{L}^{l}(n))=Tr(\boldsymbol{L}^{l}(0))$. Then, $\mathscr{I}_{l}\equiv Tr(\boldsymbol{L}^{l}(n))$, where $l=1,2,...,N$, are the invariants in the discrete case. The same conclusion, with those in the continuous case, can be drawn that the existence of discrete Lax pair allows us to construct all invariants for the discrete system.
\subsection{Example: The rational Calogero-Moser system}
In this section, we provide an explicit example of the integrable system. We choose a one-dimensional $N$-body system with pairwise interaction called the rational Calogero-Moser system \cite{CF1,JM1} which is a well-known example of the finite-dimensional integrable Hamiltonian system. The equations of motion are given by \cite{F}
\begin{eqnarray}
\ddot{X}_i = g^2 \sum\limits_{\mathop {j=1}\limits_{i \ne j}}^{N} \frac{1}{(X_i-X_j)^3}\;, \quad i = 1, 2, ..., N\;,
\end{eqnarray}
where $g^2$ is a coupling constant and $X_i=X_i(t)$ is the position of the $i^{\text{th}}$ particle. In the continuous-time case, Lax matrices are given by\cite{LMCM1}
\begin{eqnarray}
\boldsymbol L&=&\sum_{i=1}^{N} p_i E_{ii}+{\sum\limits_{\mathop {i,j=1}\limits_{i \ne j}}^{N}}\mathsf{i}a\frac{E_{ij}}{X_i-X_j}\;,\\
\boldsymbol M&=& a\sum\limits_{\mathop {i, k = 1}\limits_{i \ne k}}^{N}\frac{E_{ii}}{(X_i -X_k)^2} -a{\sum\limits_{\mathop {i,j=1}\limits_{i \ne j}}^{N}}\frac{E_{ij}}{(X_i-X_j)^2}\;,
\end{eqnarray}
where $a$ is a constant related to the coupling constant $g$ defined by $g \equiv a^2-a$, $p_i$ is the momentum of the $i^{\text{th}}$ particle and $E_{ij}$ are entries defined by ($E_{ij})_{kl} = \delta_{ij}\delta_{kl} $. The first three constants of motion are 
\begin{eqnarray}
I_1 &=& Tr\boldsymbol L = \sum_{i=1}^{N} p_i \;,\\
I_2 &=& \frac{1}{2}Tr(\boldsymbol L^2) =\sum_{i=1}^{N} \frac{p_i^2}{2}+\sum\limits_{\mathop {i,j = 1}\limits_{i \ne j}}^{N} \frac{g}{2}\frac{1}{(X_i-X_j)^2}\equiv \mathscr H \;,\label{HamilconL}\\
I_3 &=&  \frac{1}{3}Tr(\boldsymbol L^3)= \sum_{i=1}^{N} \frac{p_i^3}{3} + g\sum\limits_{\mathop {i,j = 1}\limits_{i \ne j}}^{N} \frac{p_j}{(X_i-X_j)^2} \;,
\end{eqnarray}
where \eqref{HamilconL} is the Hamiltonian of the continuous-time rational Calogero-Moser system. 
\\ 
\\  
In the discrete case, the equations of motion are given by \cite{NP}
\begin{eqnarray}
\frac{1}{x_i-\mathsf{T}x_i}+\frac{1}{x_i-\mathsf{T}^{-1}x_i}+\sum\limits_{\mathop {j = 1}\limits_{i \ne j}}^{N}\bigg( \frac{1}{x_i-\mathsf{T}x_j} +  \frac{1}{x_i-\mathsf{T}^{-1}x_j} - \frac{2}{x_i-x_j} \bigg) = 0\;, \; i = 1, 2, ..., N\;,
\end{eqnarray}
where $x_i=x_i(n)$ is the position of the $i^{\text{th}}$ particle. The discrete Lax matrices are given by \cite{NP}
\begin{eqnarray}
\boldsymbol{L}&=&\sum_{i=1}^{N}p_i E_{ii}-\sum\limits_{\mathop {i,j=1}\limits_{i \ne j}}^{N}\frac{1}{x_i -x_j}E_{ij}\label{dlexL}\;,\\
\boldsymbol{M}&=&-\sum_{i,j=1}^{N}\frac{1}{\mathsf{T}x_i -x_j}E_{ij}\label{dlexM}\;,
\end{eqnarray} 
where $x_i$ is the position of $i^{\text{th}}$ particle and the variables $p_i$ are given by
\begin{eqnarray}
p_i=\sum_{j=1}^{N}\frac{1}{x_i -\mathsf{T}x_j}-\sum\limits_{\mathop {j=1}\limits_{i \ne j}}^{N}\frac{1}{x_i -x_j},\;\;i=1,2,...,N\;.
\end{eqnarray}
We see that \eqref{dlexL} is exactly the same as the continuous case. Therefore, all invariants can be immediately obtained from $\mathscr{I}_{l}\equiv Tr\boldsymbol{L}^{l}$, where $l=1,2,...,N$.
\section{Lagrangian 1-form and closure relation} \label{section3}
In the previous section, the integrability criterion is constructed on the Hamiltonian structure of the system. The Poisson commuting of invariants plays an important role to exhibit the commuting Hamiltonian flows. In this section, we set out to construct a mathematical relation to indicate the integrability from the Lagrangian point of view. The variational principle is a main mathematical process that will be used throughout the whole section.
\subsection{Discrete-time Lagrangian 1-form structure}
In this section, we would like to introduce the notion of the discrete-time Lagrangian 1-form. First, let us define a set of discrete-time generalised coordinates (dependent variables) $x(n)$, where $x\equiv (x_1,x_2,x_3,...,x_N)$, $N$ is the number of degrees of freedom and $n \equiv (n_1, n_2, n_3, ..., n_M)$ is the set of the discrete-time variables (independent variables). Next, let us define the time shift operators as follows
\begin{eqnarray}
\mathsf{(T_i)}^{\mu}\mathsf{(T_j)}^{\nu} x\equiv x(n_1,n_2,...,n_i +\mu, ...,n_j+\nu,....,n_M)\; ,\;\; \mu  \;, \nu \in \mathbb{Z} \;, \nn
\end{eqnarray}
where $\mathsf T_j$ is a time shift operator in $j$-direction. 
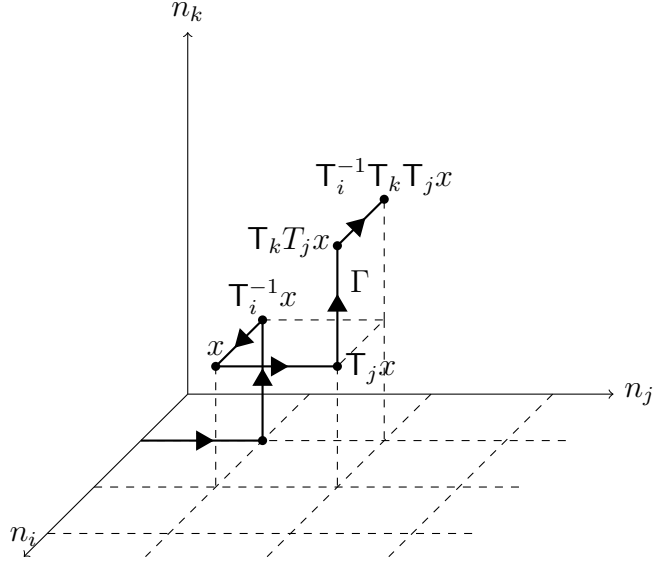
\begin{figure}[h]
\begin{center}
{
\tikzset{middlearrow/.style={
        decoration={markings,
            mark= at position 0.6 with {\arrow{#1}} ,
        },
        postaction={decorate}
    }
}
\begin{tikzpicture}[scale=0.4]
\draw[->] (0,0,0) -- (14,0,0) node[anchor=west] {$n_j$};
 \draw[->] (0,0,0) -- (0,12,0) node[anchor=south] {$n_k$};
 \draw[->] (0,0,0) -- (0,0,14) node[anchor=south] {$n_i$};
 \draw[dashed,black] (0,0,4)--(14,0,4);
 \draw[dashed,black] (0,0,8)--(14,0,8);
 \draw[dashed,black] (0,0,12)--(14,0,12);
 \draw[dashed,black] (4,0,0)--(4,0,14);
 \draw[dashed,black] (8,0,0)--(8,0,14);
 \draw[dashed,black] (12,0,0)--(12,0,14);
\draw[dashed,black] (4,4,4)--(8,4,4)--(8,4,8);
\draw[dashed,black] (8,4,4)--(8,0,4);
\draw[dashed,black] (8,4,8)--(8,0,8);
\draw[dashed,black] (4,4,8)--(4,0,8);
\draw[dashed,black] (8,4,4)--(8,8,4);
\draw[middlearrow={{triangle 60}},thick] (0,0,4)--(4,0,4);
\draw[middlearrow={{triangle 60}},thick] (4,0,4)--(4,4,4); 
\draw[middlearrow={{triangle 60}},thick] (4,4,4)--(4,4,8); 
\draw[middlearrow={{triangle 60}},thick] (4,4,8)--(8,4,8); 
\draw[middlearrow={{triangle 60}},thick] (8,4,8)--(8,8,8); 
\draw[middlearrow={{triangle 60}},thick] (8,8,8)--(8,8,4); 
 \fill (4,0,4) circle (0.15);
 \fill (4,4,4) circle (0.15);
 \fill (4,4,8) circle (0.15);
 \fill (8,4,8) circle (0.15);
 \fill (8,8,8) circle (0.15);
 \fill (8,8,4) circle (0.15);
 \draw (4,5.2,8) node[anchor=north] {${x}$};
 \draw (4,5.7,4) node[anchor=north] {${\mathsf{T}_i^{-1} x}$};
 \draw (9.1,4.8,8) node[anchor=north] {${\mathsf T_j  x}$};
 \draw (6.4,9,8) node[anchor=north] {${\mathsf T_k T_j  x}$};
 \draw (8,9.7,4) node[anchor=north] {${\mathsf{T}_i^{-1} \mathsf T_k \mathsf T_j x}$};
 \draw (7.2,6,4) node[anchor=north] {$\Gamma$};
 \end{tikzpicture}}
\end{center}
\caption{Discrete curve ${\Gamma}$ on the space of discrete independent variables.}\label{p.0}
\end{figure}
\\
\\
Suppose that the system evolves along an arbitrary discrete curve $\Gamma$ embedded on $M$-dimensional space of discrete independent variables, see figure \ref{p.0}. Therefore, 1-form \footnote{The term ``1-form" here may be not yet relevant at this discrete level. It will become clear in later section on the continuous-time level. However, we can think that the term ``1-form" indicates the fact that there is only 1 discrete variable active on each discrete element.} Lagrangian is a two-point function between two end points of the discrete line element given by
\begin{eqnarray}
\mathscr L_{(i)}(n_i)=\mathscr L_{(i)}(x,\mathsf T_i x)\;\; , i=1, 2,..., M\;.\label{disLa}
\end{eqnarray}
The Lagrangian \eqref{disLa} possesses an antisymmetric property under an interchange of arguments, i.e., $\mathscr L(x,y)=-\mathscr L(y,x)$.\footnote{The antisymmetric property indicates the direction of evolution on discrete element.} The action of discrete curve ${\Gamma}$ can then be expressed in the form      
\begin{eqnarray} \label{disLa11}
S_\Gamma[x(n)]=\sum_{\sigma (n) \in\Gamma} \mathscr L_{(i)}(x,\mathsf T_i x) \;.
\end{eqnarray}
Equation \eqref{disLa11} is nothing but the sum of all discrete elements $\sigma (n)= \{\sigma_i (n)=(n,\mathsf T_i n)\;,i=1, 2,..., M \}$.
\subsubsection{Discrete variation on the dependent variables}\label{secds1}
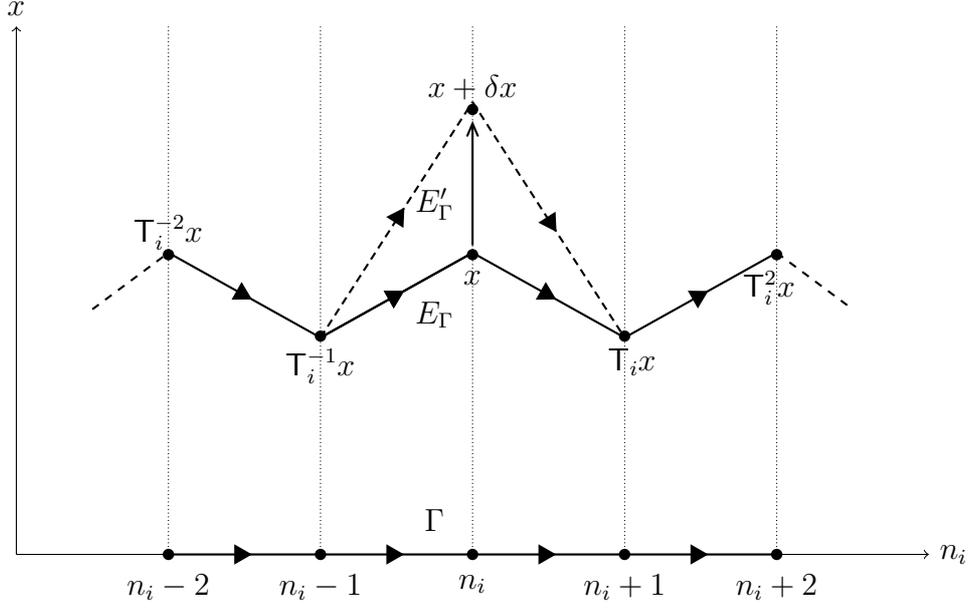
\begin{figure}[h]
\begin{center}
{
\tikzset{middlearrow/.style={
        decoration={markings,
            mark= at position 0.55 with {\arrow{#1}} ,
        },
        postaction={decorate}
    }
}
\begin{tikzpicture}[scale=0.5]
\draw[->] (0,0) --(24,0) node[anchor=west] {$n_i$};
 \draw[->] (0,0) -- (0,14) node[anchor=south] {$x$};
\draw[densely dotted,black] (4,0)--(4,14);
\draw[densely dotted,black] (8,0)--(8,14);
\draw[densely dotted,black] (12,0)--(12,14);
\draw[densely dotted,black] (16,0)--(16,14);
\draw[densely dotted,black] (20,0)--(20,14);
\draw[middlearrow={{triangle 60}},thick] (4,8)--(8,5.75);
\draw[middlearrow={{triangle 60}},thick] (8,5.75)--(12,8);
\draw[middlearrow={{triangle 60}},thick] (12,8)--(16,5.75);
\draw[middlearrow={{triangle 60}},thick] (8,5.75)--(12,8);
\draw[middlearrow={{triangle 60}},thick] (16,5.75)--(20,8);
\draw[thick,dashed,black] (2,6.5)--(4,8);
\draw[thick,dashed,black] (20,8)--(22,6.5);
\draw[middlearrow={{triangle 60}}, densely dashed, thick,black] (8,5.75)--(12,12);
\draw[middlearrow={{triangle 60}}, densely dashed, thick,black] (12,12)--(16,5.75);
\draw[-angle 45 ,thick,black] (12,8.2)--(12,11.5);
\draw[middlearrow={{triangle 60}}, thick,black] (4,0)--(8,0);
\draw[middlearrow={{triangle 60}}, thick,black] (8,0)--(12,0);
\draw[middlearrow={{triangle 60}}, thick,black] (12,0)--(16,0);
\draw[middlearrow={{triangle 60}}, thick,black] (16,0)--(20,0);
\draw (4,-0.25) node[anchor=north] {$n_i -2$};
\draw (8,-0.25) node[anchor=north] {$n_i -1$};
\draw (12,-0.25) node[anchor=north] {$n_i$};
\draw (16,-0.25) node[anchor=north] {$n_i +1$};
\draw (20,-0.25) node[anchor=north] {$n_i +2$};
\draw (4,9.3) node[anchor=north] {${\mathsf{T}_i^{-2} x}$};
\draw (8,5.75) node[anchor=north] {${\mathsf{T}_i^{-1} x}$};
\draw (12,7.8) node[anchor=north] {${ x}$};
\draw (12,13) node[anchor=north] {${ x + \delta x}$};
\draw (16.2,5.75) node[anchor=north] {${\mathsf{T}_i x}$};
\draw (19.8,7.8) node[anchor=north] {${\mathsf{T}_i^{2} x}$};
\draw (11,7) node[anchor=north] {$E_{\Gamma} $};
\draw (11,10) node[anchor=north] {$E_{\Gamma}^\prime $};
\draw (11,1.5) node[anchor=north] {${\Gamma} $};
\fill (4,7.95) circle (0.15);
\fill (8,5.79) circle (0.15);
\fill (12,7.95) circle (0.15);
\fill (12,11.8) circle (0.15);
\fill (16,5.79) circle (0.15);
\fill (20,7.95) circle (0.15);
\fill (4,0) circle (0.15);
\fill (8,0) circle (0.15);
\fill (12,0) circle (0.15);
\fill (16,0) circle (0.15);
\fill (20,0) circle (0.15);
 \end{tikzpicture}}
\end{center}
\caption{The local deformation at point $x$ on discrete curve $E_\Gamma$.}\label{p.1}
\end{figure}
\begin{figure}[h]
\begin{center}
\tikzset{middlearrow/.style={
        decoration={markings,
           mark= at position 0.5 with {\arrow{#1}} ,
       },
        postaction={decorate}
    }
}
\begin{tikzpicture}[scale=0.35]
\draw[->] (-8,-2,0) -- (24,-2,0) node[anchor=west] {$n_i$};
 \draw[->] (-8,-2,0) -- (-8,20,0) node[anchor=south] {$n_j$};
 \draw[->] (-8,-2,0) -- (-8,-2,15) node[anchor=south] {$x$};
\draw[middlearrow={{triangle 60}} , thick] (6,7,12)--(7,17,11);
\draw[middlearrow={{triangle 60}} , thick] (17,8,11)--(16,16,11);
\draw[middlearrow={{triangle 60}} , thick] (7,17,11)--(16,16,11);
 \draw [middlearrow={{triangle 60}} ,thick] (6,7,12)--(17,8,11);
\draw[->, thick](17,8,11)--(16,6.5,12.5);
\draw[middlearrow={{triangle 60}} ,dashed, thick] (6,7,12)--(16,6.5,13);
\draw[middlearrow={{triangle 60}} ,dashed, thick] (16,6.5,13)--(16,16,11);
\draw[->, thick](7,17,11)--(5,15,10.5); 
\draw[middlearrow={{triangle 60}} ,dashed, thick] (6,7,12)--(5,15,11); 
\draw[middlearrow={{triangle 60}} ,dashed, thick] (5,15,11)--(16,16,11); 
\draw [thick,loosely dotted] (6,7,12)--(6,6,0);
\draw [thick,loosely dotted] (7,17,11)--(6,16,0);
\draw [thick,loosely dotted] (16,16,11)--(15,16,0);
\draw [thick,loosely dotted] (17,8,11)--(15,6,0);
\draw [middlearrow={{triangle 60}},densely dotted] (6,6,0)--(15,6,0);
\draw [middlearrow={{triangle 60}},densely dotted] (15,6,0)--(15,16,0);
\draw [middlearrow={{triangle 60}},densely dotted] (6,16,0)--(15,16,0);
\draw [middlearrow={{triangle 60}},densely dotted] (6,6,0)--(6,16,0);
\draw (13,9.5,11) node[anchor=north west] {${{E} _{{\Gamma}_2}}$};
\draw (6.5,13,11) node[anchor=north west] {${{E} _{{\Gamma}_1}}$};
\draw (12.5,7.5,11) node[anchor=north west] {${{E'} _{{\Gamma}_2}}$};
\draw (2.5,12,11) node[anchor=north west] {${{E'} _{{\Gamma}_1}}$};
\draw (6,14.5,0) node[anchor=north west] {${\Gamma_1}$};
\draw (15,14,0) node[anchor=north west] {${\Gamma_2}$};
\draw (5,7,12) node[anchor=north west] {$ x$};
\draw (17,8.5,11) node[anchor=north west] {${\mathsf{T}_i x}$};
\draw (13,6,12.5) node[anchor=north west] {${\mathsf{T}_i x + \delta\mathsf{T}_i x}$};
\draw (15,18,11) node[anchor=north west] {${\mathsf{T}_i \mathsf{T}_j x} $};
\draw (5.5,19,11) node[anchor=north west] {${\mathsf{T}_j x}$};
\draw (-2,15.7,10.5) node[anchor=north west] {${\mathsf{T}_j x + \delta\mathsf{T}_j x}$};
\draw (15,7,0) node[anchor=north west] {${(n_i +1,n_j)}$};
\draw (5,6,0) node[anchor=north west] {${(n_i,n_j)}$};
\draw (15,17.5,0) node[anchor=north west] {${(n_i +1,n_j +1)}$};
\draw (4,18,0) node[anchor=north west] {${(n_i ,n_j +1)}$};
\fill (6,7,12) circle (0.2);
\fill (16,16,11) circle (0.2);
\fill (7,17,11) circle (0.2);
\fill (17,8,11) circle (0.2);
\fill (6,6,0) circle (0.2);
\fill (15,6,0) circle (0.2);
\fill (15,16,0) circle (0.2);
\fill (6,16,0) circle (0.2);
\fill (5,15,11) circle (0.2);
\fill (16,6.5,13) circle (0.2);
\end{tikzpicture}
\end{center}
\caption{The local deformation of corner curves on the space of dependent variables.}\label{p.3}
\end{figure}
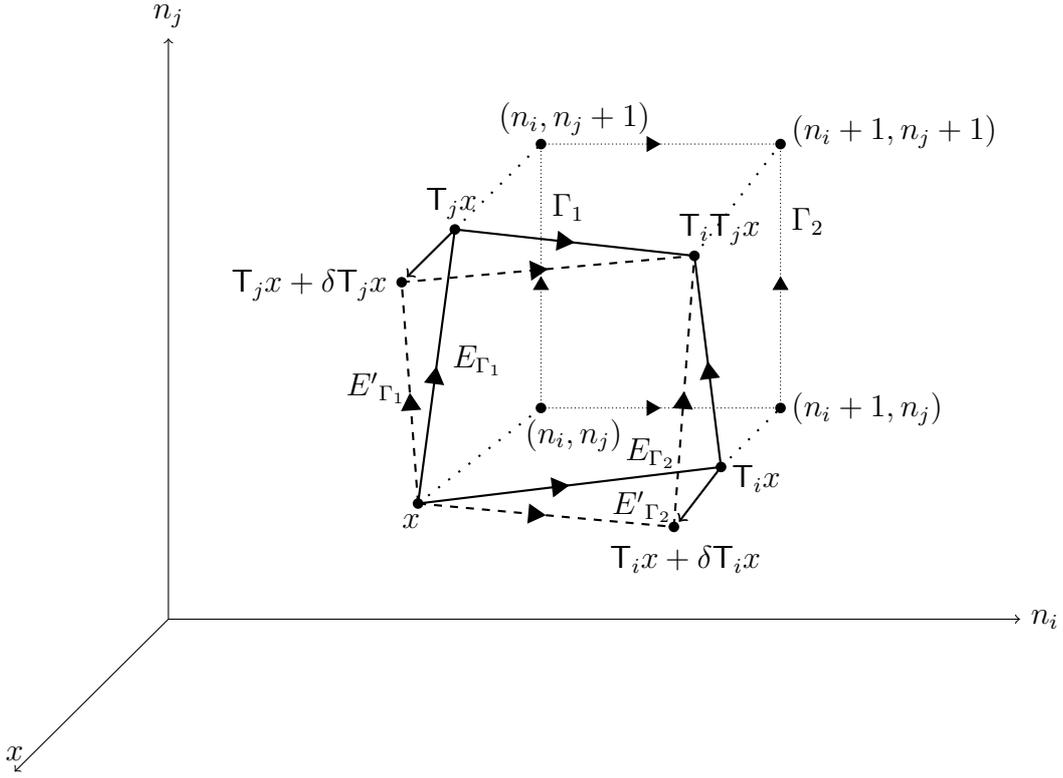

Now, we consider the discrete curve $E_\Gamma$ along the $i$-direction, see figure \ref{p.1}, and the action functional is given by 
\begin{eqnarray}
S_{E_{\Gamma}}[x(n)]=  ...+\mathsf T_{i}^{-1} \mathscr L_{(i)} +  \mathscr L_{(i)} +...\; \label{ds1} ,
\end{eqnarray}
where $\mathsf T_{i}^{-1} \mathscr L_{(i)} \equiv \mathscr L_{(i)}(\mathsf T_{i}^{-1} x,x)$ and $\mathscr L_{(i)} \equiv \mathscr L_{(i)}(x,\mathsf T_i x)$. Under local deformation $E_\Gamma \rightarrow E'_\Gamma $ such that $x\rightarrow x+\delta x$, the action functional of a new discrete curve $E_{\Gamma}^\prime$ is given by
\begin{eqnarray}
S_{E_{\Gamma}^\prime}[x(n)]=  ...+\mathsf T_{i}^{-1} \mathscr L_{(i)}^\prime +  \mathscr L_{(i)} ^\prime+...\; \label{ds12} ,
\end{eqnarray}
where $\mathsf T_{i}^{-1} \mathscr L_{(i)}^\prime \equiv \mathscr L_{(i)}(\mathsf T_{i}^{-1} x,x+\delta x)$ and $\mathscr L_{(i)}^\prime \equiv \mathscr L_{(i)}(x+\delta x,\mathsf T_i x)$. The variation of action between these two actions is
\begin{eqnarray}
S_{E_{\Gamma}^\prime}[x(n)]-S_{E_{\Gamma}}[x(n)]=  \mathsf T_{i}^{-1} \mathscr L_{(i)}^\prime +  \mathscr L_{(i)} ^\prime-\mathsf T_{i}^{-1} \mathscr L_{(i)} - \mathscr L_{(i)}\; \label{ds123} .
\end{eqnarray}
Using Taylor expansion with respect to $\delta x$ and keeping only the first-order contribution, we obtain 
\begin{eqnarray}
S_{E_{\Gamma}^\prime}[x(n)]-S_{E_{\Gamma}}[x(n)] \equiv\delta S_{E_{\Gamma}}[x(n)]= \delta x\left(\frac{\partial \mathsf T_{i}^{-1} \mathscr L_{(i)}}{\partial x}+\frac{\partial \mathscr L_{(i)}}{\partial x}\right)\label{dsL}\; \label{ds1234} .
\end{eqnarray}
According to the least action principle: $\delta S_{E_{\Gamma}}=0$, with the conditions $\delta \mathsf T_{i}^{-1} x =\delta  \mathsf T_{i}x = 0$, then we find that
\begin{eqnarray} 
\frac{\partial \mathsf T_{i}^{-1} \mathscr L_{(i)}}{\partial x}+\frac{\partial \mathscr L_{(i)}}{\partial x} =0 \label{dsL}\;,
\end{eqnarray} 
since $\delta x\neq 0$. This is the \emph{discrete-time Euler-Lagrange equation}. In fact, we could have $M$ ($i=1,2,3,...,M$) copies of \eqref{dsL} for each discrete direction.
\\
\\
Next, we consider a bit more complicate discrete curve $E_{\Gamma_1}$ that lives on ($N+M$)-dimensional space of dependent variables, see figure \ref{p.3}. 
\\
\\
The action functional is given by 
\begin{eqnarray}
S_{E_{\Gamma_1}}[x(n)] =  \mathscr L_{(j)}(x,\mathsf T_j x)+  \mathscr L_{(i)}(\mathsf T_j x,\mathsf T_i \mathsf T_j x )\;  \label{ds14}.
\end{eqnarray}
Then we consider the local deformation such that $\mathsf T_jx\rightarrow \mathsf T_jx+\delta \mathsf T_jx$ 
producing a new discrete curve $E_{\Gamma_1}^\prime$ with the action functional 
\begin{eqnarray}
S_{E_{\Gamma_1}^\prime}[x(n)] =  \mathscr L_{(j)}(x,\mathsf T_j x+\delta \mathsf T_jx)+  \mathscr L_{(i)}(\mathsf T_j x+\delta \mathsf T_j x,\mathsf T_i \mathsf T_j x )\;  \label{ds140}.
\end{eqnarray}
We do the Taylor expansion with respect to $\delta \mathsf T_jx$ in \eqref{ds140} and the variation of the action $\delta S_{E_{\Gamma_1}}[x(n)]\equiv S_{E_{\Gamma_1}^\prime}[x(n)] -S_{E_{\Gamma_1}}[x(n)]$ is 
\begin{eqnarray}
\delta S_{E_{\Gamma_1}}[x(n)] =\delta \mathsf T_jx\left(  \frac{\partial \mathscr L_{(j)}(x,\mathsf T_j x) }{\partial \mathsf T_j x } + \frac{\partial \mathscr L_{(i)}(\mathsf T_j x+,\mathsf T_i \mathsf T_j x ) }{\partial \mathsf T_j x }\right) \;  \label{ds1400}.
\end{eqnarray}
Again, the least action principle requires $\delta S_{E_{\Gamma_1}}=0$ with the end point conditions $\delta x=\delta\mathsf T_i\mathsf T_j x=0$, resulting in 
\begin{eqnarray}
\frac{\partial \mathscr L_{(j)}(x,\mathsf T_j x) }{\partial \mathsf T_j x } + \frac{\partial \mathscr L_{(i)}(\mathsf T_j x,\mathsf T_i \mathsf T_j x ) }{\partial \mathsf T_j x }=0\;,\; i,j=1,2,3,...,M\;,
\end{eqnarray}
which are the \emph{corner Euler-Lagrange equations}. In fact,  these Euler-Lagrange equations produce the equations of motion, called the \emph{constraint equations}, which tell us how the system evolves from discrete $i$-direction to the discrete $j$-direction. Similarly, the variation of action functional of $E_{\Gamma_2}$ expressed as
\begin{eqnarray}
S_{E_{\Gamma_2}}[x(n)]=  \mathscr L_{(i)}(x,\mathsf T_i x)+  \mathscr L_{(j)}(\mathsf T_i x,\mathsf T_i \mathsf T_j x )\;  \label{ds14-1}
\end{eqnarray}
produces constraint equations in the form of
\begin{eqnarray}
\frac{\partial \mathscr L_{(i)}(x,\mathsf T_i x) }{\partial \mathsf T_i x } + \frac{\partial \mathscr L_{(j)}(\mathsf T_i x,\mathsf T_i \mathsf T_j x ) }{\partial \mathsf T_i x } =0 \;\label{ds15} \;,\; i,j=1,2,3,...,M\;,
\end{eqnarray}
describing the evolution from discrete $j$-direction to discrete $i$-direction. These corner equations first appeared in \cite{Sikarin1,Sikarin3}.
\subsubsection{Discrete variation on the independent variables}\label{secds2}
In the previous section, we consider the variational principle of the discrete curve $E_\Gamma$ with respect to dependent variables and we obtain two types of discrete-time Lagrangian equations. The first one, Euler-Lagrange equation, tells us how the system interpolates on a certain discrete direction and the second one, constraint equation, tells us how the system changes the discrete direction from one to another. 
\\
\\
Here, in this section, we will consider the discrete curve $\Gamma$ which lives on the space of independent variables, see figure \ref{p.1}. Actually, we can see that the discrete curve $\Gamma$ is the projection of the discrete curve $E_\Gamma$. In figure \ref{p.3}, the action functional of the curve $\Gamma_1$ and $\Gamma_2$ are given by
\begin{eqnarray}
S_{{\Gamma_1}}[x(n)] =  \mathscr L_{(j)}(x,\mathsf T_j x)+  \mathscr L_{(i)}(\mathsf T_j x,\mathsf T_i \mathsf T_j x )\;  \label{ds1401},
\end{eqnarray}
and 
\begin{eqnarray}
S_{{\Gamma_2}}[x(n)]=  \mathscr L_{(i)}(x,\mathsf T_i x)+  \mathscr L_{(j)}(\mathsf T_i x,\mathsf T_i \mathsf T_j x )\;  \label{ds14-12}.
\end{eqnarray}
We find that the discrete curve $\Gamma_2$ can be obtained by locally deforming the curve $\Gamma_1$ such that $(n_i+1,n_j)\rightarrow (n_i,n_j+1)$, and vice versa. The least action principle, $\Delta S_\Gamma=S_{{\Gamma_2}}-S_{{\Gamma_1}}=0$, gives us immediately 
\begin{eqnarray}  
0=\mathscr L_{(j)}(x,\mathsf T_j x)-\mathscr L_{(i)}(x,\mathsf T_i x)- \mathscr L_{(j)}(\mathsf T_i x,\mathsf T_i \mathsf T_j x ) +\mathscr L_{(i)}(\mathsf T_j x,\mathsf T_i \mathsf T_j x ) \;, \label{ds17}
\end{eqnarray}
or in short
\begin{eqnarray}  
0=\mathscr L_{(j)}-\mathscr L_{(i)}- \mathsf T_i \mathscr L_{(j)} +\mathsf T_j \mathscr L_{(i)} \;,\; i\ne j= 1,2,...M \;, \label{ds18}
\end{eqnarray}
where $\mathsf T_i \mathscr L_{(j)} \equiv \mathscr L_{(j)}(\mathsf T_i x,\mathsf T_i \mathsf T_j x )$ and $\mathsf T_j \mathscr L_{(i)} \equiv \mathscr L_{(i)}(\mathsf T_j x,\mathsf T_i \mathsf T_j x )$. Equations \eqref{ds18} are called the \emph{1-form discrete closure relations}. These equations ensure the invariance of action between two arbitrary discrete curves sharing the same end points on the space of independent variables.  
\subsubsection{Example: The discrete-time rational Calogero-Moser system}\label{secEx1}
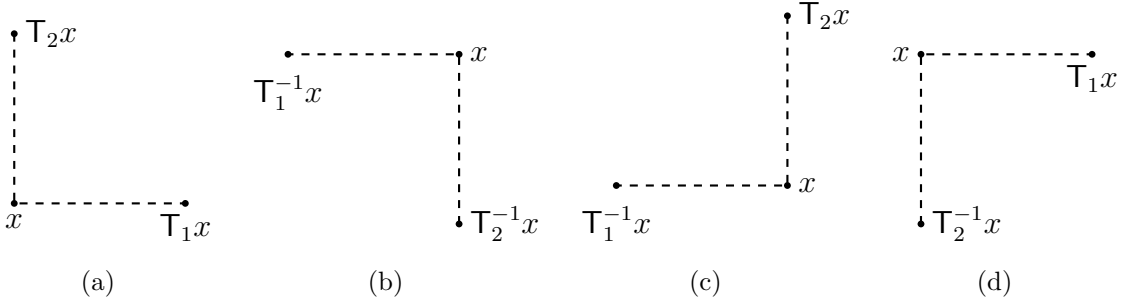
\begin{figure}[h]
\begin{center}
\subfigure[]{
\begin{tikzpicture}[scale=0.45]
 \draw[thick,dashed,black] (7,12)--(7,7)--(12,7);
 \fill (7,12) circle (0.1);
 \fill (7,7) circle (0.1);
 \fill (12,7) circle (0.1);
 \draw (7,12) node[anchor=west] {$\mathsf T_{2} { x}$};
 \draw (7,7) node[anchor=north] {${{ x}}$};
 \draw (12,7) node[anchor=north] {${\mathsf T_{1}{{ x}}}$};
 \end{tikzpicture}}
\subfigure[]{
\begin{tikzpicture}[scale=0.45]
\draw[thick,dashed,black] (7,2)--(7,7)--(2,7);
 \fill (7,2) circle (0.1);
 \fill (7,7) circle (0.1);
 \fill (2,7) circle (0.1);
 \draw (7,2) node[anchor=west] {$\mathsf T_{2}^{-1} { x}$};
 \draw (7,7) node[anchor=west] {${{ x}}$};
 \draw (2,6.7) node[anchor=north] {$\mathsf T_{1}^{-1} { x}$};
\end{tikzpicture}}
\subfigure[]{
\begin{tikzpicture}[scale=0.45]
\draw[thick,dashed,black] (7,12)--(7,7)--(2,7);
 \fill (7,12) circle (0.1);
 \fill (7,7) circle (0.1);
 \fill (2,7) circle (0.1);
 \draw (7,12) node[anchor=west] {$\mathsf T_{2} { x}$};
 \draw (7,7) node[anchor=west] {${{ x}}$};
 \draw (2,6.7) node[anchor=north] {$\mathsf T_{1}^{-1} { x} $};
\end{tikzpicture}}
\subfigure[]{
\begin{tikzpicture}[scale=0.45]
\draw[thick,dashed,black] (7,2)--(7,7)--(12,7);
 \fill (7,2) circle (0.1);
 \fill (7,7) circle (0.1);
 \fill (12,7) circle (0.1);
 \draw (7,2) node[anchor=west] {$\mathsf T_{2}^{-1} { x}$};
 \draw (7,7) node[anchor=east] {${{ x}}$};
 \draw (12,7) node[anchor=north] {$\mathsf T_{1} { x}$};
\end{tikzpicture}}
\end{center}
\caption{Discrete-time evolutions around the corners.}\label{HL1}
\end{figure}

In this section, we give a concrete example of the Lagrangian 1-form structure of the discrete-time Calogero-Moser system \cite{Sikarin1}. The Lagrangian is given by
\begin{eqnarray}
\mathscr{L}_{(j)}(x,\mathsf T_j{x})&=&\sum_{m,l=1}^{N} \log \vert x_m-\mathsf T_j{x}_l \vert -\frac{1}{2}{\sum\limits_{\mathop {m,l=1}\limits_{m \ne l}}^{N}}\bigg[ \log \vert x_m-x_l \vert + \log \vert \mathsf T_j{x}_m-\mathsf T_j{x}_l \vert \bigg] \nonumber\\
&&-p_{j}\sum_{m=1}^{N} \left(x_m -\mathsf T_j{x}_m \right)\;,\;\;\;j=1,2\;,\label{dtni}
\end{eqnarray}
where $x_m (n_1 , n_2)$ is the position of the $m{^\text{th}}$ particle, $N$ is the number of the particles in the system and $p_{j}$ is the lattice parameter.
\\
\\
\textbf{Equation of motion}: Suppose that the action of two discrete elements along the $j$-direction is given by
\begin{eqnarray}
S_{n_j}=\mathsf T_j ^{-1} \mathscr L_{(j)} + \mathscr L_{(j)} \;\label{dse2},\;\;\;j=1,2\;\;,
\end{eqnarray} 
where $\mathscr L_{(j)}\equiv \mathscr L_{(j)}\left(x, \mathsf T_j x\right)$ and $\mathsf T_j ^{-1} \mathscr L_{(j)}\equiv \mathscr L_{(j)}(\mathsf T_j ^{-1} x,  x)$, resulting in the Euler-Lagrange equation for $j$-direction
\begin{eqnarray}
\frac{\partial \mathsf T_{j}^{-1} \mathscr L_{(j)}}{\partial x_{m}}+\frac{\partial \mathscr L_{(j)}}{\partial x_{m}} =0\;.\label{dse3}
\end{eqnarray}
Substituting \eqref{dtni} into \eqref{dse3}, we obtain 
\begin{eqnarray}
\sum_{l=1}^{N} \left[\frac{1}{x_m - \mathsf T_j{x}_l }+\frac{1}{x_m - \mathsf{T}_j^{-1} {x}_l} \right]-{\sum\limits_{\mathop {l = 1}\limits_{m \ne l}}^{N}} \frac{2}{x_m -x_l} =0 \;,\;j=1,2\;,m=1,2,...,N\;, \label{dteqmotni}
\end{eqnarray}
which are the discrete-time equations of motion for Calogero-Moser system along the $j$-direction \cite{NP}.
\\
\\
\textbf{Constraint equation}: There are four different types of discrete trajectories around the corners, see figure \ref{HL1}.
\\
\\
The actions of each configurations around point $x$ are given by
\begin{eqnarray}
S_a &=& \mathscr{L}_{(2)}(\mathsf T_{2}x,x)+\mathscr{L}_{(1)}(x,\mathsf T_{1}x)\;, \\
S_b &=& \mathscr{L}_{(1)}(\mathsf T_{1}^{-1}x,x)+\mathscr{L}_{(2)}(x,\mathsf T_{2}^{-1}x)\;, \\
S_c &=& \mathscr{L}_{(1)}(\mathsf T_{1}^{-1}x,x)+\mathscr{L}_{(2)}(x,\mathsf T_{2}x)\;, \\
S_d &=& \mathscr{L}_{(2)}(\mathsf T_{2}^{-1}x,x)+\mathscr{L}_{(1)}(x,\mathsf T_{1}x)\;. 
\end{eqnarray}
The local variation at point $x$  of each action gives us
\begin{subequations} \label{gELC}
\begin{eqnarray}
\frac{\partial \mathscr{L}_{(2)}(\mathsf T_{2}x,x)}{\partial x}+\frac{\partial \mathscr{L}_{(1)}(x,\mathsf T_{1}x)}{\partial x}=0\;,\label{ELC1}\\
\frac{\partial \mathscr{L}_{(1)}(\mathsf T_{1}^{-1}x,x)}{\partial x}+\frac{\partial \mathscr{L}_{(2)}(x,\mathsf T_{2}^{-1}x)}{\partial x}=0\;,\label{ELC2}\\
\frac{\partial \mathscr{L}_{(1)}(\mathsf T_{1}^{-1}x,x)}{\partial x}+\frac{\partial \mathscr{L}_{(2)}(x,\mathsf T_{2}x)}{\partial x}=0\;,\label{ELC3}\\
\frac{\partial \mathscr{L}_{(2)}(\mathsf T_{2}^{-1}x,x)}{\partial x}+\frac{\partial \mathscr{L}_{(1)}(x,\mathsf T_{1}x)}{\partial x}=0\;,\label{ELC4}
\end{eqnarray}\\
\end{subequations}
respectively. Equation \eqref{gELC} gives us a set of equations
\begin{subequations} \label{gdtconseq}
\begin{eqnarray}
p_1 -p_2 &=& \sum_{l=1}^{{N}} \left(\frac{1}{x_m - \mathsf{T_1}  x_l }-\frac{1}{x_m - \mathsf{T_2} x_l} \right)\;,\label{dtconseq}
\end{eqnarray}
\begin{eqnarray}
-(p_1 -p_2) &=& \sum_{l=1}^{{N}} \left(\frac{1}{x_m - \mathsf T_{1}^{-1} x_l}-\frac{1}{x_m - \mathsf T_{2}^{-1} x_l} \right)\;,\label{secdtconseq}
\end{eqnarray} 
\begin{eqnarray}
-(p_1 -p_2)&=&\sum_{l=1}^{N}\left(\frac{1}{x_m - \mathsf T_{1}^{-1} x_l}+\frac{1}{x_m - \mathsf T_{2} x_l} \right)-{\sum\limits_{\mathop {l = 1}\limits_{m \ne l}}^{N}} \frac{2}{x_m -x_l}\;,\label{thirdtconseq} \\
p_1 -p_2&=&\sum_{l=1}^{N}\left(\frac{1}{x_m - \mathsf T_{2}^{-1} x_l}+\frac{1}{x_m - \mathsf T_{1} x_l} \right)-{\sum\limits_{\mathop {l = 1}\limits_{m \ne l}}^{N}} \frac{2}{x_m -x_l}\;.\label{fourthtconseq}
\end{eqnarray} 
\end{subequations}
Equation \eqref{gdtconseq} is a set of constraint equations corresponding to discrete evolutions in figure \ref{HL1}, respectively.  
\\
\\
\textbf{Legendre transformation}: With the Lagrangian defined in \eqref{dtni}, we find that their associated momentum variable is given by
\begin{eqnarray}
\mathsf{T}_{j}P_{ml}&=&-\frac{\partial\mathscr L_{(j)}}{\partial \mathsf T_j{x}_l} = \sum_{m=1}^{N} \frac{1}{x_m -\mathsf T_j{x}_l}-\sum\limits_{\mathop {m = 1}\limits_{m \ne l}}^{N}\frac{1}{\mathsf T_j{x}_m - \mathsf T_j{x}_l} -p_j 
\;. \label{Ndtmo}
\end{eqnarray} 
We now introduce two extra variables $\mathcal{P}_{ml}$ and ${\rho}_{ml}$ such that
\begin{eqnarray}\label{MOMEN}
\mathsf{T}_j{\mathcal{P}}_{ml}\equiv\frac{1}{x_m - \mathsf{T}_j{x}_l}\; \text{and} \;\; \mathsf{T}_j{\rho}_{ml} \equiv-\frac{1}{\mathsf{T}_j{x}_m -\mathsf{T}_j{x}_l}\;,
\end{eqnarray}
together with the Legendre transformation
\begin{eqnarray}
\mathscr{H}_{(j)}(\mathsf{T}_j{\mathcal{P}}_{ml},\mathsf{T}_j{\rho}_{ml},x)&=&\sum_{m,l=1}^{N}\mathsf{T}_j{\mathcal{P}}_{ml}(x_m - \mathsf{T}_j{x}_l)+\frac{1}{2}{\sum\limits_{\mathop {m,l = 1}\limits_{m \ne l}}^{N}}\mathsf{T}_j{\rho}_{ml}(\mathsf{T}_j{x}_m - \mathsf{T}_j{x}_l) \nonumber \\
&&- \mathscr{L}_{(j)}(x,\mathsf{T}_j{x})\;. \label{dtNTLF}
\end{eqnarray}
Obviously, the relation \eqref{dtNTLF} is not the same with those in \cite{NP} since the extra term appears in the Lagrangian \eqref{dtni}. Furthermore, Nijhoff and Pang pointed out that $\mathscr{H}_{(j)}$ is not the Hamiltonian in the usual sense, see \cite{NP}, but rather be the Hamiltonian generator, see section 2.3. However, we insist to use \eqref{dtNTLF} as the Legendre transformation. Substituting \eqref{dtni} into \eqref{dtNTLF}, we obtain
\begin{eqnarray}
\mathscr{H}_{(j)}(\mathsf{T}_j{\mathcal{P}}_{ml},\mathsf{T}_j{\rho}_{ml},x)&=&\sum_{m,l=1}^{N} \log \vert\mathsf{T}_j{\mathcal{P}}_{ml} \vert - \frac{1}{2}{\sum\limits_{\mathop {m,l = 1}\limits_{m \ne l}}^{N}} \log \vert \mathsf{T}_j{\rho}_{ml} \vert +\frac{1}{2}{\sum\limits_{\mathop {m,l = 1}\limits_{m \ne l}}^{N}} \log \vert x_m-x_l \vert \nonumber \\
&& + p_j\sum_{m=1}^{N} \frac{1}{\mathsf{T}_j{\mathcal{P}}_{mm}}\;.\label{dtH1}
\end{eqnarray}
The action can be rewritten in terms of Hamiltonian as
\begin{eqnarray}
{S}_{n_j}&=&\sum_{m,l=1}^{N}\mathsf{T}_j{\mathcal{P}}_{ml}(x_m - \mathsf{T}_j{x}_l)+\frac{1}{2}{\sum\limits_{\mathop {m,l = 1}\limits_{m \ne l}}^{N}}\mathsf{T}_j{\rho}_{ml}(\mathsf{T}_j{x}_m - \mathsf{T}_j{x}_l)\nonumber \\ 
&&- \mathscr{H}_{(j)}(\mathsf{T}_j{\mathcal{P}}_{ml},\mathsf{T}_j{\rho}_{ml},x)\;.
\end{eqnarray}
We consider the variation of action with respect to dependent variables, i.e., $\mathsf{T}_j{\mathcal{P}}_{ml} \rightarrow \mathsf{T}_j{\mathcal{P}}_{ml}+\delta \mathsf{T}_j{\mathcal{P}}_{ml}$, $\mathsf{T}_j{\rho}_{ml} \rightarrow \mathsf{T}_j{\rho}_{ml} + \delta \mathsf{T}_j{\rho}_{ml}$ and $x_{m} \rightarrow x_{m}+\delta x_{m}$, resulting in 
\begin{eqnarray}
\delta {S}_{n_j}&=&\sum_{m,l=1}^{N} \delta \mathsf{T}_j{\mathcal{P}}_{ml} \left[(x_m - \mathsf{T}_j{x}_l ) - \frac{\partial \mathscr{H}_{(j)}}{\partial \mathsf{T}_j{\mathcal{P}}_{ml}} \right] \nonumber\\
&&+ {\sum\limits_{\mathop {m,l = 1}\limits_{m \ne l}}^{N}} \delta \mathsf{T}_j{\rho}_{ml} \left[\frac{1}{2}(\mathsf{T}_j{x}_m - \mathsf{T}_j{x}_l )- \frac{\partial \mathscr{H}_{(j)}}{\partial \mathsf{T}_j{\rho}_{ml}} \right] \nonumber\\
&&+ \sum_{m=1}^{N}\delta{x_m} \left[ \sum_{l=1}^{N} \left(\mathsf{T}_j{\mathcal{P}}_{ml}- {\mathcal{P}}_{ml}\right)+\frac{1}{2}{\sum\limits_{\mathop {l = 1}\limits_{m \ne l}}^{N}}\left( {\rho}_{ml}-{\rho}_{lm} \right) -\frac{\partial\mathscr{H}_{(j)}}{\partial {x}_{m}}\right]\;.
\end{eqnarray}
Imposing the least action condition: $\delta {S}=0$, we obtain
\begin{eqnarray}
\frac{\partial \mathscr{H}_{(j)}}{\partial \mathsf{T}_j{\mathcal{P}}_{ml}}&=& x_m - \mathsf{T}_j{x}_l \;,\label{dtHEP} \\
\frac{\partial \mathscr{H}_{(j)}}{\partial \mathsf{T}_j{\rho}_{ml}}&=&\frac{1}{2}(\mathsf{T}_j{x}_m - \mathsf{T}_j{x}_l )\; , m\neq l \;,\label{dtHErho} \\
\frac{\partial\mathscr{H}_{(j)}}{\partial {x}_{m}}&=&\sum_{l=1}^{{N}} \left(\mathsf{T}_j{\mathcal{P}}_{ml}- {\mathcal{P}}_{ml}\right)+\frac{1}{2}{\sum\limits_{\mathop {l = 1}\limits_{m \ne l}}^{N}}\left( {\rho}_{ml}-{\rho}_{lm} \right)\;.\label{dtHEx}
\end{eqnarray}
These equations are the discrete-time Hamilton's equations. Equations \eqref{dtHEP} and \eqref{dtHErho} give us back the definition of the extra variables in \eqref{MOMEN}. Equation \eqref{dtHEx} gives us precisely the equation of motion \eqref{dteqmotni}. 
\\
\\
Using \eqref{dse3} and \eqref{Ndtmo}, all momentum variables are given by
\begin{subequations} \label{gPMLT}
\begin{eqnarray}
 \frac{\partial \mathscr L_{(1)}}{\partial  {x}_l}&=& P_{ml}= \sum_{l=1}^{N} \frac{1}{x_m -\mathsf T_1 {x}_l}-\sum\limits_{\mathop {l = 1}\limits_{m \ne l}}^{N}\frac{1}{{x}_m - {x}_l} - p_1 \label{PMLT1} \;,\\
\frac{\partial  \mathscr L_{(2)}}{\partial  {x}_l}&=& P_{ml}= \sum_{l=1}^{N} \frac{1}{x_m -\mathsf T_2 {x}_l}-\sum\limits_{\mathop {l = 1}\limits_{m \ne l}}^{N}\frac{1}{{x}_m - {x}_l} - p_2  \label{PMLT2} \;,\\
-\mathsf T_1 ^{-1} \left(\frac{\partial  \mathscr L_{(1)}}{\partial \mathsf T_1 {x}_l}\right)&=& P_{ml}=- \sum_{l=1}^{N} \frac{1}{x_m -\mathsf T_1 ^{-1} {x}_l}+\sum\limits_{\mathop {l = 1}\limits_{m \ne l}}^{N}\frac{1}{{x}_m - {x}_l} - p_1 \label{PMLT3} \;,\\
-\mathsf T_2 ^{-1} \left(\frac{\partial \mathsf T_2 \mathscr L_{(2)}}{\partial \mathsf T_2 {x}_l}\right)&=& P_{ml}=- \sum_{l=1}^{N} \frac{1}{x_m -\mathsf T_2 ^{-1} {x}_l}+\sum\limits_{\mathop {l = 1}\limits_{m \ne l}}^{N}\frac{1}{{x}_m - {x}_l} - p_2 \;. \label{PMLT4} 
\end{eqnarray}
\end{subequations}
Alternatively, the constraint equation \eqref{dtconseq} can also be acquired by equating \eqref{PMLT1} with \eqref{PMLT2}. Similarly, equating \eqref{PMLT3} with \eqref{PMLT4} results in another constraint equation \eqref{secdtconseq}. The last two constraint equations \eqref{thirdtconseq} and \eqref{fourthtconseq} can be obtained by the equating \eqref{PMLT3} with \eqref{PMLT2}, and \eqref{PMLT4} with \eqref{PMLT1}, respectively.
\\
\\
\textbf{Closure relation}: In \cite{Sikarin1}, it has been shown that the discrete-time rational Calogero-Moser system possesses the discrete closure relation
\begin{eqnarray}
0=\mathscr L_{(2)}-\mathscr L_{(1)}- \mathsf T_1 \mathscr L_{(2)} +\mathsf T_2 \mathscr L_{(1)} \;.\label{CR12}
\end{eqnarray} 
A crucial ingredient to show the existence of closure relation is the connection between discrete Lagrangian \eqref{dtni} and temporal part of the Lax matrices \eqref{dlexL}
\begin{eqnarray}
\mathscr{L}_{(j)}(x,\mathsf{T}_j x)&=& \ln |\text{det} \boldsymbol{M}_j | +p_j \sum_{m=1}^{N} \left( x_m - \mathsf{T}_{j}x_m \right)\, , j=1,2 \;. \label{reLandM}
\end{eqnarray} 
The compatibility between the matrices $\boldsymbol{M}_1$ and $\boldsymbol{M}_2$ results in
\begin{eqnarray}\label{TM12}
(\mathsf{T}_2 \boldsymbol{M}_1) \boldsymbol{M}_2 &=& (\mathsf{T}_1 \boldsymbol{M}_2) \boldsymbol{M}_1 \;.
\end{eqnarray}
From this equation, we fine that
\begin{eqnarray}
\ln|\text{det} \mathsf{T}_2 \boldsymbol{M}_1|+\ln|\text{det} \boldsymbol{M}_2 | &=& \ln|\text{det} \mathsf{T}_1 \boldsymbol{M}_2|+\ln|\text{det}  \boldsymbol{M}_1|\;.\label{MNNM}
\end{eqnarray}
We observe that \eqref{MNNM} is precisely the discrete closure relation with
\begin{eqnarray}
\sum_{m=1}^{N}\left( \mathsf{T}_{1}x_m - \mathsf{T}_{1}\mathsf{T}_{2}x_m -x_m +\mathsf{T}_{2}x_m\right)=0\;,
\end{eqnarray}
which is an extra relation of center of mass of the system.
\begin{remark}
Discrete multidimensional consistency: We knew that the compatibility between different temporal Lax matrices \eqref{TM12} gives the constraint equations \eqref{dtconseq} and \eqref{secdtconseq}. This relation tells us that it does not matter whether the system will go from point $x$ to point $\mathsf T_1\mathsf T_2 x$, see figure \ref{p.3}, through point $\mathsf T_1 x$ in curve $E_{\Gamma_1}$, or point $\mathsf T_2 x$ in curve $E_{\Gamma_2}$. Then, \eqref{TM12} exhibits the multidimensional consistency on the level of dependent variables. Furthermore, the projection of $E_{\Gamma_1}$ and $E_{\Gamma_2}$ gives $\Gamma_1$ and $\Gamma_2$, respectively, on the space of independent variables. Therefore, these two curves $\Gamma_1$ and $\Gamma_2$ relate to each other by the closure relation which exhibits the multidimensional consistency on the level of independent variables. Then, this seems to suggest that the existence of \eqref{TM12} implies the existence of \eqref{ds17}, and vice versa through \eqref{MNNM}
\end{remark}
\subsection{Semi-discrete time Lagrangian 1-form structure}
Suppose a set of generalised coordinates is a function of discrete and continuous variables $\mathrm x\equiv \mathrm x(n,\tau)=( \mathrm x_1, \mathrm x_2,..., \mathrm x_N)$, where $(n,\tau)=(n_1,n_2,...,\tau_1,\tau_2,....)$. The definition of discrete shift operators in the previous section is still applicable. Now, we consider curve $E_\Gamma$ that consists of one discrete element and one continuous element shown in figure \ref{psemi1}. Then, the action functional associated with this curve is given by
\begin{eqnarray}\label{eqsemi1}
S_{E_{\Gamma}}[{\mathrm x}({ n ,\tau})]&=&\sum_{n_i} \mathscr L_{(i)}(\mathrm x,\mathsf{T}_i {\mathrm x})+ \sum_{i=1} \int_{\tau^{\prime}}^{\tau^{\prime\prime}} \mathsf {d} \tau_i  \mathscr L_{(\tau_i)} \left(\mathrm x,\mathsf{T}_k{\mathrm x}, \frac{\partial \mathrm x }{\partial \tau_1},\frac{\partial \mathrm x }{\partial \tau_2},...  \right) \;,
\end{eqnarray} 
where $\mathscr{L}_{(i)}$ is the discrete-time Lagrangian defined in the previous section and $\mathscr{L}_{(\tau_i)}$ is the Lagrangian associated with the continuous time $\tau_i$ containing both discrete shifts and derivatives.
\subsubsection{Variation on dependent variables}\label{semid1}
\begin{figure}[h]
\begin{center}
{
\begin{tikzpicture}[scale=0.35]
 \tikzset{middlearrow/.style={
        decoration={markings,
            mark= at position 0.5 with {\arrow{#1}} ,
        },
        postaction={decorate}
    }
}
 \draw[->] (-2,0,0) -- (20,0,0) node[anchor=west] {$n_k$};
 \draw[->] (-2,0,0) -- (-2,20,0) node[anchor=south] {$\tau_i$};
 \draw[->] (-2,0,0) -- (-2,0,13) node[anchor=south] {$\mathrm x$};
\draw [middlearrow={{triangle 60}} , thick] (6,7,12) .. controls (4,7.5,10) and (5,9,3) ..(6,17,11);
\draw [middlearrow={{triangle 60}} , dashed] (6,7,12) .. controls (2,8,10) and (6,12,3) ..(6,17,11);
\draw [middlearrow={{triangle 60}} , thick] (18,7,15) .. controls (19,9,10) and (15,12,10) ..(18,17,13);
 \draw[middlearrow={{triangle 60}} , thick] (6,17,11)--(18,17,13);
 \draw [middlearrow={{triangle 60}} ,thick] (6,7,12)--(18,7,15);
\draw [thick,loosely dotted] (6,7,12)--(6,7,0);
\draw [thick,loosely dotted] (6,17,11)--(6,17,0);
\draw [thick,loosely dotted] (18,17,13)--(18,17,0);
\draw [thick,loosely dotted] (18,7,15)--(18,7,0);
\draw [middlearrow={{triangle 60}},densely dotted] (6,7,0)--(6,17,0);
\draw [middlearrow={{triangle 60}},densely dotted] (6,17,0)--(18,17,0);
\draw [middlearrow={{triangle 60}},densely dotted] (6,7,0)--(18,7,0);
\draw [middlearrow={{triangle 60}},densely dotted] (18,7,0)--(18,17,0);
\draw (9,6,11) node[anchor=north west] {${{E} _{\Gamma_{2}}}$};
\draw (3,13,11) node[anchor=north west] {${{E} _{\Gamma_{1}}}$};
\draw (6,12,0) node[anchor=north west] {${\Gamma_{1}}$};
\draw (18,12,0) node[anchor=north west] {${{\Gamma}_{2}}$};
\draw (4.5,7,12) node[anchor=north west] {${{\mathrm x}({ n_k ,\tau_i})}$};
\draw (16.5,6,11) node[anchor=north west] {${{\mathrm x}({ n_k  +1 ,\tau_i})}$};
\draw (15,18.3,11) node[anchor=north west] {${{\mathrm x}({ n_k+ 1 ,\tau_i +\delta \tau_i})} $};
\draw (4,19,11) node[anchor=north west] {${{
{{\mathrm x}({ n_k ,\tau_i +\delta \tau_i })}}}$};
\draw (18,8,0) node[anchor=north west] {${(n_k+1 ,\tau_i)}$};
\draw (5,7,0) node[anchor=north west] {${(n_k ,\tau_i)}$};
\draw (15,19,0) node[anchor=north west] {${(n_k +1 ,\tau_i +\delta \tau_i)}$};
\draw (3,19,0) node[anchor=north west] {${(n_k ,\tau_i +\delta \tau_i)}$};
\fill (6,7,12) circle (0.2);
\fill (6,17,11) circle (0.2);
\fill (18,17,13) circle (0.2);
\fill (18,7,15) circle (0.2);
\fill (6,7,0) circle (0.2);
\fill (6,17,0) circle (0.2);
\fill (18,17,0) circle (0.2);
\fill (18,7,0) circle (0.2);
 \end{tikzpicture}}
\end{center}
\caption{The deformation of the semi-discrete curve. Note that here we use $x(n_k,\tau_i) \equiv x(..., n_k,..., \tau_i,...)$.}\label{psemi1}
\end{figure}
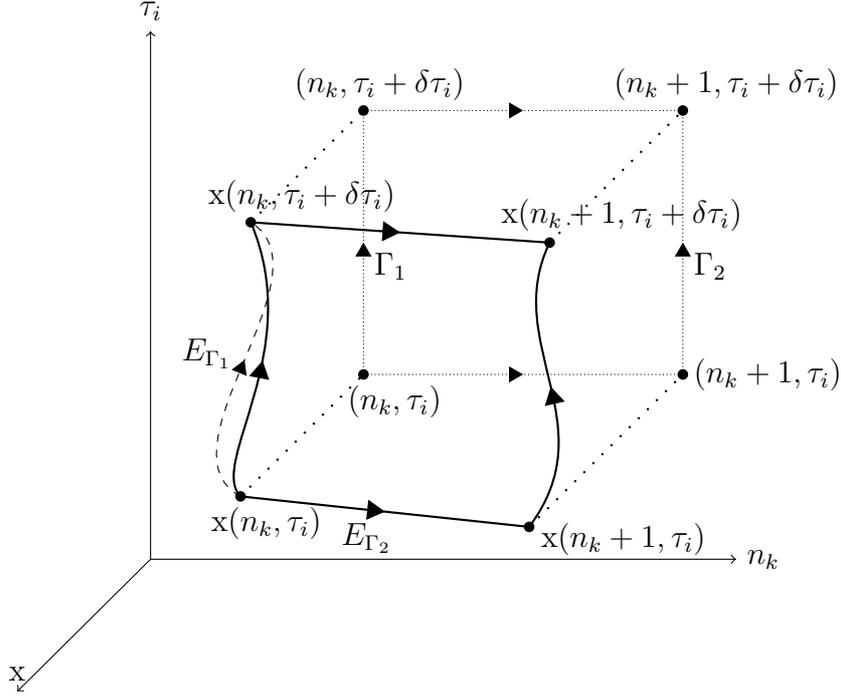

The situation now is that we have a curve constituted from discrete and continuous elements. Then, it is possible to consider the variation on discrete and continuous elements separately. Of course, if we consider only the discrete elements, we can obtain the result in the previous section and we won't repeat it here. We now start with this action functional 
\begin{eqnarray}
S[{\mathrm x}({n ,\tau})]&=&\sum_{i=1}\int_{\tau'}^{\tau ''} \mathsf {d} \tau_i  \mathscr{L}_{(\tau_i)} \left(\mathrm x,\mathsf{T}_k{\mathrm x},\left\{\frac{\partial \mathrm x }{\partial \tau_j}\right\}\right)\;,j=1,\;2,... \;.\label{ACTION1}
\end{eqnarray}
The variations of the variables $\mathrm x\rightarrow\mathrm x+ \delta \mathrm x$, $\mathsf{T}_k{\mathrm x}\rightarrow\mathsf{T}_k{\mathrm x}+\delta \mathsf{T}_k{\mathrm x}$ and $\frac{\partial \mathrm x }{\partial \tau_j}\rightarrow \frac{\partial \mathrm x }{\partial \tau_j}+\delta\frac{\partial \mathrm x }{\partial \tau_j}$, result in a curve, see figure \ref{psemi1},  with the action functional 
\begin{eqnarray}
S'[{\mathrm x}({n ,\tau})]&=&\sum_{i=1} \int_{\tau'}^{\tau ''} \mathsf {d} \tau_i  \mathscr{L}_{(\tau_i)} \left(\mathrm x+ \delta \mathrm x,\mathsf{T}_k{\mathrm x}+\delta \mathsf{T}_k{\mathrm x},\left\{\frac{\partial \mathrm x }{\partial \tau_j}+\delta\frac{\partial \mathrm x }{\partial \tau_j}\right\} \right)\;,j=1,2,...\;.\label{ACTION12}
\end{eqnarray}
We then do the Taylor expansion of \eqref{ACTION12} and keep only the first two terms in the expansion. The variation of action between \eqref{ACTION1} and \eqref{ACTION12} is
\begin{eqnarray}\label{Delta1}
\delta S &=& 
 \sum_{i=1}^{N} \int_{\tau'}^{\tau''} \mathsf {d} \tau_i \left[\delta (\mathsf{T}_k{\mathrm x}) \frac{\partial \mathscr{L}_{(\tau_i)}}{\partial \mathsf{T}_k{\mathrm x}}+\delta \mathrm x \frac{\partial \mathscr{L}_{(\tau_i)}}{\partial \mathrm x}   \right]+ \sum_{i=1}^{N}\int_{\tau'}^{\tau''} \mathsf {d} \tau_i \left[ \delta \left(\frac{\partial \mathrm x }{\partial \tau_i}\right) \frac{\partial \mathscr{L}_{(\tau_i)}}{\partial \left(\frac{\partial \mathrm x }{\partial \tau_i}\right)} \right]\nonumber \\
&&+\sum\limits_{\mathop {i,j=1}\limits_{i \ne j}}^{N}\int_{\tau'}^{\tau''} \mathsf {d} \tau_i \left[ \delta \left(\frac{\partial \mathrm x }{\partial \tau_j}\right) \frac{\partial \mathscr{L}_{(\tau_i)}}{\partial \left(\frac{\partial \mathrm x }{\partial \tau_j}\right)} \right] \;.
\end{eqnarray}
Integrating by parts the second term and imposing the end-point conditions $\delta \mathrm x (\tau')=\delta \mathrm x (\tau'')=0 $, we obtain
\begin{eqnarray}\label{eqsemitau2}
\delta S&=&  \sum_{i=1}^{N} \int_{\tau'}^{\tau''} \mathsf {d} \tau_i \left[\delta (\mathsf{T}_k{\mathrm x}) \frac{\partial \mathscr{L}_{(\tau_i)}}{\partial \mathsf{T}_k{\mathrm x}} \right]+ \sum_{i=1}^{N} \int_{\tau'}^{\tau''} \mathsf {d} \tau_i   \left[\frac{\partial \mathscr{L}_{(\tau_i)}}{\partial {\mathrm x}} - \frac{\partial \mathscr{L}_{(\tau_i)}}{\partial \left(\frac{\partial \mathrm x }{\partial \tau_i}\right)} \right]\delta {\mathrm x} \nonumber \\
&&+\sum\limits_{\mathop {i,j=1}\limits_{i \ne j}}^{N}\int_{\tau'}^{\tau''} \mathsf {d} \tau_i \left[ \delta \left(\frac{\partial \mathrm x }{\partial \tau_j}\right) \frac{\partial \mathscr{L}_{(\tau_i)}}{\partial \left(\frac{\partial \mathrm x }{\partial \tau_j}\right)} \right] \;.
\end{eqnarray}
The least action principle, $\delta S=0$, gives us
\begin{eqnarray}\label{conLaTi}
\frac{\partial\mathscr{L}_{(\tau_i)}}{\partial \mathrm x}-\frac{\partial}{\partial \tau_i}  \left(\frac{\partial \mathscr{L}_{(\tau_i)}}{\partial \left(\frac{\partial \mathrm x }{\partial \tau_i} \right)} \right)=0 \;,\;i=1,2,...\;,
\end{eqnarray}
which are the Euler-Lagrange equations. Furthermore, we also have extra equations
\begin{eqnarray}\label{concoT}
\frac{\partial \mathscr{L}_{(\tau_i)}}{\partial \mathsf{T}_k{\mathrm x}}=0\;\;\;\; \text{and} \;\;\;\;    \frac{\partial \mathscr{L}_{(\tau_i)}}{\partial \left(\frac{\partial \mathrm x }{\partial \tau_j}\right)}=0 \;,\;\text{where}\; i\neq j\;,
\end{eqnarray}
since $\delta(\mathsf{T}_k{\mathrm x})$ and $\delta \left(\frac{\partial \mathrm x }{\partial \tau_j}\right) $ are nonzero, respectively. These equations are in fact the constraints in this situation. The first one tells us how the system evolves from discrete $k$-direction to the continuous $i$-direction while the second one tells us how the system evolves from the continuous $i$-direction to the continuous $j$-direction.
\subsubsection{Variation on independent variables}\label{semid2}
The curve $\Gamma_1$ is the projection of the curve $E_{\Gamma_1}$ onto the space of independent variables and the action functional is 
\begin{eqnarray}\label{eqsemitau3}
S_{\Gamma_{1}}[\mathrm x(n,\tau)]&=& \int_{\tau_i}^{\tau_i + \delta \tau_i} \mathscr{L}_{(\tau_i)}(\mathrm x(n_k,\tau_i),\mathrm x(n_k,\tau_i + \delta \tau_i))\mathsf {d} \tau_i \;\nonumber \\
&& + \mathscr{L}_{(k)}(\mathrm x(n_k,\tau_i + \delta \tau_i),\mathrm x(n_k +1,\tau_i + \delta \tau_i) )\;,
\end{eqnarray}
where the first term represents the evolution along the vertical line and second term represents the evolution along the horizontal line of the curve $\Gamma_1$. We now introduce another curve $\Gamma_2$ which shares the end points with the previous curve $\Gamma_1$. The action functional is given by
\begin{eqnarray}\label{eqsemitau4}
S_{\Gamma_{2}} [\mathrm x(n,\tau)]&=&\mathscr{L}_{(k)}(\mathrm x(n_k,\tau_i ),\mathrm x(n_k +1,\tau_i ) ) \nonumber \\
&&+ \int_{\tau_i}^{\tau_i + \delta \tau_i} \mathscr{L}_{(\tau_i)}(\mathrm x(n_k +1,\tau_i ) ) ,\mathrm x(n_k+1,\tau_i + \delta \tau_i))\mathsf {d} \tau_i \;,
\end{eqnarray}
where the first term represents the evolution along the horizontal line, while, the second term represents the evolution along the vertical line of the curve $\Gamma_2$. 
The variation between two actions is given by
\begin{eqnarray}\label{eqsemitau5}
S_{\Gamma_2} - S_{\Gamma_1} &=&\mathscr{L}_{(k)}(\mathrm x(n_k,\tau_i ),\mathrm x(n_k+1,\tau_i ) )- \mathscr{L}_{(k)}(\mathrm x(n_k,\tau_i + \delta \tau_i),\mathrm x(n_k+1,\tau_i + \delta \tau_i) ) \nonumber \\
&&+\int_{\tau_i}^{\tau_i + \delta \tau_i} [ \mathscr{L}_{(\tau_i)}(\mathrm x(n_k+1,\tau_i )  ,\mathrm x(n_k+1,\tau_i + \delta \tau_i)) \nonumber \\
&&-\mathscr{L}_{(\tau_i)}(\mathrm x(n_k,\tau_i),\mathrm x(n_k,\tau_i + \delta \tau_i)) ] \mathsf {d} \tau_i  \;.
\end{eqnarray}
Using the Taylor expansion with respect to $\delta \tau_i$ and keeping only for the first two terms in the expansion, we have now 
\begin{eqnarray}\label{eqsemitau5}
\delta S&\equiv& S_{\Gamma_2} - S_{\Gamma_1} \nn \\  
&=& -\delta \tau_i \frac{\partial }{\partial \tau_i} \mathscr{L}_{(k)}(\mathrm x(n_k,\tau_i ),\mathrm x(n_k+1,\tau_i ) ) + \int_{\tau_i}^{\tau_i + \delta \tau_i} [ \mathscr{L}_{(\tau_i)}(\mathrm x(n_k+1,\tau_i ) ) ,\mathrm x(n_k+1,\tau_i + \delta \tau_i)) \nonumber \\
&&-\mathscr{L}_{(\tau_i)}(\mathrm x(n_k,\tau_i),\mathrm x(n_k,\tau_i + \delta \tau_i)) ] \mathsf {d} \tau_i \nonumber \\
&=& \delta \tau_i [\mathscr{L}_{(\tau_i)}(\mathrm x(n_k+1,\tau_i ) ) ,\mathrm x(n_k+1,\tau_i + \delta \tau_i))- \mathscr{L}_{(\tau_i)}(\mathrm x(n_k,\tau_i),\mathrm x(n_k,\tau_i + \delta \tau_i)) ] \nonumber \\
&&-\delta \tau_i \frac{\partial }{\partial \tau_i} \mathscr{L}_{(k)}(\mathrm x(n_k,\tau_i ),\mathrm x(n_k+1,\tau_i ) )\;.
\end{eqnarray}
By imposing the least action principle, $\delta S=0$, \eqref{eqsemitau5} holds if
\begin{eqnarray}
\frac{\partial }{\partial \tau_i} \mathscr{L}_{(k)}(\mathrm x(n_k,\tau_i ),\mathrm x(n_k+1,\tau_i ) ) &=& \mathscr{L}_{(\tau_i)}(\mathrm x(n_k+1,\tau_i ) ) ,\mathrm x(n_k+1,\tau_i + \delta \tau_i)) \nonumber \\
&&- \mathscr{L}_{(\tau_i)}(\mathrm x(n_k,\tau_i),\mathrm x(n_k,\tau_i + \delta \tau_i)) \;,\nonumber
\end{eqnarray}
or in short
\begin{eqnarray}
\frac{\partial \mathscr{L}_{(k)}}{\partial \tau_i} &=& \mathsf{T}_k{\mathscr{L}} _{(\tau_i)}-\mathscr{L}_{(\tau_i)} \;,\;i=1,2,...\;,\label{semiclosure} 
\end{eqnarray}
which is the semi-discrete-time closure relation. Again, this relation tells us that the action is invariant under the local deformation of the curve on the space of independent variables.
\subsubsection{Example: The semi-discrete time rational Calogero-Moser system}
We present the semi-discrete time rational Calogero-Moser system\cite{Sikarin1}, which possesses the semi-discrete time closure relation \eqref{semiclosure}. 
The Lagrangian associated with the continuous variable $\tau$ is given by 
\begin{eqnarray}
\mathscr{L}_{(\tau)}\left({\mathrm x},\mathsf{T}_k {{\mathrm x}},\frac{\partial \mathsf{T}_k {\mathrm x}}{\partial \tau}\right)&=&-\sum_{m,l=1}^{N}\frac{\partial \mathsf{T}_{k}{\mathrm x_l}}{\partial \tau}\frac{1}{\mathrm x_m-\mathsf{T}_{k}{ {\mathrm x}}_l}-\frac{1}{2}{\sum\limits_{\mathop {m,l = 1}\limits_{m \ne l}}^{N}}\left( \frac{\partial \mathsf{T}_{k} {\mathrm x}_m}{\partial \tau}-\frac{\partial \mathsf{T}_{k} {\mathrm x}_l}{\partial \tau}\right)\frac{1}{\mathsf{T}_{k} {\mathrm x}_m-\mathsf{T}_{k} {\mathrm x}_l}\nonumber \\
&&+\sum_{m=1}^{N} \left({ {\mathrm x}}_m - \mathsf{T}_{k} {\mathrm x}_m +\frac{\partial \mathsf{T}_{k} {\mathrm x}_m}{\partial \tau} \right)\;,\label{contisemiL}
\end{eqnarray} 
which has both discrete- and continuous-time variables.  
\\
\\
\textbf{Equation of motion}: Substituting the Lagrangian \eqref{contisemiL} into \eqref{conLaTi}, we obtain
\begin{eqnarray}
\sum_{l=1}^{N}\left[\frac{\partial \mathsf{T}_{k} {\mathrm x}_l}{\partial \tau} \frac{1}{({\mathrm x}_m -  \mathsf{T}_{k} {\mathrm x}_l)^2} -\frac{\partial \mathsf{T}_{k}^{-1}  {\mathrm x}_l}{\partial \tau} \frac{1}{({\mathrm x}_m -  \mathsf{T}_{k}^{-1}  {\mathrm x}_l)^2}\right]=0\;,\;m=1,2,...N\;,\label{semiEOMontauu} 
\end{eqnarray}
which are the equations of motion along the continuous variable $\tau$. 
\\
\\
\textbf{Constraint equation}:  Since the Lagrangian \eqref{contisemiL} contains two type of discrete variables, namely $\mathsf{T}^{-1}_k \mathrm{x}$ and $\mathsf{T}_k \mathrm{x}$, we have
\begin{eqnarray}
\frac{\partial \mathscr{L}_{\tau_i}}{\partial\mathsf{T}^{-1}_k \mathrm{x}} =0 \;, \nn
\end{eqnarray} 
which gives
\begin{eqnarray}
-1&=&\sum_{l=1}^{N}\frac{{\partial \mathsf{T}_k  {\mathrm x}_l}}{\partial \tau }\frac{1}{\left(\mathrm x_m - \mathsf{T}_k {\mathrm x}_l \right)^2}\;, \label{semiconstr1}
\end{eqnarray}
and 
\begin{eqnarray}
\frac{\partial \mathscr{L}_{\tau_i}}{\partial \mathsf{T}_k \mathrm{x}} =0 \;, \nn
\end{eqnarray}
which gives
\begin{eqnarray}
-1&=&\sum_{l=1}^{N}\frac{{\partial \mathsf{T}^{-1}_k  {\mathrm x}_l}}{\partial \tau }\frac{1}{\left(\mathrm x_m - \mathsf{T}^{-1}_k {\mathrm x}_l \right)^2}\;. \label{semiconstr2}
\end{eqnarray}
Equations \eqref{semiconstr1} and \eqref{semiconstr2} are the constraints on semi-discrete time level. Combining \eqref{semiconstr1} and \eqref{semiconstr2} together, we obtain the semi-discrete equations of motion \eqref{semiEOMontauu}.
\\
\\
\textbf{Legendre transformation}: In the semi-discrete situation, we have both discrete and semi-discrete Lagrangian. The Legendre transformation for discrete Lagrangian has already been given in \eqref{dtNTLF} and Legendre transformation for semi-discrete Lagrangian is given by  
\begin{eqnarray}
\mathscr{H}_{(\tau)}(\mathsf{T}_{k}{\mathcal{P}}_{ml},\mathsf{T}_{k}{\rho}_{ml},\mathsf{T}_{k}{\nu}_{ml},\mathsf{T}_{k}{\Omega}_{ml},{\mathrm x})&=&\sum_{m,l=1}^{N}  \left[ \mathsf{T}_{k}{\nu}_{ml} ({\mathrm x}_m - \mathsf{T}_{k} {\mathrm x}_l)-\mathsf{T}_{k}{\mathcal{P}}_{ml}\frac{\partial \mathsf{T}_{k} {\mathrm x}_l}{\partial \tau} \right] \nn \\
&&+\frac{1}{2}{\sum\limits_{\mathop {m,l = 1}\limits_{m \ne l}}^{N}} \Bigg[\mathsf{T}_{k}{\Omega}_{ml}(\mathsf{T}_{k} {\mathrm x}_m -\mathsf{T}_{k} {\mathrm x}_l)\nonumber \\
&& + \mathsf{T}_{k}{\rho}_{ml}\left(\frac{\partial \mathsf{T}_{k} {\mathrm x}_m}{\partial \tau}-\frac{\partial \mathsf{T}_{k} {\mathrm x}_l}{\partial \tau} \right) \Bigg] \nonumber \\
&&-\mathscr{L}_{(\tau)}\left({\mathrm x},\mathsf{T}_k{{\mathrm x}},\frac{\partial \mathsf{T}_{k} {\mathrm x}}{\partial \tau}\right)\;,\label{semiLFontau}
\end{eqnarray}
where
\begin{subequations}\label{gsemiLT}
\begin{eqnarray}
\mathsf{T}_{k}{\mathcal{P}}_{ml} &=&\frac{1}{\mathrm x_m - \mathsf{T}_{k}{\mathrm x}_j}\;, \label{semiLTFN}\\
 \mathsf{T}_{k}{\rho}_{ml} &=&-\frac{1}{\mathsf{T}_{k} {\mathrm x}_m -\mathsf{T}_{k} {\mathrm x}_l}\;, \label{semiLTFT}\\
 \mathsf{T}_{k}{\nu}_{ml}&=& \frac{\partial \mathsf{T}_{k} {\mathrm x}_l}{\partial \tau}\frac{1}{\left(\mathrm x_m - \mathsf{T}_{k} {\mathrm x}_l \right)^2}\;, \label{semiLTSN} \\
 \mathsf{T}_{k}{\Omega}_{ml}&=&\left(\frac{\partial \mathsf{T}_{k} {\mathrm x}_m}{\partial \tau}-\frac{\mathsf{T}_{k} {\mathrm x}_l}{\partial \tau} \right)\frac{1}{(\mathsf{T}_{k} {\mathrm x}_m -\mathsf{T}_{k} {\mathrm x}_l)^2}\;.\label{semiLTST}  
\end{eqnarray}
\end{subequations}
Using \eqref{semiLFontau} with \eqref{contisemiL}, we obtain 
\begin{eqnarray}
\mathscr{H}_{(\tau)}(\mathsf{T}_{k}{\mathcal{P}}_{ml},\mathsf{T}_{k}{\rho}_{ml},\mathsf{T}_{k}{\nu}_{ml},\mathsf{T}_{k}{\Omega}_{ml},{\mathrm x})&=&\sum_{m,l=1}^{N}\frac{\mathsf{T}_{k}{\nu}_{ml}}{\mathsf{T}_{k}{\mathcal{P}}_{ml}}-\frac{1}{2}{\sum\limits_{\mathop {m,l = 1}\limits_{m \ne l}}^{N}} \frac{\mathsf{T}_{k}{\Omega}_{ml}}{\mathsf{T}_{k}{\rho}_{ml}} \nn \\
&&-\sum_{m=1}^{N}\Bigg[\frac{1}{\mathsf{T}_{k}{\mathcal{P}}_{mm}} +p\frac{\mathsf{T}_{k}{\nu}_{mm}}{(\mathsf{T}_{k}{\mathcal{P}}_{mm})^2} \Bigg]\;,\label{semiHamiltau} 
\end{eqnarray}
which is the semi-discrete time Hamiltonian associated with the continuous time variable $\tau$.
\\
\\
The action associated with the continuous curve is given by
\begin{eqnarray}
{S}_{(\tau)}&=&\int_{\tau_0}^{\tau_{1}}\mathsf{d} \tau \Bigg[ \sum_{m,l=1}^{N}  \left[ \mathsf{T}_{k}{\nu}_{ml} ({\mathrm x}_m - \mathsf{T}_{k} {\mathrm x}_l)-\mathsf{T}_{k}{\mathcal{P}}_{ml}\frac{\partial \mathsf{T}_{k} {\mathrm x}_l}{\partial \tau} \right] \nn \\
&&+\frac{1}{2}{\sum\limits_{\mathop {m,l = 1}\limits_{m \ne l}}^{N}} \Bigg[\mathsf{T}_{k}{\Omega}_{ml}(\mathsf{T}_{k} {\mathrm x}_m -\mathsf{T}_{k} {\mathrm x}_l)+ \mathsf{T}_{k}{\rho}_{ml}\left(\frac{\partial \mathsf{T}_{k} {\mathrm x}_m}{\partial \tau}-\frac{\partial \mathsf{T}_{k} {\mathrm x}_l}{\partial \tau} \right) \Bigg] \nn \\
&&-\mathscr{H}_{(\tau)}(\mathsf{T}_{k}{\mathcal{P}}_{ml},\mathsf{T}_{k}{\rho}_{ml},\mathsf{T}_{k}{\nu}_{ml},\mathsf{T}_{k}{\Omega}_{ml},{\mathrm x})\Bigg] \;. \label{semiactionTAU} 
\end{eqnarray}
The variation of action \eqref{semiactionTAU} with $\mathsf{T}_{k}{\nu}_{ml} \rightarrow \mathsf{T}_{k}{\nu}_{ml}+\delta \mathsf{T}_{k}{\nu}_{ml}$, $\mathsf{T}_{k}{\mathcal{P}}_{ml} \rightarrow \mathsf{T}_{k}{\mathcal{P}}_{ml}+\delta \mathsf{T}_{k}{\mathcal{P}}_{ml}$, $\mathsf{T}_{k}{\rho}_{ml}\rightarrow\mathsf{T}_{k}{\rho}_{ml}+ \delta \mathsf{T}_{k}{\rho}_{ml}$, $\mathsf{T}_{k}{\Omega}_{ml} \rightarrow \mathsf{T}_{k}{\Omega}_{ml}+\delta \mathsf{T}_{k}{\Omega}_{ml}$ and ${\mathrm x}_m \rightarrow {\mathrm x}_{m}+ \delta {\mathrm x}_{m}$ results in
\begin{eqnarray}
\delta {S}_{(\tau)}&=&\int_{\tau_0}^{\tau_1} \mathsf{d} \tau \Bigg(   \sum_{m,l=1}^{N}\delta\mathsf{T}_{k}{\nu}_{ml} \left[ ({\mathrm x}_m - \mathsf{T}_{k} {\mathrm x}_l)-\frac{\partial \mathscr{H}_{(\tau)}}{\partial  \mathsf{T}_{k}{\nu}_{ml}} \right]+\sum_{m,l=1}^{N}\delta \mathsf{T}_{k}{\mathcal{P}}_{ml} \left[ -\frac{\partial \mathsf{T}_{k} {\mathrm x}_l}{\partial \tau} - \frac{\partial \mathscr{H}_{(\tau)}}{\partial  \mathsf{T}_{k}{\mathcal{P}}_{ml}} \right] \nn\\
&&+ {\sum\limits_{\mathop {m,l = 1}\limits_{m \ne l}}^{N}} \delta \mathsf{T}_{k}{\Omega}_{ml} \left[ \frac{1}{2}(\mathsf{T}_{k} {\mathrm x}_m -\mathsf{T}_{k} {\mathrm x}_l)-\frac{\partial \mathscr{H}_{(\tau)}}{\partial  \mathsf{T}_{k}{\Omega}_{ml}}\right] \nn \\
&&+ {\sum\limits_{\mathop {m,l = 1}\limits_{m \ne l}}^{N}} \delta \mathsf{T}_{k}{\rho}_{ml} \left[\frac{1}{2} \left(\frac{\partial \mathsf{T}_{k} {\mathrm x}_m}{\partial \tau}-\frac{\partial \mathsf{T}_{k} {\mathrm x}_l}{\partial \tau} \right) - \frac{\partial \mathscr{H}_{(\tau)}}{\partial  \mathsf{T}_{k}{\rho}_{ml}}\right] \nn\\
&&+ \sum_{m=1}^{N}\delta {\mathrm x}_m \Bigg[\sum_{l=1}^{N} \left[\left(  \mathsf{T}_{k}{\nu}_{ml} - {\nu}_{ml}\right)-\frac{\partial {\mathcal{P}}_{ml}}{\partial \tau} \right] +\frac{1}{2}{\sum\limits_{\mathop {l = 1}\limits_{m \ne l}}^{N}} \left[\left(\Omega_{ml} - \Omega_{lm} \right)- \left(\frac{\partial \rho_{ml}}{\partial \tau}-\frac{\partial {\rho}_{lm}}{\partial \tau} \right)\right] \nn \\
&&-\frac{\partial\mathscr{H}_{(\tau)}}{\partial \mathrm{x}_{m}} \Bigg]\Bigg)\;. \label{semiactionTAUU}
\end{eqnarray} 
According to the condition $\delta {S}=0$, we obtain
\begin{subequations} \label{gsemiHamil}
\begin{eqnarray}
\frac{\partial \mathscr{H}_{(\tau)}}{\partial  \mathsf{T}_{k}{\nu}_{ml}}  &=& {\mathrm x}_m - \mathsf{T}_{k} {\mathrm x}_l\;, \\ \label{semiHamilST}
 \frac{\partial \mathscr{H}_{(\tau)}}{\partial  \mathsf{T}_{k}{\mathcal{P}}_{ml}} &=& -\frac{\partial \mathsf{T}_{k} {\mathrm x}_l}{\partial \tau} \;, \\ \label{semiHamilND}
\frac{\partial \mathscr{H}_{(\tau)}}{\partial  \mathsf{T}_{k}{\Omega}_{ml}} &=& \frac{1}{2}(\mathsf{T}_{k} {\mathrm x}_m -\mathsf{T}_{k} {\mathrm x}_l) \;, \\ \label{semiHamilRD}
 \frac{\partial \mathscr{H}_{(\tau)}}{\partial  \mathsf{T}_{k}{\rho}_{ml}} &=& \frac{1}{2} \left(\frac{\partial \mathsf{T}_{k} {\mathrm x}_m}{\partial \tau}-\frac{\partial \mathsf{T}_{k} {\mathrm x}_l}{\partial \tau} \right)\;, \\ \label{semiHamilFOUR}
\frac{\partial\mathscr{H}_{(\tau)}}{\partial \mathrm{x}_{m}} &=&\sum_{l=1}^{N} \left[\left(  \mathsf{T}_{k}{\nu}_{ml} - {\nu}_{ml}\right)-\frac{\partial {\mathcal{P}}_{ml}}{\partial \tau} \right] \nn \\ 
&&+\frac{1}{2}{\sum\limits_{\mathop {l = 1}\limits_{m \ne l}}^{N}} \left[\left(\Omega_{ml} - \Omega_{lm} \right)- \left(\frac{\partial \rho_{ml}}{\partial \tau}-\frac{\partial {\rho}_{lm}}{\partial \tau} \right)\right] \;,  \label{semiHamilFIVE}
\end{eqnarray}
\end{subequations}
which are the Hamilton's equations for $\tau$-direction in semi-discrete time level and \eqref{semiHamilFIVE} produces the equations of motion in semi-discrete time level \eqref{semiEOMontauu}.
\subsection{Continuous-time Lagrangian 1-form structure}\label{sectioncontime}
\begin{figure}[h]
\begin{center}
\tikzset{middlearrow/.style={
        decoration={markings,
            mark= at position 0.5 with {\arrow{#1}} ,
       },
        postaction={decorate}
    }
}
\begin{tikzpicture}[scale=0.5]
\draw[->] (0,0,0) -- (13,0,0) node[anchor=west] {$t_{i}$};
 \draw[->] (0,0,0) -- (0,12,0) node[anchor=south] {$X(t)$};
 \draw[->] (0,0,0) -- (0,0,12) node[anchor=south] {$t_{j}$};
\draw[middlearrow={{triangle 60}},thick] (2,9,0) .. controls (4,8,3) and (6,9,3) .. (12,11,5);
\draw[middlearrow={{triangle 60}},dashed,thick] (2,9,0) .. controls (4,6,3) and (5,7,3) .. (12,11,5);
\draw [middlearrow={{triangle 60}},dashed,thick] (2,-1,0) .. controls (4,-1,-1) and (5,-1,-2) .. (12,0,5);
\draw[middlearrow={{triangle 60}},thick](2,-1,0) .. controls (4,-1,3) and (6,-1,7) .. (12,0,5);
 \draw (4,9.5,0) node[anchor=north west] {${E} _{\Gamma}$};
\draw (2.5,6.5,0) node[anchor=north west] {${E}_{{\Gamma}^\prime}$};
\draw (4,0,3) node[anchor=north west] {${\Gamma}$};
\draw (6,-1,-0.5) node[anchor=north west] {${{\Gamma}^\prime}$};
\draw (-0.25,10.5,0) node[anchor=north west] {${X}(t(s_0))$};
\draw (12,12,5) node[anchor=north west] {${X}(t(s_1))$};
\draw (-0.5,0,-0.25) node[anchor=north west] {$t(s_0)$};
\draw (10.5,-1,3) node[anchor=north west] {$t(s_1)$};
\fill (2,9,0) circle (0.15);
\fill (12,11,5) circle (0.15);
\fill (2,-1,0) circle (0.15);
\fill (12,0,5) circle (0.15);
\draw[dashed] (2,9,0)--(2,-1,0);
\draw[dashed] (12,11,5)--(12,0,5);
\end{tikzpicture}
\end{center}
\caption{The curve $\Gamma$ and ${E}_{\Gamma}$ in the $X-t$ configuration.}\label{pcon1}
\end{figure}
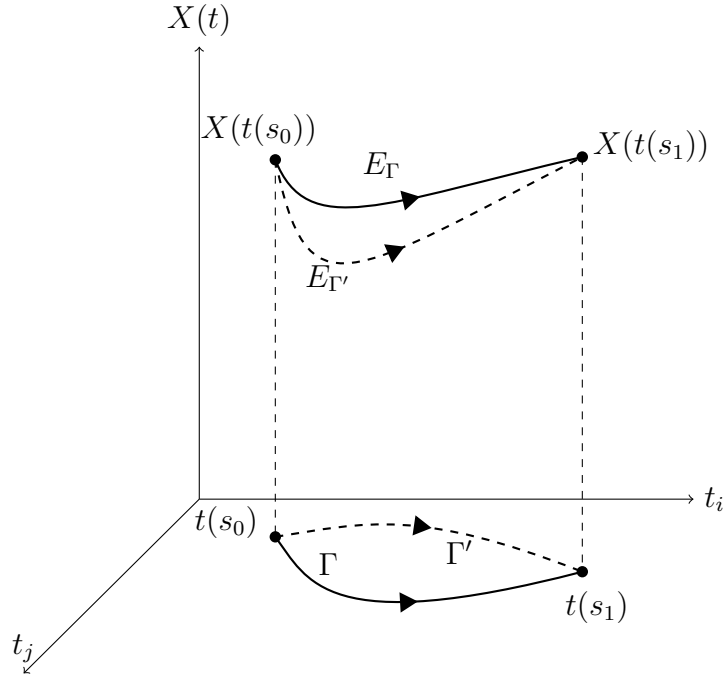

In this section, we will investigate the variational principle for the continuous-time Lagrangian 1-form structure. Suppose now we have a set of generalised coordinates $X(t)\equiv (X_1(t), X_2(t),...,X_N(t))$ and a set of time variables $t(s) \equiv (t_1 (s) ,t_2 (s), t_3 (s),...,t_N(s))$ which are parameterised by the parameter $s$ with the boundary: $s_0 \leq s \leq s_1$. The action functional for this case is written in the form
\begin{eqnarray}  
S_\Gamma[X(t)]&=& \int_{\Gamma} \left(\sum_{i=1}^N \mathscr{L}_{(t_i)}  \mathsf {d} t_i \right) = \int_{s_0}^{s_1} \mathsf {d} s L(s)\; \label{coneq1},
\end{eqnarray}
where $L(s)\equiv \mathscr{L}_{(t_i)}[X(t(s)),\{X_{(j)} (t(s))\}]\mathsf{d}t_i/\mathsf{d}s $  is the multi-time Lagrangian and $X_{(j)} (t)\equiv \mathsf {d}X/\mathsf {d}t_j$, where $j=1,2,...,N$. The curve $\Gamma$ is on the space of independent variables starting from point $t(s_0)$ to point $t(s_1)$, see figure \ref{pcon1}.
\subsubsection{Variation of independent variables}
We consider the variation of the curve $\Gamma\rightarrow\Gamma'$ on the space of independent variables while the end points are fixed as shown in figure \ref{pcon1}. By doing so, it is convenient to write the Lagrangian as a function of times, i.e., $\mathscr{L}_{(t_i)}\equiv \mathscr{L}_{(t_i)}(t)$. We then perform the variation $t(s)\rightarrow t(s)+\delta t(s)$ resulting in a new curve $\Gamma'$. A new action functional is
 \begin{eqnarray}  
S_{\Gamma'}[X(t+\delta t)]&=& \int_{s_0}^{s_1} \mathsf {d} s \left(\sum_{i=1}^N \mathscr{L}_{(t_i)}(t+\delta t) \frac{\mathsf {d} (t_i+\delta t_i)}{\mathsf {d} s} \right)\; \label{coneq8}.
\end{eqnarray}
Performing the Taylor expansion and keeping only the first two contributions in the series, we find that the variation of action is
 \begin{eqnarray}
\delta {S}_\Gamma\equiv S_{\Gamma'}-S_{\Gamma}&=& \int_{s_0}^{s_1} \mathsf {d} s \left[\sum_{i,j=1}^{N} \delta t_j  \frac{\partial \mathscr{L}_{(t_i)} }{\partial t_j}  \frac{\mathsf {d} t_i}{\mathsf {d} s}+ \sum_{i=1}^N \mathscr{L}_{(t_i)}\frac{\mathsf {d} \delta t_i}{\mathsf {d} s} \right] \; .\label{coneq9}
\end{eqnarray} 
Using integration by parts on the second term of \eqref{coneq9}, with conditions $\delta t(s_0)=\delta t(s_1)=0$, we have
\begin{eqnarray}
\delta S_\Gamma&=& \int_{s_0}^{s_1} \mathsf {d} s \left[\sum_{i,j=1}^{N} \delta t_j  \frac{\partial \mathscr{L}_{(t_i)} }{\partial t_j} \frac{\mathsf {d} t_i}{\mathsf {d} s}- \sum_{i=1}^N \frac{\mathsf {d} \mathscr{L}_{(t_i)}}{\mathsf {d} s} \delta t_i\right] \;.\label{coneq10}
\end{eqnarray}
Next, using the chain rule relation
\begin{eqnarray}
\frac{ \mathsf{d} \mathscr{L}_{(t_i)}}{\mathsf{d} s}&=& \sum_{j=1}^{N}\frac{\partial \mathscr{L}_{(t_i)}}{\partial t_j}\frac{\mathsf {d} t_j}{\mathsf {d} s}\;,
\end{eqnarray}
equation \eqref{coneq10} can be rewritten as
\begin{eqnarray}
\delta {S}_\Gamma &=& \int_{s_0}^{s_1} \mathsf {d} s \left[ {\sum\limits_{\mathop {i,j = 1}\limits_{i \ne j}}^N} \delta t_i \left(\frac{\partial \mathscr{L}_{(t_i)} }{\partial t_j}-\frac{\partial\mathscr{L}_{(t_j)} }{\partial t_i} \right)\frac{\mathsf {d} t_j}{\mathsf {d} s} \right]\;. \label{coneq11}
\end{eqnarray}
From the least action principle: $\delta{S}_\Gamma =0$, we obtain the relation 
\begin{eqnarray} \label{coneq12}
\frac{\partial \mathscr{L}_{(t_i)} }{\partial t_j}=\frac{\partial \mathscr{L}_{(t_j)} }{\partial t_i} \; , \quad i, j = 1, 2, 3, ..., N \quad \mbox{and} \quad i\neq j\;.
\end{eqnarray}
Equations \eqref{coneq12} are called the \emph{continuous-time closure relations} which guarantee the invariance of action under the local deformation of a curve $\Gamma$ on the space of independent variables. 
\\
\\
\textbf{\emph{Remark}}:\emph{
We have seen that the variation of the action gives the closure relation. Here, we are approaching the problem from different perspective, namely from geometric point of view.
\\
\\
Suppose that $\alpha$ is a differential (k-1)-form. The generalised Stokes' theorem states that \textit{the integral of its exterior derivative over the surface of smooth oriented k-dimensional manifold $\Omega$ is equal to its integral of along the boundary $\partial\Omega$ of the manifold $\Omega$ }\cite{Fortney}:
\begin{eqnarray}
\int_{\partial\Omega} \alpha = \int_\Omega d\alpha \;. \label{eq1}
\end{eqnarray}
We now introduce an object $dS$ given by 
\begin{eqnarray}
d{{S}}= \;\sum_{i=1}^{N}\mathscr{L}_{({t_i})}dt_i \;,
\end{eqnarray}
as a 1-form on the $N$-dimensional space of independent variables and, therefore, the action \eqref{coneq1} becomes $S=\int_\Gamma dS$. Applying an exterior derivative to the smooth function coefficients which, in this case, is the Lagrangian, (\ref{eq1}) becomes
\begin{eqnarray}
\begin{aligned}
\oint_{\partial\Omega}\sum_{i=1}^{N}\mathscr{L}_{(t_i)}dt_i = & \iint_{\Omega} \sum_{1\leq i < j \leq N}^{N} \bigg( \frac{\partial \mathscr{L}_{(t_j)}}{\partial t_i}-\frac{\partial \mathscr{L}_{(t_i)}}{\partial t_j}\bigg) dt_i\wedge dt_j\;. 
\end{aligned}\label{eq2}
\end{eqnarray}
The left-hand side of \eqref{eq2} is equivalent to $\int_\Gamma dS-\int_{\Gamma'} dS$, see figure \ref{pcon1}.
It is suggested in \cite{FrankB} that, since the multidimensional consistency exists on the level of equations of motion provided by the Lagrangian with the use of Euler-Lagrange equations, it is essential for such consistency to be encoded in the Lagrangian. As a result, the Lagrangian 1-form is required to be a closed form on the solutions of the system. Thus, the right-hand side of \eqref{eq2} vanishes, since the exterior derivative operating on the closed form gives vanishing result. Therefore, we obtain
\begin{eqnarray}
\frac{\partial \mathscr{L}_{(t_j)}}{\partial t_i}-\frac{\partial \mathscr{L}_{(t_i)}}{\partial t_j} = 0 \; , \quad i, j = 1, 2, 3, ..., N \quad \mbox{and} \quad i\neq j\;,
\end{eqnarray}
which are the closure relations of the system that evolves in the $N$-dimensional space of independent variables.}
\\
\\
From \eqref{coneq12}, it implies that the action functional does not depend on path sharing the end-points in the space of independent variables. Actually, we can think that the continuous path constitutes from tiny discrete elements. Then, path independent property in the continuous-time case, also known as multidimensional consistency, is a direct consequence of path independent in the discrete-time case. Furthermore, with this property, there is a family of paths(homotopy), sharing the end points, that can be continuously transformed to each other in $N$-dimensional space $\mathscr{M}$ of independent variables. 
\\
\\
\textbf{Theorem 3}(\textbf{Cauchy's Theorem}\cite{dd}). If $\mathscr{M}$ is simply connected, $\oint_\Gamma L(s)ds=0$ for every piecewisely smooth closed curve $\Gamma$ and every analytic $L$.
\\
\\
With the definition of simply connected space, every closed curve $\Gamma\Gamma'$ in such space, say $\mathscr{M}$, is $homotopic$ to zero : $\oint_{\Gamma \Gamma'} L(s)ds=\int_\Gamma L(s)ds-\int_{\Gamma'} L(s)ds$, see figure \ref{pcon1}. 
\subsubsection{Variation on dependent variables}
Now, we introduce a new curve called $E_{\Gamma}$ shown in figure \ref{pcon1}. The action functional for this curve is given by
\begin{eqnarray}  
S_{E_{\Gamma}}[X(t)]&=& \int_{s_0}^{s_1} \mathsf {d} s \left(\sum_{i=1}^N \mathscr{L}_{(t_i)}(X (t(s)),\{X_{(j)}(t(s)) \}) \frac{\mathsf {d} t_i}{\mathsf {d} s} \right)\; \label{coneq1001},\;\;\;j=1,2,3,..,N\;.
\end{eqnarray}
The variation of the dependent variables,  $X\rightarrow X +\delta X$, gives us another curve $E'_{\Gamma}$ with the action functional
\begin{eqnarray}  
S_{E'_{\Gamma}}[X (t)+\delta X (t)]&=& \int_{s_0}^{s_1} \mathsf {d} s \left(\sum_{i=1}^N \mathscr{L}_{(t_i)}(X(t)+\delta X(t),\{X_{(j)}(t)+\delta X_{(j)}(t) \}) \frac{\mathsf {d} t_i}{\mathsf {d} s} \right)\; \label{coneq10011}.
\end{eqnarray}
Next, performing Taylor expansion with respect to $(\delta X, \;\delta X_{(j)})$, and keeping only the first two contributions in a series, we obtain the variation of the action
\begin{eqnarray}
\delta S_{E_{\Gamma}}\equiv S_{E'_{\Gamma}}-S_{E_{\Gamma}} &=& \int_{s_0}^{s_1} \mathsf {d} s \left[\sum_{i=1}^{N} \left(\delta X \frac{\partial\mathscr{L}_{(t_i)}}{\partial X} +\sum_{j=1}^{N} \delta X_{(j)} \frac{\partial \mathscr{L}_{(t_i)}}{\partial X_{(j)} }\right) \frac{\mathsf {d}t_i}{\mathsf {d}s}\right] \; \label{coneq2}.
\end{eqnarray} 
Using the chain rule relation
\begin{eqnarray}
\frac{\mathsf {d}\delta X}{\mathsf {d}s} = \sum_{i=1}^{N} \delta X_{(i)}  \frac{\mathsf {d}t_i}{\mathsf {d}s}\; \label{coneq3},
\end{eqnarray}
together with new variables
\begin{eqnarray}
\delta Y_{ij}\equiv \delta X_{(i)} \frac{\mathsf {d}t_i}{\mathsf {d}s}-\delta X_{(j)} \frac{\mathsf {d}t_j}{\mathsf {d}s}\;,\,\text{where}\; i<j = 1,2,...,N\;. \label{coneq4}
\end{eqnarray}
By combining \eqref{coneq3} and \eqref{coneq4}, we obtain
\begin{eqnarray}
\frac{1}{N} \left(\frac{\mathsf {d}\delta X}{\mathsf {d}s} +{\sum\limits_{\mathop {i,j = 1}\limits_{i \ne j}}^N} \delta Y_{ij}\right) &=& \sum_{i=1}^{N}\delta X_{(i)}  \frac{\mathsf {d}t_i}{\mathsf {d}s}\;. \label{coneq5}
\end{eqnarray}
Using \eqref{coneq5}, the variation of action \eqref{coneq2} becomes
\begin{eqnarray}
\delta {S}_{E_{\Gamma}} &=& \int_{s_0}^{s_1} \mathsf {d} s\Bigg \lbrace \left[ \sum_{i=1}^{N} \frac{\partial \mathscr{L}_{(t_i)}}{\partial X}\frac{\mathsf {d}t_i}{\mathsf {d}s}\right]\delta X \nonumber \\ 
&&+ \frac{1}{N} \frac{\mathsf {d}\delta X}{\mathsf {d}s} \left[\sum_{i=1}^{N}\frac{\partial\mathscr{L}_{(t_i)}}{\partial X_{(i)}} + {\sum\limits_{\mathop {i,j = 1}\limits_{i \ne j}}^N} \frac{\partial \mathscr{L}_{(t_i)}}{\partial X_{(j)}}\frac{\mathsf {d} t_i/ \mathsf {d} s}{\mathsf {d} t_j/ \mathsf {d} s}\right] \nonumber \\ 
&&+ \frac{1}{N}{\sum\limits_{\mathop {i,j = 1}\limits_{i \ne j}}^N} \delta Y_{ij} \Bigg[\frac{\partial \mathscr{L}_{(t_i)}}{\partial X_{(i)}} -\frac{\partial \mathscr{L}_{(t_j)}}{\partial X_{(j)}} -\frac{\partial \mathscr{L}_{(t_i)}}{\partial X_{(j)}}\frac{\mathsf {d} t_i/ \mathsf {d} s}{\mathsf {d} t_j/ \mathsf {d} s}+\frac{\partial \mathscr{L}_{(t_j)}}{\partial X_{(i)}}\frac{\mathsf {d} t_j/ \mathsf {d} s}{\mathsf {d} t_i/ \mathsf {d} s} \nonumber \\
&&+{\sum\limits_{\mathop {k = 1}\limits_{k \ne i,j}}^N} \left[\frac{\partial \mathscr{L}_{(t_k)}}{\partial X_{(i)}}\frac{\mathsf {d} t_k/ \mathsf {d} s}{\mathsf {d} t_i/ \mathsf {d} s} - \frac{\partial \mathscr{L}_{(t_k)}}{\partial X_{(j)}}\frac{\mathsf {d}t_k/ \mathsf {d} s}{\mathsf {d} t_j/ \mathsf {d} s} \right]\Bigg] \Bigg\rbrace  \; \label{coneq6}.
\end{eqnarray}
Integrating by parts the second term in \eqref{coneq6} with conditions: $\delta X(s_0)=\delta X(s_1)=0$, then we have
\begin{eqnarray}
\delta {S}_{E_{\Gamma}} &=& \int_{s_0}^{s_1} \mathsf {d} s\Bigg \lbrace \delta X  \Bigg[\sum_{i=1}^{N}  \frac{\partial \mathscr{L}_{(t_i)}}{\partial X}\frac{\mathsf {d}t_i}{\mathsf {d}s} \nonumber \\ 
&&- \frac{1}{N} \frac{\mathsf {d}}{\mathsf {d}s} \left(\sum_{i=1}^{N} \frac{\partial \mathscr{L}_{(t_i)}}{\partial X_{(i)}} + {\sum\limits_{\mathop {i,j = 1}\limits_{i \ne j}}^N} \frac{\partial \mathscr{L}_{(t_i)}}{\partial X_{(j)}}\frac{\mathsf {d} t_i/ \mathsf {d} s}{\mathsf {d} t_j/ \mathsf {d} s}\right) \Bigg] \nonumber \\ 
&&+ \frac{1}{N}{\sum\limits_{\mathop {i,j = 1}\limits_{i \ne j}}^N} \delta Y_{ij} \Bigg[\frac{\partial \mathscr{L}_{(t_i)}}{\partial X_{(i)}} -\frac{\partial \mathscr{L}_{(t_j)}}{\partial X_{(j)}} -\frac{\partial \mathscr{L}_{(t_i)}}{\partial X_{(j)}}\frac{\mathsf {d} t_i/ \mathsf {d} s}{\mathsf {d} t_j/ \mathsf {d} s}+\frac{\partial \mathscr{L}_{(t_j)}}{\partial X_{(i)}}\frac{\mathsf {d} t_j/ \mathsf {d} s}{\mathsf {d} t_i/ \mathsf {d} s} \nonumber \\
&&+{\sum\limits_{\mathop {k = 1}\limits_{k \ne i,j}}^N} \left[\frac{\partial \mathscr{L}_{(t_k)}}{\partial X_{(i)}}\frac{\mathsf {d} t_k/ \mathsf {d} s}{\mathsf {d} t_i/ \mathsf {d} s} - \frac{\partial \mathscr{L}_{(t_k)}}{\partial X_{(j)}}\frac{\mathsf {d}t_k/ \mathsf {d} s}{\mathsf {d} t_j/ \mathsf {d} s} \right]\Bigg] \Bigg\rbrace  \;
 \label{coneq7.1}. 
\end{eqnarray}
Now, we have two sets of variables $\{\delta X \}$ and $\{ \delta Y_{ij} \}$ in \eqref{coneq7.1}. We can interpret that $\{\delta X_i\}$ is a set of variables in tangential direction with the curve $E_\Gamma$ while $\{ \delta Y_{ij} \}$ is a set of variables in transversal direction with the curve $E_\Gamma$. Then, imposing the least action principle: $\delta S_{E_{\Gamma}}=0$, we obtain
\begin{eqnarray}
 \sum_{i=1}^{N}\frac{\partial \mathscr{L}_{(t_i)}}{\partial X}\frac{\mathsf {d}t_i}{\mathsf {d}s}- \frac{1}{N} \frac{\mathsf {d}}{\mathsf {d}s} \left( \sum_{i=1}^{N}\frac{\partial \mathscr{L}_{(t_i)}}{\partial X_{(i)}} + {\sum\limits_{\mathop {i,j=1}\limits_{i \ne j}}^N} \frac{\partial \mathscr{L}_{(t_i)}}{\partial X_{(j)}}\frac{\mathsf {d} t_i/ \mathsf {d} s}{\mathsf {d} t_j/ \mathsf {d} s}\right) =0\;, \label{coneq7}
\end{eqnarray}
since $\delta X\neq 0$. Equations \eqref{coneq7} are called the generalised Euler-Lagrange equations for Lagrangian 1-form structure. Furthermore, we also have
\begin{eqnarray}
0 &=& {\sum\limits_{\mathop {i,j = 1}\limits_{i \ne j}}^N} \left[\left(\frac{\partial \mathscr{L}_{(t_i)}}{\partial X_{(i)}} -\frac{\partial \mathscr{L}_{(t_j)}}{\partial X_{(j)}} \right)\frac{\mathsf {d} t_i}{\mathsf {d} s} \frac{\mathsf {d} t_j}{\mathsf {d} s}-\frac{\partial \mathscr{L}_{(t_i)}}{\partial X_{(j)}}\left(\frac{\mathsf {d} t_i}{\mathsf {d} s} \right)^2+\frac{\partial \mathscr{L}_{(t_j)}}{\partial X_{(i)}}\left(\frac{\mathsf {d} t_j}{\mathsf {d} s} \right)^2\right] \nonumber \\
&&+{\sum\limits_{\mathop {k = 1}\limits_{k \ne i,j}}^N} \left[\frac{\partial \mathscr{L}_{(t_k)}}{\partial X_{(i)}}\frac{\mathsf {d} t_j}{\mathsf {d}s}- \frac{\partial \mathscr{L}_{(t_k)}}{\partial X_{(j)}}\frac{\mathsf {d} t_i}{\mathsf {d} s} \right]\frac{\mathsf {d} t_k}{\mathsf {d} s} \;,\label{coneq8}
\end{eqnarray}
since  $\delta Y_{ij} \neq 0$. Equations \eqref{coneq8} are called the constraint equations. 
%
%
\subsubsection{A set of compatible Lagrangian equations}
In the previous subsection, we have a set of Lagrangian equations including the Euler-Lagrange equation, constraint equation and closure relation. For simplification in further analysis, we will consider only for the case of two time variables. Thus, we have a set of equations as follows
\begin{subequations}\label{LE1}
\begin{eqnarray}
\sum_{i=1}^{2}\frac{\partial \mathscr{L}_{(t_i)}}{\partial X}\frac{\mathsf {d}t_i}{\mathsf {d}s}- \frac{1}{2} \frac{\mathsf {d}}{\mathsf {d}s} \left(\sum_{i=1}^{2}\frac{\partial \mathscr{L}_{(t_i)}}{\partial X_{(i)}} + {\sum\limits_{\mathop {i,j=1}\limits_{i \ne j}}^2} \frac{\partial \mathscr{L}_{(t_i)}}{\partial X_{(i)}}\frac{\mathsf {d} t_i/ \mathsf {d} s}{\mathsf {d} t_j/ \mathsf {d} s}\right) =0 \;, \label{coneqret12}
\end{eqnarray}
for the Euler-Lagrange equation,
\begin{eqnarray}
\frac{\partial \mathscr{L}_{(t_2)}}{\partial X_{(1)}}\left(\frac{\mathsf {d} t_2}{\mathsf {d} s} \right)^2 +\left(\frac{\partial \mathscr{L}_{(t_1)}}{\partial X_{(1)}} -\frac{\partial \mathscr{L}_{(t_2)}}{\partial X_{(2)}} \right)\frac{\mathsf {d} t_1}{\mathsf {d} s} \frac{\mathsf {d} t_2}{\mathsf {d} s}- \frac{\partial \mathscr{L}_{(t_1)}}{\partial X_{(2)}}\left(\frac{\mathsf {d} t_1}{\mathsf {d} s} \right)^2 = 0 \; \label{coneqconsret12}
\end{eqnarray}
for constraint equation, and
\begin{eqnarray}
\frac{\partial \mathscr{L}_{(t_1)} }{\partial t_2}=\frac{\partial \mathscr{L}_{(t_2)} }{\partial t_1} \; \label{coneqclosret12}
\end{eqnarray}
\end{subequations}
for the closure relation. An interesting fact is that this set of equations \eqref{LE1} can be used to determine the explicit form of the Lagrangian 1-form, e.g. Calogero-Moser systems and Ruijsenaars-Schneider systems, see \cite{FrankB}.
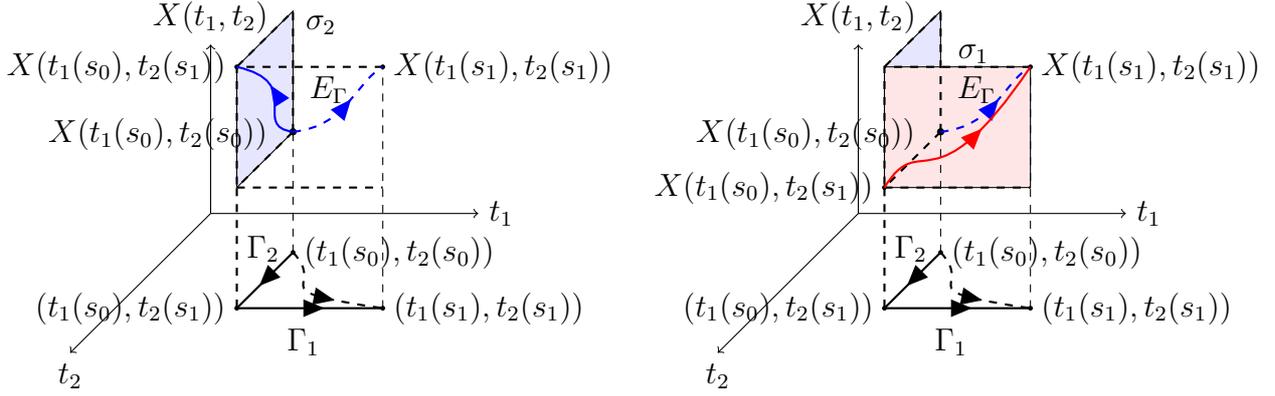
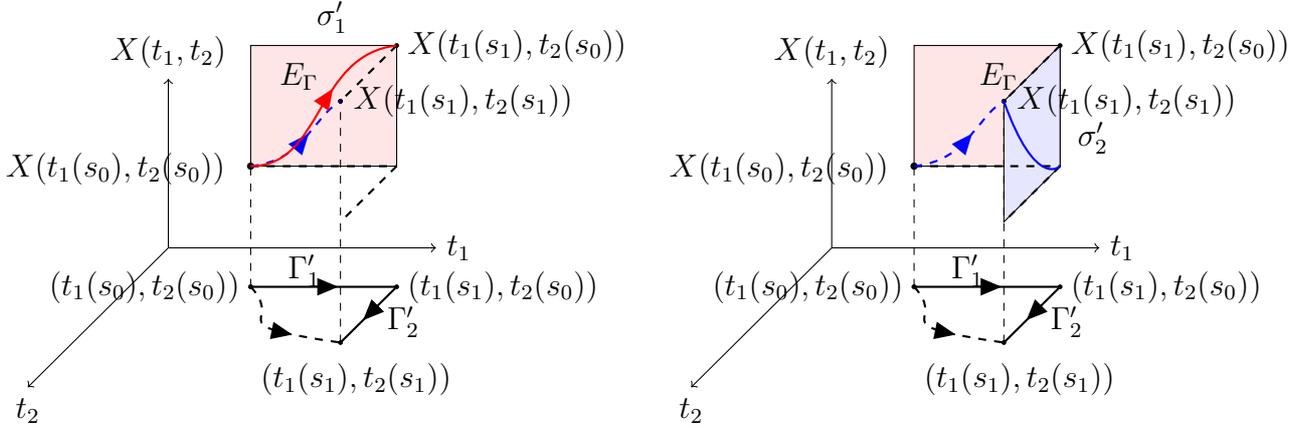
\begin{figure}[h]
\begin{center}
\tikzset{middlearrow/.style={
        decoration={markings,
            mark= at position 0.6 with {\arrow{#1}} ,
        },
        postaction={decorate}
    }
}
\subfigure[The projection of the curve ${E}_\Gamma$ on the plane $\sigma_2$ with constant $t_1(s_0)$.]{
\begin{tikzpicture}[scale=0.32]
\draw[fill=blue!10](1,5,1)--(1,5,7)--(1,10,7)--(1,10,1)--cycle;
 \draw[->] (-2,2,2) -- (9,2,2) node[anchor=west] {$t_{1}$};
 \draw[->] (-2,2,2) -- (-2,9,2) node[anchor=south] {$X(t_1,t_2)$};
 \draw[->] (-2,2,2) -- (-2,2,17) node[anchor=north] {$t_{2}$};
 \fill (1,0,1) circle (0.1) node[anchor=west]{$(t_1(s_0),t_2(s_0))$};
\draw (2,8.5,3) node[anchor=north west] {$E_{\Gamma}$};
 \draw (-2.5,0,-2) node[anchor=north west] {$\Gamma_2$};
 \draw (3,0,8) node[anchor=north west] {$\Gamma_1$};
 \draw (.8,10,.3) node[anchor=north west] {$\sigma_2$};
 \fill (7,0,7) circle (0.1) node[anchor=west]{$(t_1(s_1),t_2(s_1))$} ;
  \fill (1,0,7) circle (0.1) node[anchor=east]{$(t_1(s_0),t_2(s_1))$} ;
\draw[dashed] (1,0,1)--(1,5,1);
\draw[dashed] (7,0,7)--(7,10,7);
\fill (1,5,1) circle (0.15);
\draw (-9.5,6,1) node[anchor=north west] {$X(t_1(s_0),t_2(s_0))$};
\fill (7,10,7) circle (0.1) node[anchor=west]{$ X(t_1(s_1),t_2(s_1))$};
\fill (1,10,7) circle (0.1) node[anchor=east]{$ X(t_1(s_0),t_2(s_1))$};
\draw [thick,dashed,middlearrow={{triangle 45}}] (1,0,1) .. controls (4,0,5) and (1,0,5) .. (7,0,7);
\draw[middlearrow={{triangle 45}},thick] (1,0,1)--(1,0,7);
\draw[middlearrow={{triangle 45}},thick] (1,0,7)--(7,0,7);
\draw[thick,dashed](1,5,1)--(1,5,7)--(1,0,7);
\draw[thick,dashed](1,5,7)--(7,5,7);
\draw[thick,dashed](7,10,7)--(1,10,7)--(1,5,7);
\draw[thick,dashed](1,5,1)--(1,10,1)--(1,10,7);
\draw [middlearrow={{triangle 45}},thick,dashed,blue] (1,5,1) .. controls (4,6,3) and (2,5.5,-3) .. (7,10,7);
\draw [middlearrow={{triangle 45}},thick,blue] (1,5,1) .. controls (0,6,3) and (0,5.5,-3) .. (1,10,7);
\end{tikzpicture}}
\subfigure[The projection of the curve ${E}_\Gamma$ on the plane $\sigma_1$ with constant $t_2(s_1)$.]{
\begin{tikzpicture}[scale=0.32]
\draw[fill=blue!10](1,5,1)--(1,5,7)--(1,10,7)--(1,10,1)--cycle;
\draw[fill=red!10](1,5,7)--(1,10,7)--(7,10,7)--(7,5,7)--cycle;
 \draw[->] (-2,2,2) -- (9,2,2) node[anchor=west] {$t_{1}$};
 \draw[->] (-2,2,2) -- (-2,9,2) node[anchor=south] {$X(t_1,t_2)$};
 \draw[->] (-2,2,2) -- (-2,2,17) node[anchor=north] {$t_{2}$};
 \fill (1,0,1) circle (0.1) node[anchor=west]{$(t_1(s_0),t_2(s_0))$};
\draw (2,8.5,3) node[anchor=north west] {$E_{\Gamma}$};
 \draw (-2.5,0,-2) node[anchor=north west] {$\Gamma_2$};
 \draw (3,0,8) node[anchor=north west] {$\Gamma_1$};
  \draw (1,8.8,.3) node[anchor=north west] {$\sigma_1$};
 \fill (7,0,7) circle (0.1) node[anchor=west]{$(t_1(s_1),t_2(s_1))$} ;
  \fill (1,0,7) circle (0.1) node[anchor=east]{$(t_1(s_0),t_2(s_1))$} ;
\draw[dashed] (1,0,1)--(1,5,1);
\draw[dashed] (7,0,7)--(7,10,7);
\fill (1,5,1) circle (0.15);
\draw (-9.5,6,1) node[anchor=north west] {$X(t_1(s_0),t_2(s_0))$};
\fill (7,10,7) circle (0.1) node[anchor=west]{$ X(t_1(s_1),t_2(s_1))$};
\fill (1,5,7) circle (0.1) node[anchor=east]{$ X(t_1(s_0),t_2(s_1))$};
\draw [thick,dashed,middlearrow={{triangle 45}}] (1,0,1) .. controls (4,0,5) and (1,0,5) .. (7,0,7);
\draw[middlearrow={{triangle 45}},thick] (1,0,1)--(1,0,7);
\draw[middlearrow={{triangle 45}},thick] (1,0,7)--(7,0,7);
\draw[thick,dashed](1,5,1)--(1,5,7)--(1,0,7);
\draw[thick,dashed](1,5,7)--(7,5,7);
\draw[thick,dashed](7,10,7)--(1,10,7)--(1,5,7);
\draw[thick,dashed](1,5,1)--(1,10,1)--(1,10,7);
\draw [middlearrow={{triangle 45}},thick,dashed,blue] (1,5,1) .. controls (4,6,3) and (2,5.5,-3) .. (7,10,7);
\draw [middlearrow={{triangle 45}},thick,red] (1,5,7) .. controls (1,6,3) and (1,2,2) .. (7,10,7);
\end{tikzpicture}}
\subfigure[The projection of the curve ${E}_\Gamma$ on the plane $\sigma_1^\prime$ with constant $t_2(s_0)$.]{
\begin{tikzpicture}[scale=0.32]
\draw[fill=red!10](1,5,1)--(1,10,1)--(7,10,1)--(7,5,1)--cycle;
 \draw[->] (-2,2,2) -- (9,2,2) node[anchor=west] {$t_{1}$};
 \draw[->] (-2,2,2) -- (-2,9,2) node[anchor=south] {$X(t_1,t_2)$};
 \draw[->] (-2,2,2) -- (-2,2,17) node[anchor=north] {$t_{2}$};
 \fill (1,0,1) circle (0.1) ;
 \draw (2.5,10.5,3) node[anchor=north west] {$E_{\Gamma}$};
 \draw (6.5,0,1.9) node[anchor=north west] {$\Gamma_2^\prime$};
 \draw (0.45,0.15,-3.25) node[anchor=north west] {$\Gamma_1^\prime$};
  \draw (3,12,0.3) node[anchor=north west] {$\sigma_1^\prime$};
 \fill (7,0,7) circle (0.1);
 \draw (12,-1.5,7)node[anchor=east]{$(t_1(s_1),t_2(s_1))$} ;
  \fill (1,0,1) circle (0.1) node[anchor=east]{$(t_1(s_0),t_2(s_0))$} ;
  \fill (7,0,1) circle (0.1) node[anchor=west]{$(t_1(s_1),t_2(s_0))$} ;
  \fill (1,5,1) circle (0.15);
\draw (-9.5,6,1) node[anchor=north west] {$X(t_1(s_0),t_2(s_0))$};
\fill (7,10,1) circle (0.1);
 \draw (7,10,1) node[anchor=west]{$ X(t_1(s_1),t_2(s_0))$};
\fill (7,10,7) circle (0.1);
\draw (17,10,7)  node[anchor=east]{$ X(t_1(s_1),t_2(s_1))$};
\draw[dashed] (1,0,1)--(1,5,1);
\draw[dashed] (7,0,7)--(7,10,7);
\fill (1,5,1) circle (0.1) ;
\fill (7,10,7) circle (0.1) ;
\draw [middlearrow={{triangle 45}},thick,dashed] (1,0,1) .. controls (4,0,5) and (1,0,5) .. (7,0,7);
\draw[middlearrow={{triangle 45}},thick] (1,0,1)--(7,0,1);
\draw[middlearrow={{triangle 45}},thick] (7,0,1)--(7,0,7);
\draw[thick,dashed](7,5,1)--(7,5,7);
\draw[thick,dashed](7,10,1)--(7,10,7);
\draw[thick,dashed](1,5,1)--(7,5,1);
\draw [middlearrow={{triangle 45}},thick,dashed,blue] (1,5,1) .. controls (4,6,3) and (2,5.5,-3) .. (7,10,7);
\draw [middlearrow={{triangle 45}},thick,red] (1,5,1) .. controls (4,5,1) and (4,10,2) .. (7,10,1);
\end{tikzpicture}}
\subfigure[The projection of the curve ${E}_\Gamma$ on the plane $\sigma_2^\prime$ with constant $t_1(s_1)$.]{
\begin{tikzpicture}[scale=0.32]
\draw[fill=red!10](1,5,1)--(1,10,1)--(7,10,1)--(7,5,1)--cycle;
\draw[fill=blue!10](7,5,1)--(7,5,7)--(7,10,7)--(7,10,1)--cycle;
 \draw[->] (-2,2,2) -- (9,2,2) node[anchor=west] {$t_{1}$};
 \draw[->] (-2,2,2) -- (-2,9,2) node[anchor=south] {$X(t_1,t_2)$};
 \draw[->] (-2,2,2) -- (-2,2,17) node[anchor=north] {$t_{2}$};
 \fill (1,0,1) circle (0.1) ;
\draw (4,10.5,3) node[anchor=north west] {$E_{\Gamma}$};
 \draw (6.5,0,1.9) node[anchor=north west] {$\Gamma_2^\prime$};
 \draw (0.45,0.15,-3.25) node[anchor=north west] {$\Gamma_1^\prime$};
 \fill (7,0,7) circle (0.1);
 \draw (12,-1.5,7)node[anchor=east]{$(t_1(s_1),t_2(s_1))$} ;
  \fill (1,0,1) circle (0.1) node[anchor=east]{$(t_1(s_0),t_2(s_0))$} ;
  \fill (7,0,1) circle (0.1) node[anchor=west]{$(t_1(s_1),t_2(s_0))$} ;
\draw[dashed] (1,0,1)--(1,5,1);
\draw[dashed] (7,0,7)--(7,10,7);
\fill (1,5,1) circle (0.1) ;
\fill (7,10,7) circle (0.1) ;
 \fill (1,5,1) circle (0.15);
\draw (-9.5,6,1) node[anchor=north west] {$X(t_1(s_0),t_2(s_0))$};
\fill (7,10,1) circle (0.1);
 \draw (7,10,1) node[anchor=west]{$ X(t_1(s_1),t_2(s_0))$};
\fill (7,10,7) circle (0.1);
\draw (17,10,7)  node[anchor=east]{$ X(t_1(s_1),t_2(s_1))$};
 \draw (7,7,0.3) node[anchor=north west] {$\sigma_2^\prime$};
\draw [thick,dashed,middlearrow={{triangle 45}}] (1,0,1) .. controls (4,0,5) and (1,0,5) .. (7,0,7);
\draw[middlearrow={{triangle 45}},thick] (1,0,1)--(7,0,1);
\draw[middlearrow={{triangle 45}},thick] (7,0,1)--(7,0,7);
\draw[thick,dashed](7,5,1)--(7,5,7);
\draw[thick,dashed](7,10,1)--(7,10,7);
\draw[thick,dashed](1,5,1)--(7,5,1);
\draw [thick,blue] (7,5,1) .. controls (7,5,1) and (6,4,1) .. (7,10,7);
\draw [middlearrow={{triangle 45}},thick,dashed,blue] (1,5,1) .. controls (4,6,3) and (2,5.5,-3) .. (7,10,7);
\end{tikzpicture}}
\end{center}
\caption{According to the existence of the closure relation, the deformation of the curve $\Gamma$ can be catagorised to four cases (a), (b), (c) and (d)}\label{1234}
\end{figure}
\\
\\
Now, we consider the curve $E_\Gamma$ starting from point $X(t_1(s_0),t_2(s_0))$ to point $X(t_1(s_1),t_2(s_1))$, see figure \ref{1234}. The projection of the curve $E_\Gamma$ gives the curve $\Gamma$ on the $t_1-t_2$ plane. With the closure relation, we may deform the curve $\Gamma$ such that
$\Gamma \rightarrow \Gamma_1+\Gamma_2$ and the system evolves in the following
\begin{subequations}\label{SETL}
\begin{itemize}
\item 
Case (a): On the curve $\Gamma_2$, where the time variable $t_1$ is \emph{fixed} with the value $t_1(s_0)$, the Euler-Lagrange equation and the constraint equation are simply reduced to
\begin{eqnarray}
\mbox{Euler-Lagrange equation:}&&\frac{\partial{\mathscr{L}_{(t_2)}}}{\partial {X}}
-\frac{\mathsf {d}}{\mathsf {d} t_2}\left(\frac{\partial{\mathscr{L}_{(t_2)}}}{\partial 
{X_{(2)}}}\right)=0\;.\label{ELt1}\\
\mbox{Constraint equation:}&&\frac{\partial{\mathscr{L}_{(t_2)}}}{\partial {{X}_{(1)}}}=0\;.
\end{eqnarray}
We observe that Eq. \eqref{ELt1} is nothing that the usual Euler-Lagrange equation resulting from varying the image of the curve $E_\Gamma$ on the plane $\sigma_2$, see figure \ref{1234}(a), associated with the action functional
\begin{equation}
S_{\sigma_2}[X(t_1(s_0),t_2(s));\Gamma_2]=\int_{t_2(s_0)}^{t_2(s_1)}\mathscr{L}_{(t_2)}dt_2\;,
\end{equation}
where the Lagrangian $\mathscr{L}_{(t_2)}$ is
\begin{equation}
\mathscr{L}_{(t_2)}\equiv\mathscr{L}_{(t_2)}\left(X(t_1(s_0),t_2(s)),X_{(1)}(t_1(s_0),t_2(s)),X_{(2)}(t_1(s_0),t_2(s)) \right)\;.
\end{equation}
\item
Case (b): On the curve $\Gamma_1$, where the time variable $t_2$ is \emph{fixed} with the value $t_2(s_1)$, the Euler-Lagrange equation and the constraint equation are simply reduced to
\begin{eqnarray}
\mbox{Euler-Lagrange equation:}&&\frac{\partial{\mathscr{L}_{(t_1)}}}{\partial{X}}
-\frac{\mathsf {d}}{\mathsf {d} t_1}\left(\frac{\partial{\mathscr{L}_{(t_1)}}}{\partial 
{X}_{(1)}}\right)=0\;.\label{ELt2}\\
\mbox{Constraint equation:}&&\frac{\partial{\mathscr{L}_{(t_1)}}}{\partial {{X}_{(2)}}}=0\;.
\end{eqnarray}
We observe that Eq. \eqref{ELt2} is nothing that the usual Euler-Lagrange equation resulting from varying the image of the curve $E_\Gamma$ on the plane $\sigma_1$, see figure \ref{1234}(b), associated with the action functional
\begin{equation}
S_{\sigma_1}[X(t_1(s),t_2(s_1));\Gamma_1]=\int_{t_1(s_0)}^{t_1(s_1)}\mathscr{L}_{(t_1)}dt_1\;,
\end{equation}
where the Lagrangian $\mathscr{L}_{(t_1)}$ is
\begin{equation}
\mathscr{L}_{(t_1)}\equiv\mathscr{L}_{(t_1)}\left(X(t_1(s),t_2(s_1)),X_{(1)}(t_1(s),t_2(s_1)),X_{(2)}(t_1(s),t_2(s_1)) \right)\;.
\end{equation}
\end{itemize}
Furthermore, the closure relation also allows us to deform the curve $\Gamma$ such that $\Gamma \rightarrow \Gamma_1^\prime+\Gamma^\prime_2$. In this case, the system evolves in the following
\begin{itemize}
\item 
Case (c): On the curve $\Gamma_1^\prime$, where the time variable $t_2$ is \emph{fixed} with the value $t_2(s_0)$, the Euler-Lagrange equations and the constraint equation are simply reduced to
\begin{eqnarray}
\mbox{Euler-Lagrange equation:}&&\frac{\partial{\mathscr{L}_{(t_1)}}}{\partial{X}}
-\frac{\mathsf {d}}{\mathsf {d} t_1}\left(\frac{\partial{\mathscr{L}_{(t_1)}}}{\partial 
X_{(1)}}\right)=0\;.\label{ELt3}\\
\mbox{Constraint equation:}&&\frac{\partial{\mathscr{L}_{(t_1)}}}{\partial {X_{(2)}}}=0\;.
\end{eqnarray}
We observe that Eq. \eqref{ELt3} is nothing that the usual Euler-Lagrange equation resulting from varying the image of the curve $E_\Gamma$ on the plane $\sigma_1^\prime$, see figure \ref{1234}(c), associated with the action functional
\begin{equation}
S_{\sigma_1^\prime}[X(t_1(s),t_2(s_0));\Gamma_1^\prime]=\int_{t_1(s_0)}^{t_1(s_1)}\mathscr{L}_{(t_1)}dt_1\;,
\end{equation}
where the Lagrangian $\mathscr{L}_{(t_1)}$ is
\begin{equation}
\mathscr{L}_{(t_1)}\equiv\mathscr{L}_{(t_1)}\left(X(t_1(s),t_2(s_0)),X_{(1)}(t_1(s),t_2(s_0)),X_{(2)}(t_1(s),t_2(s_0)) \right)\;.
\end{equation}
\item
Case (d): On the curve $\Gamma_2^\prime$, where the time variable $t_1$ is \emph{fixed} with the value $t_1(s_1)$, the Euler-Lagrange equation and the constraint equation are simply reduced to
\begin{eqnarray}
\mbox{Euler-Lagrange equation:}&&\frac{\partial{\mathscr{L}_{(t_2)}}}{\partial{X}}
-\frac{\mathsf {d}}{\mathsf {d} t_2}\left(\frac{\partial{\mathscr{L}_{(t_2)}}}{\partial 
{X}_{(2)}}\right)=0\;.\label{ELt4}\\
\mbox{Constraint equation:}&&\frac{\partial{\mathscr{L}_{(t_2)}}}{\partial {{X}_{(1)}}}=0\;.
\end{eqnarray}
We observe that Eq. \eqref{ELt4} is nothing that the usual Euler-Lagrange equation resulting from varying the image of the curve ${E}_\Gamma$ on the plane $\sigma_2^\prime$, see figure \ref{1234}(d), associated with the action functional
\begin{equation}
S_{\sigma_2^\prime}[X(t_1(s_1),t_2(s));\Gamma_2^\prime]=\int_{t_2(s_0)}^{t_2(s_1)}\mathscr{L}_{(t_2)}dt_2\;,
\end{equation}
where the Lagrangian $\mathscr{L}_{(t_2)}$ is
\begin{equation}
\mathscr{L}_{(t_2)}\equiv\mathscr{L}_{(t_2)}\left(X(t_1(s_1),t_2(s)),X_{(1)}(t_1(s_1),t_2(s)),X_{(2)}(t_1(s_1),t_2(s)) \right)\;.
\end{equation}
\end{itemize}
\end{subequations}
From above results, we see that, instead of going directly along the curve $\Gamma$, the system can either start to evolve in time with $t_1$ while $t_2$ is fixed along the curve $\Gamma_1$ and then evolve in time with $t_2$, while $t_1$ is fixed along the curve $\Gamma_2$ or vice versa with the curve $\Gamma'_1$ first and later with $\Gamma'_2$, see figure \ref{1234}. Intriguingly, from this point of view, the closure relation exhibits the \emph{commuting evolutions} between $t_1$ and $t_2$ on the space of independent variables and a whole set of \eqref{SETL} exhibits the multidimensional consistency of the system.
\subsubsection{Example: The continuous-time rational Calogero-Moser system }
In this section, we choose rational Calogero-Moser system \cite{Sikarin1} as an explicit example for continuous-time Lagrangian 1-form structure. The first two Lagrangians in the hierarchy are given by 
\begin{subequations}\label{CONL1}
\begin{eqnarray}
\mathscr{L}_{(t_2)}&=&\sum_{i=1}^{N}\frac{1}{2}\left(\frac{\partial X_i}{\partial t_2} \right)^2 +{\sum\limits_{\mathop {i,j = 1}\limits_{i \ne j}}^{N}}\frac{2}{(X_i - X_j)^2}\;, \label{contLT2} \\
\mathscr{L}_{(t_3)}&=&\sum_{i=1}^{N}\left( \frac{\partial X_i}{\partial t_2}\frac{\partial X_i}{\partial t_3}+\frac{1}{4} \left(\frac{\partial X_i}{\partial t_2}\right)^3\right)-{\sum\limits_{\mathop {i,j = 1}\limits_{i \ne j}}^{N}} \frac{\partial X_i}{\partial t_2}\frac{3}{(X_i -X_j)^2} \;. \label{contLTT3}
\end{eqnarray}
\end{subequations}
The first Lagrangian \eqref{contLT2} is that of the Calogero-Moser system and the second Lagrangian \eqref{contLTT3} is the higher order in the hierarchy.
\\
\\
\textbf{Equation of motion}: The Euler-Lagrange equations are given by
\begin{subequations}\label{EL01}
\begin{eqnarray}
\frac{\partial \mathscr{L}_{(t_2)}}{\partial X_i}- \frac{\mathsf d}{\mathsf{d} t_2}\left(\frac{\partial \mathscr{L}_{(t_2)}}{\frac{\partial X_i}{\partial t_2}} \right)\;, i=1,2,...,N \;, \label{contELT2} \\
\frac{\partial \mathscr{L}_{(t_3)}}{\partial X_i}- \frac{\mathsf d}{\mathsf{d} t_3}\left(\frac{\partial \mathscr{L}_{(t_3)}}{\frac{\partial X_i}{\partial t_3}} \right)\;, i=1,2,...,N \;. \label{contELTT3} 
\end{eqnarray} 
\end{subequations}
Substituting \eqref{CONL1} into \eqref{EL01}, we obtain
\begin{subequations}
\begin{eqnarray}
 \frac{\partial^2 X_i}{\partial t_2 ^2}+{\sum\limits_{\mathop {j = 1}\limits_{i \ne j}}^N}\frac{8}{\left(X_i - X_j  \right)^3}&=&0 \;, i=1,2,...,N \;, \label{contEFMT2} \\
 \frac{\partial^2 X_i}{\partial t_2 \partial t_3} -{\sum\limits_{\mathop {j = 1}\limits_{i \ne j}}^N}\left(\frac{\partial X_i}{\partial t_2}+\frac{\partial X_j}{\partial t_2} \right)\frac{6}{\left(X_i - X_j  \right)^3} &=&0 \;, i=1,2,...,N \;, \label{contEFMTT3} 
\end{eqnarray}
\end{subequations}
where \eqref{contEFMT2} are the equations of motion for  Calogero-Moser system and of course \eqref{contEFMTT3} are the next order in the hierarchy. 
\\
\\
\textbf{Constraint equation}: Since the Lagrangian \eqref{contLTT3} contains $\partial X_i/\partial t_2$, we have
\begin{eqnarray}\label{Conconeq1}
\frac{\partial \mathscr{L}_{(t_3)}}{\frac{\partial X_i}{\partial t_2}}=0 \;, i=1,2,...,N \;.
\end{eqnarray}
Equations \eqref{Conconeq1} give
\begin{eqnarray}
\frac{1}{4}\left(\frac{\partial X_i}{\partial t_2} \right)^2 + \frac{1}{3}\frac{\partial X_i}{\partial t_3} -{\sum\limits_{\mathop {j = 1}\limits_{i \ne j}}^N}\frac{1}{(X_i - X_j)^2} &=& 0\;, i=1,2,...,N \;, \label{Conconeq}
\end{eqnarray} 
which are the constraint equations.
\\
\\
\textbf{Closure relation}: In the case of two time variables, $t_2$ and $t_3$, the closure relation is given by
\begin{eqnarray}\label{ConS}
\frac{\partial \mathscr{L}_{(t_2)}}{\partial t_3}&=&\frac{\partial \mathscr{L}_{(t_3)}}{\partial t_2} \;.
\end{eqnarray}
The proof of \eqref{ConS} has already been given in \cite{Sikarin1} so we will not repeat it here.
\section{Hamiltonian hierarchies and commuting flows}\label{section4}
In the previous section, a new feature of integrability, called the closure relation, is derived for the system with Lagrangian hierarchy from the point of view of the variational principle. We know that basically we can obtain the Hamiltonian from Lagrangian through the Legendre transformation. The action functional is then written in terms of the Hamiltonian and the variation can be performed with respect to variables on phase space resulting in the Hamilton's equations. In this section, we set out to construct the Legendre transformation to obtain the Hamiltonian hierarchy and of course consider the variational principle on the phase space.
\subsection{Legendre transformation}
To establish the Legendre transformation, we multiply $\mathsf {d}X/\mathsf {d}s$ to the Euler-Lagrange equation \eqref{coneq7}
\begin{eqnarray}
\frac{\mathsf {d}X}{\mathsf {d}s} \sum_{i=1}^{N} \frac{\partial \mathscr{L}_{(t_i)}}{\partial X}\frac{\mathsf {d}t_i}{\mathsf {d}s}- \frac{\mathsf {d}X}{\mathsf {d}s}\frac{1}{N} \frac{\mathsf {d}}{\mathsf {d}s} \left(\sum_{i=1}^{N}\frac{\partial \mathscr{L}_{(t_i)}}{\partial X_{(i)}} +  {\sum\limits_{\mathop {i,j = 1}\limits_{i \ne j}}^N} \frac{\partial \mathscr{L}_{(t_i)}}{\partial X_{(j)}}\frac{\mathsf {d} t_i/ \mathsf {d} s}{\mathsf {d} t_j/ \mathsf {d} s}\right) =0 \; \label{coneq13}.
\end{eqnarray}
and introduce the relation 
\begin{eqnarray}
\frac{\mathsf {d}X}{\mathsf {d}s} \frac{\mathsf {d}}{\mathsf {d}s} \left(\sum_{i=1}^{N}\frac{\partial \mathscr{L}_{(t_i)}}{\partial X_{(i)}} + {\sum\limits_{\mathop {i,j = 1}\limits_{i \ne j}}^N}\frac{\partial \mathscr{L}_{(t_i)}}{\partial X_{(j)}}\frac{\mathsf {d} t_i/ \mathsf {d} s}{\mathsf {d} t_j/ \mathsf {d} s}\right)&=&\frac{\mathsf {d}}{\mathsf {d}s} \left[\frac{\mathsf {d}X}{\mathsf {d}s} \left(\sum_{i=1}^{N}\frac{\partial \mathscr{L}_{(t_i)}}{\partial X_{(i)}} +{\sum\limits_{\mathop {i,j = 1}\limits_{i \ne j}}^N} \frac{\partial \mathscr{L}_{(t_i)}}{\partial X_{(j)}}\frac{\mathsf {d} t_i/ \mathsf {d} s}{\mathsf {d} t_j/ \mathsf {d} s} \right)\right] \nonumber \\ 
&&- \frac{\mathsf {d}^2 X}{\mathsf {d}s^2} \left(\sum_{i=1}^{N}\frac{\partial \mathscr{L}_{(t_i)}}{\partial X_{(i)}} +{\sum\limits_{\mathop {i,j = 1}\limits_{i \ne j}}^N} \frac{\partial \mathscr{L}_{(t_i)}}{\partial X_{(j)}}\frac{\mathsf {d} t_i/ \mathsf {d} s}{\mathsf {d} t_j/ \mathsf {d} s} \right)\nonumber
 \;.\label{DAS}\\
\end{eqnarray}
Using \eqref{DAS}, \eqref{coneq13} can be rewritten as
\begin{eqnarray}
0&=&\frac{\mathsf {d}X}{\mathsf {d}s} \sum_{i=1}^{N} \frac{\partial \mathscr{L}_{(t_i)}}{\partial X}\frac{\mathsf {d}t_i}{\mathsf {d}s}+\frac{1}{N} \frac{\mathsf {d}^2 X}{\mathsf {d}s^2} \left(\sum_{i=1}^{N}\frac{\partial \mathscr{L}_{(t_i)}}{\partial X_{(i)}} + {\sum\limits_{\mathop {i,j = 1}\limits_{i \ne j}}^N} \frac{\partial \mathscr{L}_{(t_i)}}{\partial X_{(j)}}\frac{\mathsf {d} t_i/ \mathsf {d} s}{\mathsf {d} t_j/ \mathsf {d} s} \right) \nonumber \\
&&-\frac{1}{N}\frac{\mathsf {d}}{\mathsf {d}s} \left[\frac{\mathsf {d}X}{\mathsf {d}s} \left(\sum_{i=1}^{N}\frac{\partial \mathscr{L}_{(t_i)}}{\partial X_{(i)}} + {\sum\limits_{\mathop {i,j = 1}\limits_{i \ne j}}^N} \frac{\partial \mathscr{L}_{(t_i)}}{\partial X_{(i)}}\frac{\mathsf {d} t_i/ \mathsf {d} s}{\mathsf {d} t_j/ \mathsf {d} s} \right)\right] \; \label{coneq14}.
\end{eqnarray}
Next we consider the relation
\begin{eqnarray}\label{FFF}
\frac{\mathsf {d}}{\mathsf {d}s}\left( \sum_{i=1}^{N} \mathscr{L}_{(t_i)}  \frac{\mathsf {d} t_i}{ \mathsf {d} s}\right) &=&\sum_{i=1}^{N}\left(\frac{\partial \mathscr{L}_{(t_i)}}{\partial X}\frac{\mathsf {d}X}{\mathsf {d}s}\frac{\mathsf {d}t_i}{\mathsf {d}s}+ \frac{\partial \mathscr{L}_{(t_i)}}{\partial X_{(i)}}\frac{\mathsf {d}X_{(i)}}{\mathsf {d}s}\frac{\mathsf {d}t_i}{\mathsf {d}s}\right)+{\sum\limits_{\mathop {i,j = 1}\limits_{i \ne j}}^N} \frac{\partial \mathscr{L}_{(t_i)}}{\partial X_{(j)}}\frac{\mathsf {d}X_{(j)}}{\mathsf {d}s}\frac{\mathsf {d}t_i}{\mathsf {d}s} \nonumber \\
&&+\sum_{i=1}^{N} \mathscr{L}_{(t_i)} \frac{\mathsf {d}^2 t_i}{ \mathsf {d} s^2} \;.
\end{eqnarray}
Imposing $\mathsf {d}^2 t_i / \mathsf{d} s^2 =0 $ and using constraint \eqref{coneq8}, equation \eqref{FFF} becomes
\begin{eqnarray}
\frac{\mathsf {d}}{\mathsf {d}s}\left( \sum_{i=1}^{N} \mathscr{L}_{(t_i)}  \frac{\mathsf {d} t_i}{ \mathsf {d} s}\right) = \frac{\mathsf {d}X}{\mathsf {d}s} \sum_{i=1}^{N} \frac{\partial \mathscr{L}_{(t_i)}}{\partial X}\frac{\mathsf {d}t_i}{\mathsf {d}s}+\frac{1}{N} \frac{\mathsf {d}^2 X}{\mathsf {d}s^2} \left(\sum_{i=1}^{N}\frac{\partial \mathscr{L}_{(t_i)}}{\partial X_{(i)}} + {\sum\limits_{\mathop {i,j = 1}\limits_{i \ne j}}^N} \frac{\partial\mathscr{L}_{(t_i)}}{\partial X_{(j)}}\frac{\mathsf {d} t_i/ \mathsf {d} s}{\mathsf {d} t_j/ \mathsf {d} s} \right)\;. \label{coneq15}
\end{eqnarray} 
Substituting \eqref{coneq15} into \eqref{coneq14}, we obtain
\begin{eqnarray}
\frac{\mathsf {d}}{\mathsf {d}s} \left[\frac{1}{N}\frac{\mathsf {d}X}{\mathsf {d}s} \left(\sum_{i=1}^{N}\frac{\partial \mathscr{L}_{(t_i)}}{\partial X_{(i)}} + {\sum\limits_{\mathop {i,j = 1}\limits_{i \ne j}}^N} \frac{\partial \mathscr{L}_{(t_i)}}{\partial X_{(j)}}\frac{\mathsf {d} t_i/ \mathsf {d} s}{\mathsf {d} t_j/ \mathsf {d} s} \right)- \sum_{i=1}^{N} \mathscr{L}_{(t_i)}  \frac{\mathsf {d} t_i}{ \mathsf {d} s} \right] =0 \;.\label{coneq16} 
\end{eqnarray}
We observe that the term inside the square bracket must be a constant with respect to the parameter $s$. We then define the momentum variable as
\begin{eqnarray}
P\equiv\frac{1}{N} \left(\sum_{i=1}^{N}\frac{\partial \mathscr{L}_{(t_i)}}{\partial X_{(i)}} + {\sum\limits_{\mathop {i,j = 1}\limits_{i \ne j}}^N} \frac{\partial \mathscr{L}_{(t_i)}}{\partial X_{(j)}}\frac{\mathsf {d} t_i/ \mathsf {d} s}{\mathsf {d} t_j/ \mathsf {d} s} \right)\;, \label{conmomentum}
\end{eqnarray}
 and introduce the energy function
\begin{eqnarray}
E \equiv \sum_{i=1}^{N} \mathscr{H}_{(t_i)}(X,P)  \frac{\mathsf {d} t_i}{ \mathsf {d} s} \;,
\end{eqnarray}
which is conserved under parametrised time translation, i.e.,
\begin{eqnarray}
\frac{\mathsf{d}}{\mathsf{d}s}\left(\sum_{i=1}^{N} \mathscr{H}_{(t_i)}(X,P)  \frac{\mathsf {d} t_i}{ \mathsf {d} s}\right)=0\;.
\end{eqnarray}
Therefore, what we have now are
\begin{eqnarray}
 \mathscr{H}_{(t_i)}   &=&  X_{(i)} P -  \mathscr{L}_{(t_i)}   \;, i=1,2,...,N\;, \label{coneq17}
\end{eqnarray}
which are the Legendre transformations. This means that, with explicit form of the Lagrangian hierarchy, one can obtain the Hamiltonian hierarchy.
\subsection{Variational principle on phase space}
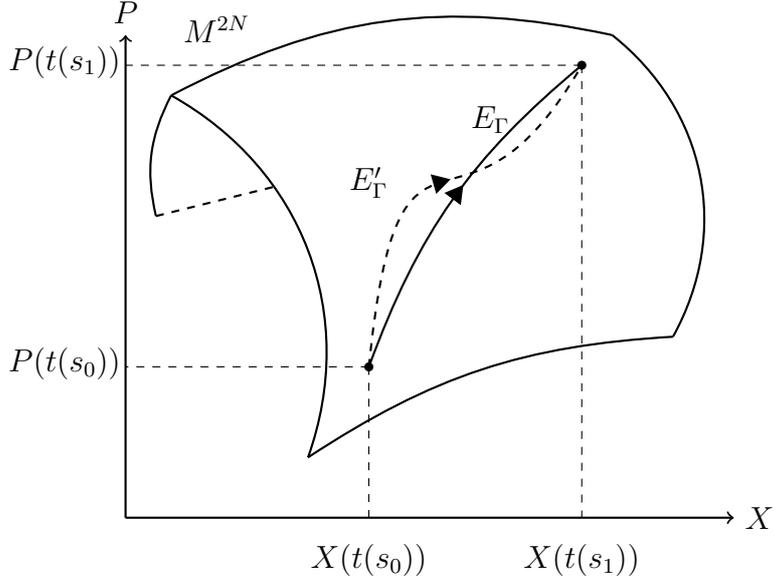
\begin{figure}[h]
\begin{center}
{
\tikzset{middlearrow/.style={
        decoration={markings,
            mark= at position 0.55 with {\arrow{#1}} ,
        },
        postaction={decorate}
    }
}
\begin{tikzpicture}[scale=0.4]
\draw[->,thick] (0,0) -- (20,0) node[anchor=west] {$X$};
 \draw[->,thick] (0,0) -- (0,16) node[anchor=south] {$P$};
\draw[thick](6,2)to [bend left=15](18,6);
\draw[thick](6,2)to [bend right=40](1.5,14);
\draw[thick](1.5,14)to [bend left=20](16,16);
\draw[thick](16,16)to [bend left=40](18,6);
\draw[thick](1.5,14)to [bend right=20](1,10);
\draw[dashed,thick](1,10)--(5,11);
\draw[middlearrow={{triangle 60}},thick](8,5)to [bend left=15](15,15);
\draw[middlearrow={{triangle 60}},thick,dashed](8,5).. controls (9,15) and (11,8) ..(15,15);
\draw[dashed](8,5)--(8,0);
\draw[dashed](15,15)--(15,0);
\draw[dashed](8,5)--(0,5);
\draw[dashed](15,15)--(0,15);
\fill (8,5) circle (0.15);
\fill (15,15) circle (0.15);
\draw (3,17) node[anchor=north] {${{M}^{2N}}$};
\draw (12,14) node[anchor=north] {${E_{\Gamma}}$};
\draw (8,12) node[anchor=north] {${E'_{\Gamma}}$};
\draw (8,-0.5) node[anchor=north] {${X(t(s_0))}$};
\draw (15,-0.5) node[anchor=north] {${X(t(s_1))}$};
\draw (-2,6) node[anchor=north] {${P(t(s_0))}$};
\draw (-2,16) node[anchor=north] {${P(t(s_1))}$};
\end{tikzpicture}
}
\caption{The trajectory of the system defined by a curve $E_\Gamma$ on $2N$-dimensional manifold $M^{2N}$ embedded in the phase space.}\label{PQ1}
\end{center}
\end{figure}

With the Legendre transformation \eqref{coneq17}, the action functional becomes
\begin{eqnarray}
{S}_{E_\Gamma}[X(t),P(t)]=\int_{s_0}^{s_1} \mathsf {d} s \sum_{i=1}^N  \left(X_{(i)} P -  \mathscr{H}_{(t_i)} (X,P)  \right)\frac{\mathsf {d} t_i}{\mathsf {d} s}\;\label{eqh1},
\end{eqnarray}
where the curve $E_\Gamma$ is defined on the phase space, see figure \ref{PQ1}.
\subsubsection{Variation on dependent variables}
In this situation, we have $2N$ variables to work with. We then consider the variations $X \rightarrow X+\delta X$ and $P \rightarrow P+\delta P$ with the fixed end-point conditions 
resulting in a new curve $E'_\Gamma$ with the action functional
\begin{eqnarray}
{S}_{E'_\Gamma}[X+\delta X,P+\delta P]=\int_{s_0}^{s_1} \mathsf {d} s \sum_{i=1}^N  \left((X_{(i)}+\delta X_{(i)}) (P+\delta P) -  \mathscr{H}_{(t_i)} (X+\delta X,P+\delta P)  \right)\frac{\mathsf {d} t_i}{\mathsf {d} s}\;.\label{eqh1}
\end{eqnarray}
Performing the Taylor expansion and keeping the first two contributions in the series, the variation of action is given by
\begin{eqnarray}
\delta {S}_{E_\Gamma}\equiv\delta {S}_{E'_\Gamma}-\delta {S}_{E_\Gamma} =\int_{s_0}^{s_1} \mathsf {d} s  \sum_{i=1}^N  \left(\delta X_{(i)} P + X_{(i)} \delta P  - \delta P\frac{\partial \mathscr{H}_{(t_i)}  }{\partial P} - \delta X_{(i)}   \frac{\partial \mathscr{H}_{(t_i)}  }{\partial X_{(i)}}\right)\frac{\mathsf {d} t_i}{\mathsf {d} s}\;\label{eqh2}. 
\end{eqnarray}
Integrating by parts the first and fourth terms in the bracket and using the end-point conditions, $\delta X(t(s_0))=\delta X(t(s_1))= 0$, we obtain
\begin{eqnarray}\label{HT}
\delta {S}_{E_\Gamma} &=&\int_{s_0}^{s_1} \mathsf {d} s  \sum_{i=1}^{N} \left[ \delta P \left(\frac{\mathsf {d}X}{\mathsf {d}s} - \frac{\partial \mathscr{H}_{(t_i)}  }{\partial P} \frac{\mathsf {d}t_i}{\mathsf {d}s} \right) - \delta X \left(\frac{\mathsf {d}P}{\mathsf {d}s} + \frac{\partial \mathscr{H}_{(t_i)}  }{\partial P} \frac{\mathsf {d}t_i}{\mathsf {d}s} \right)  \right]\;.
\end{eqnarray}
Imposing the least action principle: $\delta S_{E_\Gamma} = 0$, we obtain
\begin{subequations}\label{geqh}
\begin{eqnarray}
\frac{\mathsf {d}X}{\mathsf {d}s}&=&  \sum_{i=1}^{N} \frac{\partial \mathscr{H}_{(t_i)}  }{\partial P}\frac{\mathsf {d}t_i}{\mathsf {d}s}  \;\label{eqh3},\\
-\frac{\mathsf {d}P}{\mathsf {d}s} &=& \sum_{i=1}^{N} \frac{\partial \mathscr{H}_{(t_i)}  }{\partial X}\frac{\mathsf {d}t_i}{\mathsf {d}s} \; \label{eqh4},
\end{eqnarray}
\end{subequations}
since $\delta X\neq 0$ and $\delta P\neq 0$ and they are all independent to each other. Equations \eqref{geqh} are nothing but the generalised Hamilton's equations. 
\subsubsection{Variation on independent variables}
In this case, the time variables are embedded on the $2N$-dimensional manifold $M^{2N}$ and, then, they cannot be visualised explicitly. However, we still can consider the variation of the time variables such that $t(s)\rightarrow t(s)+\delta t(s)$ with conditions $t(s_0)=t(s_1)=0$. The variation of the action is given by
\begin{eqnarray}
\delta {S} &=& {S}\left[X(t(s)+\delta t(s)), P(t(s)+\delta t(s)) \right]-{S}\left[X(t(s), P(t(s))\right] \; \nonumber \\
&=& \int_{s_0}^{s_1} \mathsf {d}s \Bigg[\sum_{i=1}^{N} \left(P\frac{\mathsf {d}}{\mathsf {d}s}\left( X_{(i)} \delta t_i  \right) + \frac{\mathsf {d}X}{\mathsf {d}s}\left(\delta t_i \frac{\partial P}{\partial t_i}  \right)\right) \nonumber \\
&&- \sum_{i=1}^{N}\left(\delta t_{i}\frac{\partial \mathscr{H}_{(t_i)}  }{\partial t_i}+  \mathscr{H}_{(t_i)} \frac{\mathsf {d}\delta t_i}{\mathsf {d}s} \right)- {\sum\limits_{\mathop {i,j = 1}\limits_{i \ne j}}^N} \delta t_{j} \frac{\partial \mathscr{H}_{(t_i)}  }{\partial t_j}\frac{\mathsf {d}\delta t_i}{\mathsf {d}s}   \Bigg] \; \label{eqh5}. 
\end{eqnarray} 
Integrating by parts the first and the fourth terms, we obtain 
\begin{eqnarray}
\delta {S} &=& \int_{s_0}^{s_1} \mathsf {d}s \sum_{i=1}^{N}\delta t_i\Bigg[ {\sum\limits_{\mathop {j = 1}\limits_{i \ne j}}^N}  \left(\frac{\partial P}{\partial t_i}\frac{\partial X}{\partial t_j}-\frac{\partial P}{\partial t_j}\frac{\partial X}{\partial t_i}+\frac{\partial \mathscr{H}_{(t_i)}}{\partial t_j}-\frac{\partial \mathscr{H}_{(t_j)}  }{\partial t_i}\right) \frac{\mathsf {d} t_j}{\mathsf {d}s} \Bigg] \;  \label{eqh6}.   
\end{eqnarray} 
From the relation
\begin{eqnarray}
\frac{\partial P}{\partial t_i}\frac{\partial X}{\partial t_j}-\frac{\partial P}{\partial t_j}\frac{\partial X}{\partial t_i} &=&  \frac{\partial \mathscr{H}_{(t_i)}}{\partial t_j}-\frac{\partial \mathscr{H}_{(t_j)}  }{\partial t_i}\;,
\end{eqnarray}
and, imposing the least action principle: $\delta {S}= 0$, we obtain
\begin{eqnarray}\label{eqh67}
\frac{\partial \mathscr{H}_{(t_i)}}{\partial t_j}-\frac{\partial \mathscr{H}_{(t_j)}  }{\partial t_i} =0 \;,\;i\neq j =1,2,...,N\;.
\end{eqnarray}
Equations \eqref{eqh67} give  the characteristic feature of the evolution on the phase space called the commuting Hamiltonian flows. Here we show that we can obtain the involution directly from the variational principle instead of using Poisson structure and Lax matrices \cite{Babelon}.
\subsection{Example: The continuous-time rational Calogero-Moser system for Hamiltonian structure} \label{section5}
In this section, we work out explicitly on the Legendre transformation and variation of the action on phase space for the rational Calogero-Moser system.
\\
\\
\textbf{Legendre transformation}: 
With the first two Lagrangians \eqref{CONL1} in the hierarchy, the Legendre transformations are
\begin{subequations}\label{LG1}
\begin{eqnarray}
\mathscr{H}_{(t_2)}&=&\sum_{i=1}^{N} P_i \frac{\partial X_i}{\partial t_2}-\mathscr{L}_{(t_2)}\;, \label{contLGT2} \\
\mathscr{H}_{(t_3)}&=&\sum_{i=1}^{N}P_i \frac{\partial X_i}{\partial t_3}-\mathscr{L}_{(t_3)}\;, \label{contLGTT3}
\end{eqnarray}
\end{subequations}
where $P_i =\partial X_i/\partial t_2$ is the momentum. From the structure of \eqref{LG1}, we note that the Lagrangians \eqref{CONL1} share the momentum variable. Substituting \eqref{contELT2} and \eqref{contELTT3} into Legendre transformations, we obtain
\begin{subequations}\label{HTO1}
\begin{eqnarray}
\mathscr{H}_{(t_2)}&=& \sum_{i=1}^{N}\frac{1}{2}{P_i}^2 -{\sum\limits_{\mathop {i,j = 1}\limits_{i \ne j}}^{N}}\frac{2}{(X_i - X_j)^2}\;,\label{contHT2} \\
\mathscr{H}_{(t_3)}&=& \sum_{i=1}^{N}\frac{1}{3}{P_i}^3 -{\sum\limits_{\mathop {i,j = 1}\limits_{i \ne j}}^{N}} \frac{4 P_i}{(X_i -X_j)^2}\;, \label{contHTT3}
\end{eqnarray}
\end{subequations}
which are the first two Hamiltonians in the hierarchy. 
\\
\\
\textbf{Equation of motion}: 
With Hamiltonians in \eqref{HTO1}, we do have the Hamilton's equations as follows. For the first Hamiltonian \eqref{contHT2}, we have
\begin{subequations}\label{KKK1}
\begin{eqnarray}
\frac{\mathsf {d}X_i}{\mathsf {d}t_2}&=&   \frac{\partial \mathscr{H}_{(t_2)}  }{\partial P_i} \;\label{eqh3t2},\\
-\frac{\mathsf {d}P_i}{\mathsf {d}t_2} &=&  \frac{\partial \mathscr{H}_{(t_2)}  }{\partial X_i} \; \label{eqh4t2},
\end{eqnarray}
and, for the second Hamiltonian \eqref{contHTT3}, we have
\begin{eqnarray}
\frac{\mathsf {d}X_i}{\mathsf {d}t_3}&=&   \frac{\partial \mathscr{H}_{(t_3)}  }{\partial P_i}  \;\label{eqh3t3},\\
-\frac{\mathsf {d}P_i}{\mathsf {d}t_3} &=&  \frac{\partial \mathscr{H}_{(t_3)}  }{\partial X_i} \; \label{eqh4t3}.
\end{eqnarray}
\end{subequations}
Substituting the Hamiltonians \eqref{contHT2} and \eqref{contHTT3} into \eqref{KKK1}, we obtain \eqref{contEFMT2} and \eqref{contEFMTT3}, respectively.
\\
\\
\textbf{Commuting flows}: 
We now would like to show explicit proof of the relation
\begin{eqnarray}
\frac{\partial \mathscr{H}_{(t_2)}}{\partial t_3}&=&\frac{\partial \mathscr{H}_{(t_3)}}{\partial t_2} \;. \label{CSHamil}
\end{eqnarray}
With Hamiltonians \eqref{contHT2} and \eqref{contHTT3}, we find that
\begin{eqnarray}
\frac{\partial \mathscr{H}_{(t_2)}}{\partial t_{3}}&=&\sum_{i=1}^{N} {P_i} \frac{\partial P_i}{\partial t_3}+{\sum\limits_{\mathop {i,j = 1}\limits_{i \ne j}}^N}\left[\frac{\partial X_i}{\partial t_3}\frac{8}{(X_i - X_j)^3} \right]\;,\label{Diff21}\\
\frac{\partial \mathscr{H}_{(t_3)}}{\partial t_{2}}&=&\sum_{i=1}^{N} {P_i}^2 \frac{\partial P_i}{\partial t_2}-{\sum\limits_{\mathop {i,j = 1}\limits_{i \ne j}}^N} \left[\frac{\partial P_i}{\partial t_2}\frac{1}{(X_i -X_j)^2}-\frac{8 {P_i}^2}{(X_i -X_j)^3} +\frac{8 {P_i}{P_j}}{(X_i -X_j)^3}\right]\;. \label{Diff31}
\end{eqnarray}
We use \eqref{contEFMT2}, \eqref{contEFMTT3} and \eqref{Conconeq} to simplify \eqref{Diff21} and \eqref{contEFMT2} to simplify \eqref{Diff31}. The last term of \eqref{Diff31} vanishes because of the antisymmetric property. We find that \eqref{CSHamil} is reduced to
\begin{eqnarray}
\frac{\partial \mathscr{H}_{(t_3)}}{\partial t_{2}}-\frac{\partial \mathscr{H}_{(t_2)}}{\partial t_{3}}&=&{\sum\limits_{\mathop {i,j = 1}\limits_{i \ne j}}^N}{\sum\limits_{\mathop {k = 1}\limits_{i \ne k}}^N}\frac{1}{(X_i -X_j)^2}\frac{1}{(X_i -X_k)^3}\;\label{Diff22} \\
&=&{\sum\limits_{\mathop {i,j = 1}\limits_{i \ne j}}^{N}}\frac{1}{(X_i - X_j)^5}+{\sum\limits_{\mathop {i,j,k = 1}\limits_{i \ne j \ne k}}^{N}}\frac{1}{(X_i -X_j)^2 (X_i -X_k)^3}\;.\label{Diff23}
\end{eqnarray}
The first term and the second term vanish according to the antisymmetric property, see also \cite{Sikarin1}. Here we show an alternative derivation of the involution of the system through \eqref{CSHamil} instead of using the Poisson bracket $\{ \mathscr H_{(t_i)},\mathscr H_{(t_j)}\}=0$.
\section{N\"{o}ether Charges}\label{NC}
\begin{figure}[h]
\begin{center}
{
\tikzset{middlearrow/.style={
        decoration={markings,
            mark= at position 0.6 with {\arrow{#1}} ,
        },
        postaction={decorate}
    }
}
\begin{tikzpicture}[scale=0.4]
\draw[->] (0,0,0) -- (23,0,0) node[anchor=west] {$t_{i}$};
 \draw[->] (0,0,0) -- (0,20,0) node[anchor=south] {$X$};
 \draw[->] (0,0,0) -- (0,0,18) node[anchor=south] {$t_{j}$};
 \draw[middlearrow={{triangle 60}},thick] (2,9,2) .. controls (4,9,3) and (16,18,3) .. (22,18,5);
\draw[middlearrow={{triangle 60}},thick] (2,3,2) .. controls (4,3,3) and (16,11,5) .. (19,11,5);
\draw[->](2,3.3,2)--(2,8.8,2);
\draw[->](19,11.4,5)--(19,17.5,5);
\draw[->](-0.8,2.6,0)--(-0.8,8.9,0);
\draw[->](-0.8,10.45,0)--(-0.8,16.8,0);
\draw[dashed](2,3,2)--(0,2.4,0);
\draw[dashed](4,9.6,2)--(0,9.1,0);
\draw[dashed](19,11,5)--(0,10.25,0);
\draw[dashed](21,18,5)--(0,17,0);
\fill (2,3,2) circle (0.2);
\fill (4,9.7,2) circle (0.2);
\fill (19,11,5) circle (0.2);
\fill (21,18,5) circle (0.2);
\fill (0,2.4,0) circle (0.2);
\fill (0,9.1,0) circle (0.2);
\fill (0,10.25,0) circle (0.2);
\fill (0,17,0) circle (0.2);
\draw (4.5,3,2) node[anchor=north] {${X(t_i(s_0),t_j(s_0))}$};
\draw (21,10.5,5) node[anchor=north] {${X(t_i(s_1),t_j(s_1))}$};
\draw (5.5,7,0) node[anchor=north] {${\delta X(t_i(s_0),t_j(s_0))}$};
\draw (21,13,0) node[anchor=north] {${\delta X(t_i(s_1),t_j(s_1))}$};
\draw (-5,15,0) node[anchor=north] {${\mathsf{d} X(t_i(s_1),t_j(s_1))}$};
\draw (21,20,5) node[anchor=north] {${ X(t_i(s'_1),t_j(s'_1))}$};
\draw (8,10,2) node[anchor=north] {${ X(t_i(s'_0),t_j(s'_0))}$};
\draw (-5,6,0) node[anchor=north] {${\mathsf{d} X(t_i(s_0 ),t_j(s_0))}$};
\draw (10,4.8,0) node[anchor=north] {$E_{\Gamma}$};
\draw (10,14.5,0) node[anchor=north] {$E'_{\Gamma}$};
\end{tikzpicture}}
\end{center}
\caption{The variation of the curve $E_{\Gamma}$ (classical path) to $E'_{\Gamma}$ without end-point conditions.} \label{FPN.0}
\end{figure}
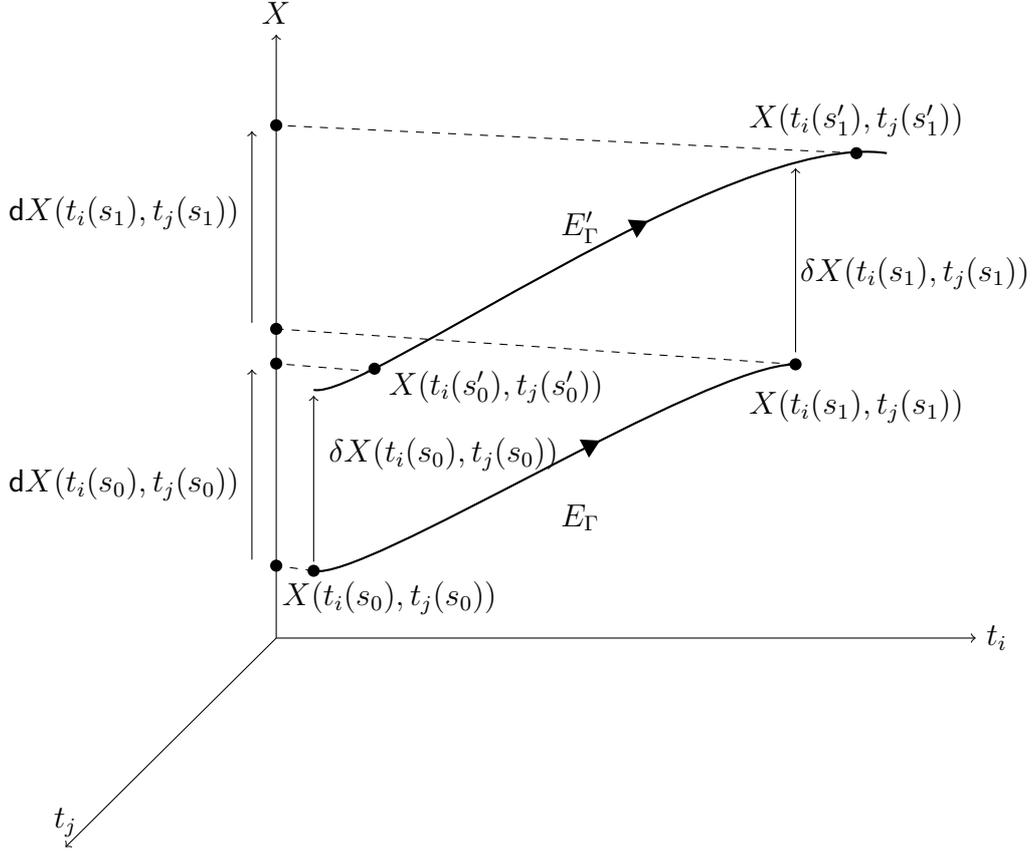

In section \ref{sectioncontime}, the local variation, where the end points of the curve are fixed, of action was performed resulting in the generalised Euler-Lagrange equation, the constraint equation and the closure relation. In this section, we again consider the variation of the action but without fixed boundary conditions. We start by considering the curve $E_\Gamma$ which is the classical trajectory of the system and the action associated with it is given by
\begin{eqnarray}
S_{E_{\Gamma}}[X(t)]&=& \int_{s_0}^{s_1} \mathsf {d} s \left(\sum_{i=1}^N \mathscr{L}_{(t_i)} \frac{\mathsf {d} t_i}{\mathsf {d} s} \right)\; \label{AcEG}\;,
\end{eqnarray}
where $\mathscr{L}_{(t_i)}\equiv \mathscr{L}_{(t_i)}(X(t(s)),\{X_{(j)}(t(s)) \})$. 
Suppose that there is a neighbouring curve called $E'_\Gamma$, which is a deformation of the curve $E_\Gamma$ such that 
${X(t_i(s_0),t_j(s_0))}\rightarrow  X(t_i(s'_0),t_j(s'_0))$ and ${X(t_i(s_1),t_j(s_1))}\rightarrow   X(t_i(s'_1 ),t_j(s'_1))$, where $s'_0 \equiv s_0+ \mathsf{d}s_0$ and $s'_1 \equiv s_1+ \mathsf{d}s_1 $, see figure \ref{FPN.0}. A new action is given by
\begin{eqnarray}  
S_{E'_{\Gamma}}[X(t)+\delta X(t)]&=& \int_{s'_0 }^{s'_1 } \mathsf {d} s \left(\sum_{i=1}^N \mathscr{L}'_{(t_i)} \frac{\mathsf {d} t_i}{\mathsf {d} s} \right)\; \label{AcEG'},
\end{eqnarray}
where $\mathscr{L}'_{(t_i)}\equiv \mathscr{L}_{(t_i)}(X(t)+\delta X(t),\{X_{(j)}(t)+\delta X_{(j)}(t) \})$. 
The variation of the action between two curves is
\begin{eqnarray}
\delta S_{E_{\Gamma}}&=& S_{E'_{\Gamma}}[X(t)+\delta X(t)]-S_{E_{\Gamma}}[X(t)] \nn\\
&=&\int_{s'_0 }^{s'_1 } \mathsf {d} s \left(\sum_{i=1}^N \mathscr{L}'_{(t_i)} \frac{\mathsf {d} t_i}{\mathsf {d} s} \right)-\int_{s_0}^{s_1} \mathsf {d} s \left(\sum_{i=1}^N \mathscr{L}_{(t_i)} \frac{\mathsf {d} t_i}{\mathsf {d} s} \right) \nn\\
&=&\int_{s_0}^{s_1} \mathsf {d} s \sum_{i=1}^N \left(\mathscr{L}'_{(t_i)}-\mathscr{L}_{(t_i)} \right)\frac{\mathsf {d} t_i}{\mathsf {d} s} +\int_{s_1}^{s'_1} \mathsf {d} s \sum_{i=1}^N \mathscr{L}'_{(t_i)}\frac{\mathsf {d} t_i}{\mathsf {d} s} \nn\\
&&-\int_{s_0}^{s'_0} \mathsf {d} s \sum_{i=1}^N  \mathscr{L}_{(t_i)}\frac{\mathsf {d} t_i}{\mathsf {d} s}\;.\label{DifACEG}
\end{eqnarray}
Using Taylor expansion and substituting \eqref{coneq5} in \eqref{DifACEG}, we have
\begin{eqnarray}
\delta {S}_{E_{\Gamma}} &=& \int_{s_0}^{s_1} \mathsf {d} s\Bigg \lbrace \left[\sum_{i=1}^{N} \frac{\partial \mathscr{L}_{(t_i)}}{\partial X}\frac{\mathsf {d}t_i}{\mathsf {d}s}\right]\delta X + \frac{1}{N} \frac{\mathsf {d}\delta X}{\mathsf {d}s} \Bigg[\sum_{i=1}^{N}\frac{\partial\mathscr{L}_{(t_i)}}{\partial X_{(i)}} + {\sum\limits_{\mathop {i,j = 1}\limits_{i \ne j}}^N} \frac{\partial \mathscr{L}_{(t_i)}}{\partial X_{(j)}}\frac{\mathsf {d} t_i/ \mathsf {d} s}{\mathsf {d} t_j/ \mathsf {d} s}\Bigg] \nonumber \\ 
&&+ \frac{1}{N}{\sum\limits_{\mathop {i,j = 1}\limits_{i \ne j}}^N} \delta Y_{ij} \Bigg[\frac{\partial \mathscr{L}_{(t_i)}}{\partial X_{(j)}} -\frac{\partial \mathscr{L}_{(t_j)}}{\partial X_{(j)}} -\frac{\partial \mathscr{L}_{(t_i)}}{\partial X_{(j)}}\frac{\mathsf {d} t_i/ \mathsf {d} s}{\mathsf {d} t_j/ \mathsf {d} s}+\frac{\partial \mathscr{L}_{(t_j)}}{\partial X_{(i)}}\frac{\mathsf {d} t_j/ \mathsf {d} s}{\mathsf {d} t_i/ \mathsf {d} s} \nonumber \\
&&+{\sum\limits_{\mathop {k = 1}\limits_{k \ne i,j}}^N} \left[\frac{\partial \mathscr{L}_{(t_k)}}{\partial X_{(i)}}\frac{\mathsf {d} t_k/ \mathsf {d} s}{\mathsf {d} t_i/ \mathsf {d} s} - \frac{\partial \mathscr{L}_{(t_k)}}{\partial X_{(j)}}\frac{\mathsf {d}t_k/ \mathsf {d} s}{\mathsf {d} t_j/ \mathsf {d} s} \right]\Bigg] \Bigg\rbrace  + \mathsf {d} s \sum_{i=1}^N  \mathscr{L}_{(t_i)}\frac{\mathsf {d} t_i}{\mathsf {d} s} \Bigg\vert_{s_0}^{s_1} \label{DifACEG1}\;.
\end{eqnarray}
Integrating by parts the second term in \eqref{DifACEG1}, we obtain
\begin{eqnarray}
\delta {S}_{E_{\Gamma}} &=& \int_{s_0}^{s_1} \mathsf {d} s\Bigg \lbrace \delta X  \Bigg[ \sum_{i=1}^{N} \frac{\partial \mathscr{L}_{(t_i)}}{\partial X}\frac{\mathsf {d}t_i}{\mathsf {d}s} - \frac{1}{N} \frac{\mathsf {d}}{\mathsf {d}s} \Bigg(\sum_{i=1}^{N}\frac{\partial \mathscr{L}_{(t_i)}}{\partial X_{(i)}} + {\sum\limits_{\mathop {i,j = 1}\limits_{i \ne j}}^N} \frac{\partial \mathscr{L}_{(t_i)}}{\partial X_{(j)}}\frac{\mathsf {d} t_i/ \mathsf {d} s}{\mathsf {d} t_j/ \mathsf {d} s}\Bigg) \Bigg] \nonumber \\ 
&&+ \frac{1}{N}{\sum\limits_{\mathop {i,j = 1}\limits_{i \ne j}}^N} \delta Y_{ij} \Bigg[\frac{\partial \mathscr{L}_{(t_i)}}{\partial X_{(i)}} -\frac{\partial \mathscr{L}_{(t_j)}}{\partial X_{(j)}} -\frac{\partial \mathscr{L}_{(t_i)}}{\partial X_{(j)}}\frac{\mathsf {d} t_i/ \mathsf {d} s}{\mathsf {d} t_j/ \mathsf {d} s}+\frac{\partial \mathscr{L}_{(t_j)}}{\partial X_{(i)}}\frac{\mathsf {d} t_j/ \mathsf {d} s}{\mathsf {d} t_i/ \mathsf {d} s} \nonumber \\
&&+{\sum\limits_{\mathop {k = 1}\limits_{k \ne i,j}}^N} \left[\frac{\partial \mathscr{L}_{(t_k)}}{\partial X_{(i)}}\frac{\mathsf {d} t_k/ \mathsf {d} s}{\mathsf {d} t_i/ \mathsf {d} s} - \frac{\partial \mathscr{L}_{(t_k)}}{\partial X_{(j)}}\frac{\mathsf {d}t_k/ \mathsf {d} s}{\mathsf {d} t_j/ \mathsf {d} s} \right]\Bigg] \Bigg\rbrace \nn \\
&&+\delta X \frac{1}{N}\Bigg(\sum_{i=1}^{N}\frac{\partial \mathscr{L}_{(t_i)}}{\partial X_{(i)}} + {\sum\limits_{\mathop {i,j = 1}\limits_{i \ne j}}^N} \frac{\partial \mathscr{L}_{(t_i)}}{\partial X_{(j)}}\frac{\mathsf {d} t_i/ \mathsf {d} s}{\mathsf {d} t_j/ \mathsf {d} s}\Bigg) \Bigg\vert_{s_0}^{s_1}
+ \mathsf {d} s \sum_{i=1}^N  \mathscr{L}_{(t_i)}\frac{\mathsf {d} t_i}{\mathsf {d} s} \Bigg\vert_{s_0}^{s_1} \label{DifACEG2}\;.
\end{eqnarray}
\\
\\
The coefficients of $\delta X$ are generalised Euler-Lagrange equations for Lagrangian 1-form structure \eqref{coneq7} and the coefficients of $\delta Y_{ij}$ are the constraint equations \eqref{coneq8}. Thus, \eqref{DifACEG2} becomes
\begin{eqnarray}
\delta {S}_{E_{\Gamma}} &=& \left[P \delta X  
+ \mathsf {d} s \sum_{i=1}^N  \mathscr{L}_{(t_i)}\frac{\mathsf {d} t_i}{\mathsf {d} s}\right] \Bigg\vert_{s_0}^{s_1} \label{DifACEG3}\;,
\end{eqnarray}
where $P$ is the momentum as given in \eqref{conmomentum}.
In figure \ref{FPN.0}, we find that
\begin{eqnarray} \label{ccc1}
\mathsf{d}X  &=&\delta X +\sum_{i=1}^N  X_{(i)} \frac{\mathsf{d} t_i}{\mathsf{d}s} \mathsf{d}s\;.
\end{eqnarray}
Inserting \eqref{ccc1} into \eqref{DifACEG3}, we obtain 
\begin{eqnarray}
\delta {S}_{E_{\Gamma}} &=& \left[P\mathsf{d}X  - 
 \mathsf {d} s  \sum_{i=1}^N \left[  X_{(i)} P -\mathscr{L}_{(t_i)}\right]\frac{\mathsf {d} t_i}{\mathsf {d} s}\right] \Bigg\vert_{s_0}^{s_1} \label{DifACEG4}\;.
\end{eqnarray}
Since the second term of \eqref{DifACEG4} is the Legendre transformation \eqref{coneq17} , we obtain
\begin{eqnarray}
\delta {S}_{E_{\Gamma}} &=& \left[ P\mathsf{d}X  - 
 \mathsf {d} s  \sum_{i=1}^N \mathscr{H}_{(t_i)}\frac{\mathsf {d} t_i}{\mathsf {d} s}\right] \Bigg\vert_{s_0}^{s_1} \label{DifACEG5}\;.
\end{eqnarray}
The least action principle $\delta {S}_{E_{\Gamma}}=0$ implies that the right-hand side of \eqref{DifACEG5} is a constant quantity defined as 
\begin{eqnarray}\label{CHN}
\mathcal{Q}&\equiv&  P\mathsf{d}X- \mathsf{d} s \sum_{i=1}^N \mathscr{H}_{(t_i)}\frac{\mathsf {d} t_i}{\mathsf {d} s}\;,
\end{eqnarray}
which is called the generalised N\"{o}ether charge.
\\
\\ 
In the case that the system goes under parametised time invariant: $s\rightarrow s+\mathsf{d}s$, the N\"{o}ether charge \eqref{CHN} becomes
\begin{eqnarray}
\mathcal{Q}_{s}&\equiv&  \sum_{i=1}^N \mathscr{H}_{(t_i)}\frac{\mathsf {d} t_i}{\mathsf {d} s} \label{QS}\;.
\end{eqnarray} 
Since any arbitrary curve $\Gamma$ on the space of independent variables can be deformed such that $\Gamma=\sum_{j=1}^N\Gamma_j$, where $\Gamma_j$ is the curve that only $t_j$ is active and $\mathsf{d}t_k / \mathsf{d}s=0$ with $k\neq j$. Then the N\"{o}ether charge \eqref{QS} is just the linear combination of the N\"{o}ether charge $\mathcal{Q}_{t_i}=\mathscr{H}_{(t_i)}$ in all possible time directions. What we have here is a set of the N\"{o}ether charges: \{$\mathcal{Q}_{t_1}, \mathcal{Q}_{t_2},..., \mathcal{Q}_{t_N}$\} which is nothing but a set of Hamiltonians:\{$\mathscr{H}_{(t_1)}, \mathscr{H}_{(t_2)},...,\mathscr{H}_{(t_N)}$\}. Alternative method on deriving N\"{o}ether charges, based on the notion of the variational symmetries \cite{LAMBOOK} for the pluri-Lagrangian structure for one-dimensional systems (a.k.a Lagrangian 1-form) can be found in \cite{LAMS7}.
\section{Conclusion}
We present a recent development for a notion of integrability from the Lagrangian perspective, called Lagrangian 1-form structure which is the simplest case of the Lagrangian multiforms, in both discrete- and continuous-time cases. The variation of the action with respect to dependent variables gives the generalised Euler-Lagrange equations as well as the constraints. While, the variation with respect to independent variables gives the closure relations which guarantees the invariant of the action under the local deformation of curve in the space of independent variables. An important feature for integrability in this context is of course the closure relation which can be considered to be a Lagrangian analogue of the commuting Hamiltonian flows (involution). Furthermore, a set of Lagrangian equations, e.g. Euler-Lagrange equations, constraints and closure relations, forms a compatible system of equations delivering the multidimensional consistency on the level of Lagrangians. One also find that this set of Lagrangian equations can be used to determine the explicit form of the Lagrangians for the integrable one-dimensional many-body systems, e.g. Calogero-Moser and Ruijsenaars-Schneider systems, in discrete-time case \cite{FrankB}. Moreover, we demonstrate that, actually, instead of using the variational principle, one can use the generalised Stokes' theorem as an alternative way to establish the closure relation for both continuous- and, consequently, discrete-time cases as the Lagrangian multiform is required to be a closed form on the solutions of equations of motion. The Lagrangian 1-form must be closed resulting in path independent property of the action with the fixed endpoints on the space of independent variables. This means that there exists a bunch of homotopic paths  which can be continuously transformed to each other. This implies that the space of independent variables must be smooth and simply connected or cannot contain hole-like objects. We also present the Legendre transformation for Lagrangian hierarchy. With this Legendre transformation, the Hamiltonians (invariances) can be obtained directly from the Lagrangians. The action can be rewritten in terms of phase space variables and the variational principle can be considered. The variation of the action with respect to dependent variables gives the generalised Hamilton's equations and the variation with respect to independent variables gives the involution condition leading to the commuting Hamiltonian flows. This means that one can obtain Liouville integrability in the language of variational principle instead of Poisson bracket and Lax pair. We go further on relaxing the local deformation of the curve and the variation of the action results in what we called the generalised N\"{o}ether charge. Of course all invariants (action variables) can be obtained from the N\"{o}ether charge under the translation in their associated angle variables.
\\
\\
The question of how to capture a new notion of integrability, namely multidimensional consistency in quantum level, is natural to be asked. From the Hamiltonian point of view, Liouville integrability, we find that naive transformation from Poisson bracket to quantum bracket is not applicable since the encounter example was purposed by Weigert \cite{SW1}. From the Lagrangian point of view, a recent work has been done by King and Nijhoff \cite{SKNF} for the discrete harmonic oscillator (quadratic Lagrangian) which is obtained from the periodic reduction of the lattice KdV equation. They found that the propagator for loop time path makes no contribution and, with fixed end points, the propagator is path independent. This, of course, can be considered as a direct consequence of the closure relation. However, the question of how can we extend the idea to the case beyond the quadratic Lagrangian, e.g. Calogero-Moser systems \cite{Sikarin1}, is not trivial. Further investigation is needed.
%
%
\begin{acknowledgements}
The authors acknowledge the financial support provided by King Mongkut's University of Technology Thonburi through the ``KMUTT $55^{th}$ Anniversary Commemorative Fund" 
\end{acknowledgements}


\begin{thebibliography}{}
%
\bibitem{ARNO} V. I. Arnold: Mathematical methods of classical mechanics, {\em Springer}, (1978).  
%
%
%
\bibitem{IN1} A. P. Veselov: What is an integral mapping?, {\em Russian Mathematical Surveys}, {\bf 46(5)}, 1-51, (1991).   
%
\bibitem{IN2} P. D. Lax: Integrals of nonlinear equation of evolution and solitary waves, {\em Communications on Pure and Applied Mathematics}, {\bf 21}, 467-490, (1968). 
%
\bibitem{IN3} B. Grammaticos, A. Ramani and V. Papageorgiou: Do integrable mappings have the Painlev\'{e} property?, {\em Physics Review Letters}, {\bf 67}, 1825-1827, (1991).  
%
\bibitem{IN4} M. P. Bellon and C. M. Viallet: Algebraric entropy,  {\em Communications in Mathematical Physics}, {\bf 204(2)}, 425-437, (1999).
%
\bibitem{FrankB} J. Hietarinta, N. Joshi and F. W. Nijhoff: Discrete Systems and Integrability, {\em Cambridge University Press}, (2016).
%
\bibitem{F20011} F. W. Nijhoff and A. J. Walker: The discrete and continuous Painlev\'{e} VI hierarchy and the Garnier systems, {\em Glasgow Mathematical Journal}, {\bf 43A}, 109-123, (2001). 
%
\bibitem{F20012} F. W. Nijhoff, R. Ramani, B. Grammaticos and Y. Ohta: On discrete Painleve Equations associated with the lattice KdV systems and the Painlev\'{e} VI equation, {\em Studies in Applied Mathematics}, {\bf 106(3)}, 261-314, (2001).     
%
\bibitem{F20021}  F. W. Nijhoff: Lax pair for the Adler (lattice Krichever-Novikov) system, {\em Physics Letters A}, {\bf 297(1-2)}, 49-58, (2002).
%
\bibitem{RB} R. Boll: On multidimensional consistent systems of asymmetric quad-equations, arXiv:1201.1203v1, (2012).
%
\bibitem{RB2} R. Boll and Yu. B. Suris: On the Lagrangian structure of 3D consistent systems of asymmetric quad-equations, {\em Journal of Physics A: Mathematical and Theoretical}, {\bf 45(11)}, 115-201, (2012). 
%
\bibitem{SF1} S. B. Lobb and F. W. Nijhoff: Lagrangian multiforms and multidimensional consistency, {\em Journal of Physics A: Mathematical and Theoretical}, {\bf 42(45)}, (2009). 
%
\bibitem{FSU1} V. E. Adler, A. I. Bobenko and Yu. B. Suris: Classification of integrable equations on quad-graphs. The consistency approach, {\em Communications in Mathematical Physics}, {\bf 233(3)}, 513-543, (2003).
%
%
%
%
%
%
%
%
\bibitem{Sikarin1} S. Yoo-Kong, S. B. Lobb and F. W. Nijhoff: Discrete-time Calogero-Moser system and Lagrangian 1-form structure, {\em Journal of Physics A: Mathematical and Theoretical}, {\bf 44(36)}, (2011).
%
\bibitem{Sikarin2} S. Yoo-Kong and F. W. Nijhoff: Discrete-time Ruijsenaars-Schneider system and Lagrangian 1-form structure, arXiv:1112.4576v2 {nlin.SI}, (2013).
%
\bibitem{LAMS3} Yu. B. Suris: Variational formulation of commuting Hamiltonian flows: multi-time Lagrangian 1-forms, {\em Journal Geometric Mechanics}, {\bf 5(3)}, 365-379, (2013).
%
\bibitem{LAMS4} R. Boll, M. Petrera and Yu. B. Suris: Multi-time Lagrangian 1-forms for families of B\"{a}cklund transformations Toda-type systems, {\em Journal of Physics A: Mathematical and Theoretical}, {\bf 46(27)}, (2013).
%
\bibitem{LAMS5} R. Boll, M. Petrera and Yu. B. Suris: Multi-time Lagrangian 1-forms for families of B\"{a}cklund transformations Relativistic Toda-type systems, {\em Journal of Physics A: Mathematical and Theoretical}, {\bf 48(8)}, (2015).
%
\bibitem{Umpon} U. Jairuk, S. Yoo-Kong and M. Tanasittikosol: On the Lagrangian structure of Calogero's Goldfish model, {\em Theoretical and Mathematical Physics}, {\bf 183(2)},  665-683, (2013).
%
\bibitem{Umpon1} U. Jairuk, S. Yoo-Kong and M. Tanasittikosol: On the Lagrangian structure of the Hyperbolic Calogero-Moser System, {\em Reports on Mathematical Physics}, {\bf 79(3)}, 299-330, (2017).
%
\bibitem{Umpon2} U. Jairuk: Lagrangian 1-form structure for Calogero-Moser type system, PhD. Thesis, King Mongkut's University of Technology Thonburi, (2016).
%
\bibitem{LAMS6} Yu. B. Suris: Variational symmetries and pluri-Lagrangian systems, {\em Dynamical Systems, Number Theory and Applications A Festschrift in Honor of Armin Leutbecher’s 80th Birthday, Eds. Th. Hagen,
F. Rupp and J. Scheurle, World Scientific}, 255-266, (2016).
%
\bibitem{LAMS7} M. Petrera and Yu. B. Suris: Variational symmetries and pluri-Lagrangian systems in classical mechanics, {\em Journal of Nonlinear Mathematical Physics}, {\bf 24}, 121-145, (2017)
%
\bibitem{VES1} A. P. Veselov: What is an integral mapping? in What is Integrability? (ed V. E. Zakharov), {\em Springer-Verlag}, 251-272, (1991).
%
\bibitem{BRST} M. Bruschi, O. Ragnisco, P. M. Santini, and G.Z. Tu: Integrable symplectic maps, {\em Physica D Nonlinear Phenomena}, {\bf 49(3)}, 273-294, (1991).
%
\bibitem{CF1} F. Calogero: Solution of a three-body problem in one dimension, {\em Journal of Mathematical Physics}, {\bf 10(12)}, 2791-2196, (1969).  
%
\bibitem{JM1} J. Moser : Three integrable Hamiltonian systems connected with isospectral deformations, {\em Advances in Mathematics}, {\bf 16(2)}, 197-220, (1975).  
%
\bibitem{F} F. Calogero : On a technique to identify solvable discrete-time many-body problems, {\em Theoretical and Mathematical Physics}, {\bf 172(2)}, 1052-1072, (2012).
%
\bibitem{LMCM1} H. Jino, M. Wadati and K. Hikami: The Quantum Calogero-Moser Model: Algebraic Structures, {\em Journal of the Physical Society of Japan}, {\bf 62(9)}, 3035-3043, (1993).
%
\bibitem{NP} F. W. Nijhoff and G. D. Pang: A time-discretized version of the Calogero-Moser model, {\em Physics Letters}, {\bf191A}, 101-107, (1994).
%
\bibitem{Sikarin3} S. Yoo-Kong: Calogero-Moser type systems,
associated KP systems, and Lagrangian structures, PhD. Thesis, University of Leeds, (2011).
%
\bibitem{Fortney} J. P. Fortney: A Visual Introduction to Differential Forms and Calculus on Manifolds, {\em Birkh\"{a}user}, 337, (2018).
%
\bibitem{dd} W. Rudin: Real and Complex Analysis, {\em McGraw-HiII Book Company}, 218-222, (1987).
%
\bibitem{Babelon} O. Babelon, D Bernard and M. Talon: Introduction to Classical Integrable Systems, {\em Cambridge University Press}, 11-17, (2003).
%
\bibitem{LAMBOOK} P. J. Olver: Applications of Lie Groups to Differential Equation, Graduate Texts in Mathematics, 2nd Edition, {\em Spinger}, 288-368, (1993).
%
\bibitem{SW1} S. Weigert: The problem of quantum integrability, {\em Physica D: Nonlinear Phenomena}, {\bf 56(1)}, 107-109, (1992).
%
\bibitem{SKNF} S. D. King and F. W. Nijhoff: Quantum Variational Principle and quantum multiform structure: the case of quadratic Lagrangians, arXiv:1702.08709, (2017).
%
\end{thebibliography}
\end{document}